\documentclass[12pt]{article}
\pdfoutput=1
\newcommand{\bxi}{\boldsymbol{\xi}}
\DeclareMathAlphabet{\bi}{OML}{cmm}{b}{it}

\newcommand{\rmd}{\mathrm{d}}

\usepackage{authblk} % author and affiliation blocks
\usepackage{amsmath}
\usepackage{amsthm, amsfonts}
\usepackage{mathrsfs, mathtools}

\usepackage[table,xcdraw]{xcolor} % For coloring
\usepackage{booktabs}

\usepackage{siunitx} % Used for SI units
\usepackage{graphicx} % For figures
\usepackage{float}

\usepackage{subcaption}
\usepackage{setspace}
\usepackage[a4paper, margin=1in]{geometry}

\usepackage[comma]{natbib}

\usepackage{url}

\newcommand{\mycomment}[1]{}

\title{\LARGE{\textbf{Probabilistic Proton Treatment Planning: a novel approach for optimizing underdosage and overdosage probabilities of target and organ structures}}}

\author{Jelte R. de Jong$^1$, Sebastiaan Breedveld$^3$,
Steven J. M. Habraken$^{2,4}$, Mischa S. Hoogeman$^{2,3}$, Danny Lathouwers$^1$ and Zoltán Perkó}

\affil[1]{\footnotesize{Delft University of Technology, Department of Radiation, Science and Technology, Delft, The Netherlands}}
\affil[2]{\footnotesize{Holland Proton Therapy Centre Delft, Delft, the Netherlands}}
\affil[3]{\footnotesize{Erasmus MC Cancer Institute, University Medical Center Rotterdam, Department of Radiotherapy, Rotterdam, the Netherlands}}
\affil[4]{\footnotesize{Leiden University Medical Center, Department of Radiation Oncology, Leiden, the Netherlands}}

\date{\small{\today}}

\begin{document}

\bibliographystyle{dcu}
\def\newblock{\ }%
\setcitestyle{authoryear,open={(},close={)}}

\maketitle

\begin{abstract}
Uncertainties in treatment planning are typically managed using either margin-based or robust optimization. Margin-based methods expand the clinical target volume (CTV) towards a planning target volume (PTV), which is generally unsuited for proton therapy. Robust optimization considers worst-case scenarios, but its quality depends on the chosen \textit{uncertainty (scenario) set}: excluding extremes reduces robustness, while including too many make plans overly conservative. Probabilistic optimization overcomes these limitations by modeling a continuous scenario distribution, enabling the use of statistical measures.

We propose a novel approach to probabilistic optimization that steers plans towards individualized probability levels, to control CTV and organs-at-risk (OARs) under- and overdosage. Voxel-wise dose percentiles ($d$) are estimated by expected value ($\mathbb{E}$) and standard deviation (SD) as $\mathbb{E}[d] \pm \delta \cdot SD[d]$, where $\delta$ is iteratively tuned to match the target percentile of the underlying probability distribution (given setup and range uncertainties). The approach involves an inner optimization of $\mathbb{E}[d] \pm \delta \cdot SD[d]$ for fixed $\delta$, and an outer optimization loop that updates $\delta$. Polynomial Chaos Expansion (PCE) provides accurate and efficient dose estimates during optimization. We validated the method on a spherical CTV (prescribed \qty{60}{\gray}) abutted by an OAR in different directions and a horseshoe-shaped CTV surrounding a cylindrical spine, under Gaussian-distributed setup (\qty{3}{\milli\meter}) and range (3\%) uncertainties.

For spherical cases with similar CTV coverage, $P(D_{2\%} > \SI{30}{\gray})$ dropped by 10-15\%; for matched OAR dose, $P(D_{98\%} > \qty{57}{\gray})$ increased by 67.5-71\%. In spinal plans, $P(D_{98\%} > \qty{57}{\gray})$ increased by 10-15\% while $P(D_{2\%} > \qty{30}{\gray})$ dropped by 24-28\% in the same plan. Probabilistic and robust optimization times were comparable for spherical (hours) but longer for spinal cases (7.5 - 11.5 \unit{\hour} vs. 9–20 \unit{\minute}). 

Compared to discrete scenario-based optimization, the probabilistic approach offered better OAR sparing or target coverage, depending on individualized priorities.

\end{abstract}

%\vspace{2pc}

\section{Introduction} \label{sec:Introduction}

Intensity-Modulated Proton Therapy (IMPT) has demonstrated improved sparing of organs at risk (OARs) and normal tissue compared to intensity-modulated radiation therapy (IMRT) for various treatment sites \citep{vandeSande2016,Nguyen2021,Stuschke2012}. However, the precise dose delivery of IMPT makes it more sensitive to uncertainties that may occur during treatment. Uncertainties include patient setup misalignment, anatomical changes during the treatment, or range uncertainties due to CT to stopping power conversions \citep{Lomax2008,Schaffner1998}. To ensure an effective treatment, it is crucial that treatment plans are robust against such uncertainties. \\

In photon therapy, uncertainties are typically managed using planning target volume (PTV) margins around the CTV \citep{vanHerk2000}. However, as the proton dose distribution is more sensitive to geometrical shifts, applying PTV margins is often ineffective in IMPT. Robust optimization offers an alternative by considering a predefined set of uncertainty scenarios, in practice corresponding to errors of a fixed magnitude. Multiple robust approaches exist \citep{Unkelbach2018}, but most commonly used is mini-max robust optimization \citep{Fredriksson2011}. It treats all scenarios in the uncertainty set as equally important and, for each iteration, optimizes for the worst-case among the scenario set. Depending on the desired conservativeness of the plan, one can choose to optimize using voxel-wise \citep{Pflugfelder2008,Liu2012}, objective-wise \citep{Chen2012} or composite-wise \citep{Janson2024,vanDijk2016} worst-case objectives. While effective with a well-defined uncertainty set, robust optimization can lead to overly conservative plans if extreme scenarios dominate or to insufficient robustness if the uncertainty set is too narrow \citep{vanderVoort2016}. Unlike mini-max robust optimization, stochastic programming \citep{Unkelbach2007,Unkelbach2008} assigns a probability weight to each scenario in the set. However, as the number of scenarios is still limited in this approach, it is not straightforward how the probability weights can be effectively assigned to a single scenario. \\

\begin{figure}[]
    \centering
    \includegraphics[width=0.8\textwidth]{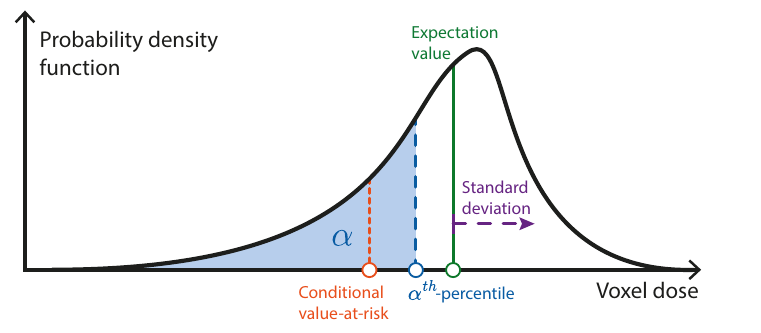}
    \caption{Illustration of the probability density function of the voxel dose, being a result of the statistical nature of the considered uncertainties. Shown are the expectation value (solid green), standard deviation (purple arrow) and the $\alpha^{th}$-percentile of the voxel dose (dotted blue), which is the dose value for which $\alpha \%$ of the error scenarios (area under the curve) result in lower dose values. The conditional value-at-risk (dotted orange) corresponding to probability $\alpha$ is the mean of the lowest $\alpha \%$ of scenarios. This example is used to control target underdosage probability, while equivalent representations can be made for (target and OAR) overdosage probability.}
    %Dose threshold $\gamma$ (gray) is a reference level that is intended to be met or exceeded by the $\alpha^{th}$-percentile after optimization. 
    \label{fig:percentileIllustration}
\end{figure}

% Gordon2010: photon. Mescher extended dose coverage to constraints \cite{Mescher2017}. Tilly2019 for photons. Chan2014: IMRT in IMPT.
Probabilistic optimization presents a promising alternative by defining uncertainty as a distribution, so that it simultaneously accounts for a full spectrum of possible errors. As quantities of interest (e.g., voxel dose) depend on the uncertainties, they become random variables as well, whose statistical effect can be presented by a probability density function (PDF), as shown in Figure \ref{fig:percentileIllustration}. The impact of uncertainty can be quantified by stochastic metrics (e.g., using expectation values, variances and percentiles), which provide consistent and statistically interpretable results, thereby helping to reduce inter-patient variation \citep{RojoSantiago2023_2,deJongJI_2025submission}. The $\alpha^{th}$-percentile is the dose level such that $\alpha \%$ of the scenarios result in lower dose values. For small or large $\alpha$, the percentile allows to quantify the most extreme scenarios, which in the context of radiotherapy can be used to respectively minimize for under- and overdosage probability. The Conditional Value-at-Risk (CVaR) can be used alternatively, which quantifies the average of the worst $\alpha \%$ of scenarios. CVaR has been applied before in photon radiotherapy \citep{Tilly2019}, IMRT \citep{Chan2014} and IMPT \citep{An2017}, using a discrete number of error scenarios.

The calculation of percentiles and other probabilistic metrics require the evaluation of the PDF of the response of interest (e.g., voxel dose). This is challenging because these PDFs typically do not have an analytical form. A common approximation is to assume that response PDFs have Gaussian-distributed tails, which is reasonable for e.g., the total dose in treatments involving a sufficient number of fractions \citep{Sobotta2010}, but may be inadequate for other metrics. \citet{Chu2005} approximated voxel-wise dose percentiles by expectation value ($\mathbb{E}$) and standard deviation ($\text{SD}$) as $\mathbb{E} \pm \delta \cdot \text{SD}$, where $\delta$ is a constant that quantifies the displacement from the mean in units of standard deviation. \citet{Fabiano2022} optimized for biological effective dose (BED) using $\delta = 2$, estimating the $2^{nd}$ and $98^{th}$ percentiles as $\mathbb{E} - 2 \text{SD}$ and $\mathbb{E} + 2 \text{SD}$, respectively. These correspond to the dose values below which 2\% (or 98\%) of the scenarios have lower values (if the response PDF would be truly Gaussian distributed). Although the Gaussian approximation potentially allows the PDF to be steered into the desired direction \citep{Sobotta2010}, these methods do not allow to tune exactly for desired probability levels and dose thresholds. In fact, distribution shapes may be complex in practice and there is no guarantee that they are even near Gaussian distributed. \\ 

Besides voxel-wise probabilistic optimization, other studies have been done on target coverage objectives \citep{Gordon2010} and constraints \citep{Mescher2017} for photon plans, e.g., requiring that 90\% of the scenarios lead to a near-minimum dose (i.e., $D_{98\%}$) greater than 95\% of the prescribed dose. These approaches use approximate DVH penalties \citep{Wu2000}, only including voxels within a rim around the CTV. This rim was defined using the Van Herk margin recipe (VHMR) \citep{vanHerk2000}, and dose-volume histograms (DVHs) were sampled under the static dose cloud approximation. Such margin recipes and the static dose cloud approximation do not hold in general for proton therapy, limiting the applicability of these approximations in proton plans. \\

These limitations motivate the need for a flexible probabilistic approach that can be applied to proton therapy as well, which allows us to optimize for individualized probabilities of under- and overdosage, regardless of the PDF shape. Accurate probabilities are obtained by sampling thousands of error scenarios from the Gaussian distribution, resulting in the same number of dose distributions. As it is computationally expensive to perform these calculations by Monte-Carlo sampling, we use Polynomial Chaos Expansion (PCE). It has proven to be an accurate meta-model of the dose-engine, also for probabilistic evaluation of treatment plans \citep{Perk2016,RojoSantiago2023}. Once constructed, PCE facilitates the efficient sampling of dose distributions in various error scenarios.

The probabilistic approach consists of an inner optimization and an outer optimization loop. The inner part optimizes the pencil-beam weights for a fixed set of $\delta$-factors (as done by \citet{Fabiano2022}). Qualitatively, the $\delta$-factor is a bridge between the probability levels and dose thresholds, even when the underlying distribution is non-Gaussian. In other words, a probabilistic goal (e.g., at most 10\% probability of underdosing) is translated into a specific dose threshold used in the optimization. The outer optimization loop updates the $\delta$-factors to improve the $\mathbb{E} \pm \delta \cdot \text{SD}$ approximation, effectively rescaling it based on the updated PDF. We use a voxel-wise objective so that spatial information of the regions of interest (ROIs) remains. 

The approach eliminates the optimization of nonphysical distributions, which is particularly the case for voxel-wise mini-max optimized plans \citep{McGowan2013}. Regarding plan evaluation, often worst-case evaluation metrics (e.g., the $D_{98\%}$ of the voxel-wise minimum) are used \citep{Korevaar2019}, which lack statistical insight into the plan quality \citep{Park2013,Sterpin2021}. Here, as we optimize for statistical objectives, the step towards plan evaluation with statistically meaningful metrics is a straightforward consequence. \\

This paper presents a proof-of-principle of a novel probabilistic approach to probabilistic treatment planning for systematic setup and range errors. In Section \ref{sec:Methods} we discuss the PCE method that is used for the scenario sampling, the probabilistic approach and the (homogeneous) phantom geometries used: a simple spherical CTV with surrounding OARs, and a more complex horseshoe-shaped CTV around a cylindrical spine. In Section \ref{sec:Results}, probabilistic VHMR equivalence with the probabilistic approach is validated for setup errors in a spherical CTV, after which comparisons to composite-wise mini-max robust plans are done, matching either CTV coverage or OAR sparing. In Section \ref{sec:Discussion} and Section \ref{sec:Conclusions}, a discussion and conclusion on the results are respectively presented.

\section{Methods and Materials} \label{sec:Methods}

\subsection{Polynomial Chaos Expansion} \label{subsec:Methods_PCE}
In this work, we consider Gaussian-distributed setup (in $x$ and $y$) and range errors (denoted as $\bxi$) with respective standard deviations of $\sigma_{setup} = \qty{3}{\mm}$ and $\sigma_{range} = 3\%$. The Gaussian distribution is truncated such that the combined setup and range errors is cut at the 99\% confidence level to include the majority of the uncertainty space used for sampling. Setup errors are modeled by shifting all pencil-beam spots with respect to the dose distribution in the shift-direction. Range errors are modeled by scaling all pencil-beam spots in the beam direction (i.e., the $\pm z$-direction) with the relative range uncertainty.

For the uncertainty quantification of the response, Polynomial Chaos \citep{Wiener1938} is used. The response is estimated as a Polynomial Chaos Expansion (PCE) that is a function of the $N$ uncertainty variables, in this work $N=3$. As we assume these to be independent and Gaussian-distributed, their joint PDF $p(\bxi)$ can be described as the product of the one-dimensional PDFs, such that $p(\bxi) = \prod_{j=1}^N p(\xi_j)$. The PCE is an expansion using multi-dimensional basis vectors $\Psi_k(\bxi)$, where the PCE of response $R(\bxi)$ is defined as
\begin{equation} \label{eq:PCE_truncated}
     R(\bxi) = \sum_{k=0}^{P} r_k \Psi_k(\bxi),
\end{equation}
where $r_k$ are the polynomial coefficients, such that $P + 1$ basis vectors are used. The type of the basis vector is chosen based on the uncertainty distribution. For Gaussian input variables, the Wiener-Askey scheme \citep{Xiu2002} proposes to use probabilists' Hermite polynomials $He_{\gamma_{k,j}}(\xi_j)$, where $\gamma_{k,j} = (\gamma_{k,1}, \ldots, \gamma_{k,N} )$ denotes the polynomial order of the $j^{th}$ polynomial corresponding to basis vector $k$. The multi-dimensional basis vectors are thus given by $\Psi_k(\bxi) = \prod_{j=1}^N He_{\gamma_{k,j}}(\xi_j)$. The PCE corresponding to order $O$ with a full basis set is defined by including the multi-dimensional polynomials for which $\sum_{j=1}^N \gamma_{k,j} \leq O$. As a result, the PCE in Equation \ref{eq:PCE_truncated} has $P + 1 = (N + O)!/(N!O!)$ basis vectors. In this work the full basis set is used.

Constructing the PCE comes down to determining the polynomial coefficients $r_k$. In this work we use spectral projection, such that
\begin{equation}
    r_k = \frac{\bigl \langle R(\bxi) \Psi_k(\bxi) \bigr \rangle}{\bigl \langle \Psi_k(\bxi) \Psi_k(\bxi) \bigr \rangle} = \frac{\int R(\bxi) \Psi_k(\bxi) p(\bxi) d\bxi }{\int \Psi_k(\bxi) \Psi_k(\bxi) p(\bxi) d\bxi} = \frac{1}{h_k^2} \int R(\bxi) \Psi_k(\bxi) p(\bxi) d\bxi,
\end{equation}
where $\bigl \langle \cdot \bigr \rangle$ denotes the inner product and $h_k^2 = \bigl \langle \Psi_k(\bxi) \Psi_k(\bxi) \bigr \rangle$ is the norm of basis vector $k$. The integral in the nominator is determined by Gauss-Hermite cubature, for which a defined set of cubature points $\bxi_l$ is used with corresponding weights $w_l$ to yield 
\begin{equation} \label{eq:cubature}
    \int R(\bxi) \Psi_k(\bxi) p(\bxi)d\bxi=\sum_lR(\bxi_l) \Psi_k(\bxi_l) p(\bxi_l)w_l.
\end{equation}
For this computation, the exact response $R(\bxi_l)$ only has to be calculated on these cubature points, which is done using the dose engine. Instead of using full cubature grids we use Smolyak sparse grids \citep{smolyak}, in which higher-order cubature points that simultaneously occur in multiple dimensions are neglected, essentially reducing the number of terms in Equation \ref{eq:cubature} and therefore the number of necessary dose computations without compromising accuracy. A more advanced form is to use extended Smolyak sparse grids, where the grid level along the single dimensions is increased by $lev_{extra}$ levels, as often a significant increase in PCE accuracy can be obtained by only limited number of extra calculations. To reduce memory cost, we neglect voxels that have a dose lower than \qty{0.01}{\gray} in all scenarios that are displaced $3 \sigma$ from the nominal scenario along the principal axes. Further details about PCE construction and numerical integration with sparse grids can be found in the supplementary material of \citep{Perk2016}.

Through PCE construction, we obtain a meta-model of the exact response that can be used for efficient sampling. Moreover, obtaining the first two moments from Equation \ref{eq:PCE_truncated} is computationally simple. The mean of the response $\mu_R$ is equal to the zeroth polynomial coefficient $r_0$ and its variance is $\sigma_R^2 = \sum_{k=1}^{\infty} r_k^2 h_k^2 \approx \sum_{k=1}^P r_k^2 h_k^2$. These metrics become useful in the probabilistic evaluation of the treatment plans.

\subsection{The probabilistic approach} \label{subsec:Methods_optimizationScheme}
\begin{figure}[]
    \centering
    \includegraphics[width=\textwidth]{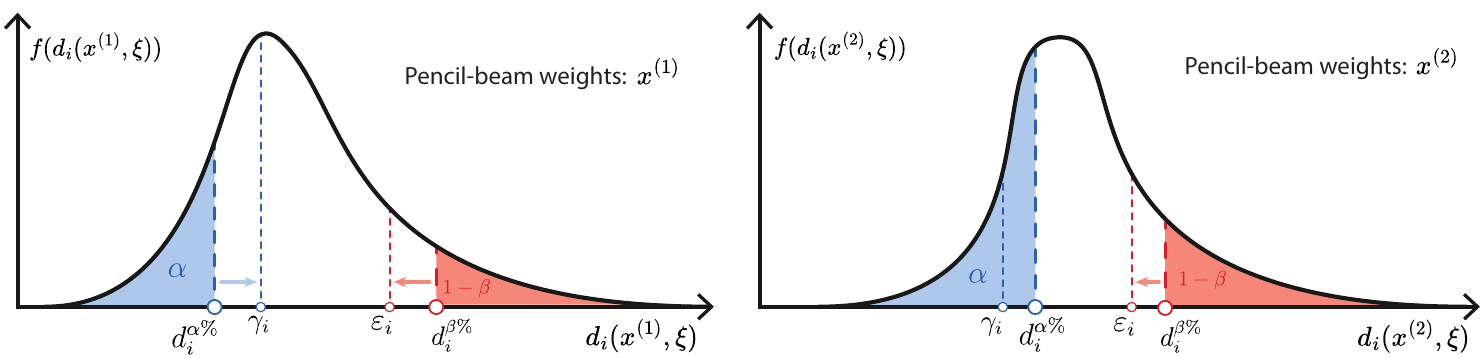}
    \caption{Probability density functions $f$ of the voxel dose $d_i(\bi{x}^{(k)},\bxi)$ for two different pencil-beam weights (e.g., at iterations $k$) during the optimization. Target underdosage probability is optimized for $P(d_i(\bi{x},\bxi) \leq \gamma_i) \leq \alpha$, or equivalently $d_i^{\alpha\%}(\bi{x}) \geq \gamma_i$. Target and OAR overdosage can be optimized by $P(d_i(\bi{x},\bxi) \leq \epsilon_i) \geq \beta$, i.e., $d_i^{\beta\%}(\bi{x}) \leq \epsilon_i$. The PDF shape has changed for iteration 2, resulting in sufficient target coverage in this voxel ($d_i^{\alpha\%}(\bi{x}^{(2)}) > \gamma_i$).}
    \label{fig:percentileDescription}
\end{figure}

Underdosage and overdosage probability can be quantified using percentiles, which define the worst voxel dose scenarios that may occur during a treatment. In the context of underdosing the CTV, we specifically aim to limit the probability (or in simpler terms, the fraction of scenarios) $\alpha$ where the dose falls below a voxel dose threshold $\gamma_i$, such that 
\begin{equation}
\label{eq:Methods_probObjective}
    P(d_i(\bi{x},\bxi) \leq \gamma_i) \leq \alpha \quad \forall i\in CTV.
\end{equation}
For example, one may aim for the voxel dose to fall below $\gamma_i = 0.95 \cdot d_i^p$ (i.e., underdosing) in at most $\alpha = 10\%$ of the error scenarios. In this work, we aim to reformulate Equation \ref{eq:Methods_probObjective} into an objective, such that dose threshold $\gamma_i$ and probability $\alpha$ are quantities that can be controlled as part of the objective in our optimization approach. For this purpose, we define the $\alpha^{th}$-percentile of the voxel dose PDF as $d_i^{\alpha \%}(\bi{x})$. Its definition is such that $\alpha \%$ of the error scenarios lead to voxel doses smaller than $d_i^{\alpha \%}(\bi{x})$, such that
\begin{align} \label{eq:percentile}
    \alpha = P(d_i(\bi{x},\bxi) \leq d_i^{\alpha}(\bi{x})) = \int_{-\infty}^{d_i^{\alpha}(\bi{x})} f(d_i(\bi{x},\bxi)) \rmd(d_i(\bi{x},\bxi)),
\end{align}
where the voxel dose PDF $f(d_i(\bi{x},\bxi))$ is analytically unknown in general. Substituting Equation \ref{eq:percentile} into Equation \ref{eq:Methods_probObjective} for voxel $i$ gives
\begin{equation}
    P(d_i(\bi{x},\bxi) \leq \gamma_i) \leq \alpha = P(d_i(\bi{x},\bxi) \leq d_i^{\alpha}(\bi{x})),
\end{equation}
which holds if $\alpha$ is such that $d_i^{\alpha}(\bi{x}) \geq \gamma_i$. This implies that reducing $P(d_i(\bi{x},\bxi) \leq \gamma_i)$ below $\alpha$ can be achieved by increasing $d_i^{\alpha}(\bi{x})$ above $\gamma_i$. This is visualized in Figure \ref{fig:percentileDescription}, where the PDFs corresponding to two example pencil-beam weight vectors ($\bi{x}^{(1)}$ and $\bi{x}^{(2)}$) during the optimization are illustrated. For pencil-beam weights $\bi{x}^{(2)}$, the target underdosage is lower than the desired level ($d_i^{\alpha}(\bi{x}) > \gamma_i$), so $P(d_i(\bi{x},\bxi) \leq \gamma_i) < \alpha$. 

Similarly, for overdosing structure $\Sigma$ (e.g., for CTV or OAR), we aim to limit the probability $1 - \beta$ where the dose exceeds voxel dose threshold $\epsilon_i$, such that $P(d_i(\bi{x},\bxi) \leq \epsilon_i) \geq \beta \quad \forall i \in \Sigma$. Equivalently, we use $d_i^{\beta \%}(\bi{x}) \leq \epsilon_i$. In this context, one may aim for the voxel dose to exceed $\epsilon_i = 1.07 \cdot d_i^p$ (i.e., overdosing) in at most $1 - \beta = 10\%$ of the error scenarios. Or to put it differently, we aim for the voxel dose to fall below $\epsilon_i = 1.07 \cdot d_i^p$ in at least $\beta = 90\%$ of the error scenarios. \\

% Explain how to determine the percentile by PCE sampling.
In order to determine the voxel dose percentiles, $d_i(\bi{x},\bxi)$ must be computed in a large number of uncertainty scenarios. The voxel dose $d_i(\bi{x},\bxi) = \sum_{j \in \mathbb{B}} D_{ij}(\bxi) x_j$ is determined by summing over all physical proton pencil-beams ($j \in \mathbb{B}$) with corresponding intensity $x_j$ (i.e., pencil-beam weights, also referred to as beam weights). Each pencil-beam’s Bragg peak is positioned at a predefined spot position in the pencil-beam grid. Its contribution to each voxel $i \in \mathbb{V}$ is defined by the dose-influence matrix $D_{ij}(\bxi)$. A PCE of the dose-influence matrix is approximated as
\begin{equation} \label{eq:PCE_Dij}
    D_{ij}(\bxi) \approx \sum_{k=0}^{P} R_{ij}^{(k)} \Psi_k(\bxi),
\end{equation}
with $P + 1$ number of basis vectors, $k^{th}$ coefficient $R_{ij}^{(k)}$ and multi-dimensional Hermite basis vector $\Psi_k(\bxi)$. Sampling directly from Equation \ref{eq:PCE_Dij} is possible, but requires the construction of maximally $N_v \cdot N_b$ (the number of $D_{ij}(\bxi)$ elements) PCEs. The number of needed PCEs can be reduced to only $N_v$ (the number of $d_i(\bi{x},\bxi)$ elements) by converting Equation \ref{eq:PCE_Dij} to a PCE of the voxel dose $d_i(\bi{x},\bxi)$ for all voxels $i=1,...,N_v$ as
\begin{align}\label{eq:Methods_PCEdose}
    d_i(\bi{x},\bxi) & = \sum_{j \in \mathbb{B}} D_{ij}(\bxi) x_j \approx \sum_{j \in \mathbb{B}} \left( \sum_{k=0}^{P} R_{ij}^{(k)} \Psi_k(\bxi) \right) x_j \\
    & = \sum_{k=0}^{P} \left( \sum_{j \in \mathbb{B}} R_{ij}^{(k)} x_j \right) \Psi_k(\bxi) = \sum_{k=0}^{P} q_{i}^{(k)} \Psi_k(\bxi),
\end{align}
where the PCE coefficients of the voxel dose $q_{i}^{(k)} = \sum_{j \in \mathbb{B}} R_{ij}^{(k)} x_j$ can be obtained from the PCE coefficients of $D_{ij}$. Since Equation \ref{eq:PCE_Dij} is independent of the beam weights, constructing it once before the optimization is sufficient (as opposed to constructing the voxel dose PCE for each iteration). The fact that PCE can be used to quickly sample a response of interest, opens the possibility to incorporate percentiles in the optimization.

\begin{figure}[]
    \centering
    \includegraphics[width=0.7\textwidth]{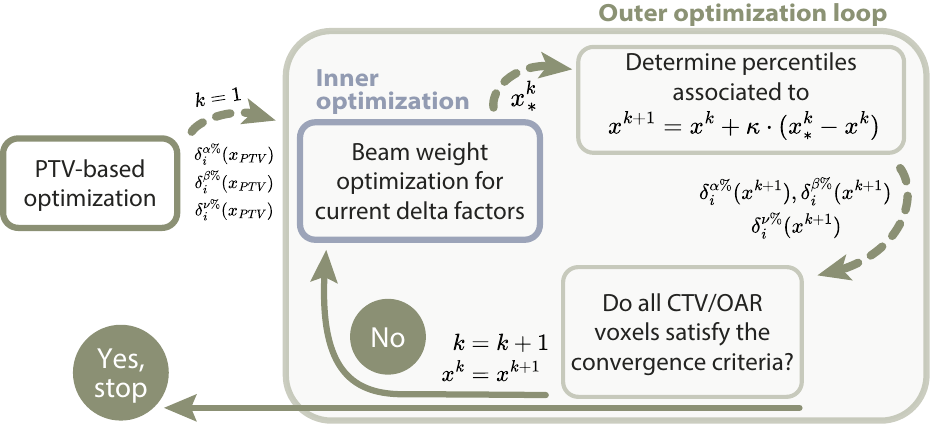}
    \caption{The proposed probabilistic optimization approach has a nested structure: the inner optimization optimizes pencil-beam weights, while the outer optimization loop ensures convergence to desired probability levels.}
    \label{fig:ProbMethodology}
\end{figure}

The remainder of Section \ref{subsec:Methods_optimizationScheme} introduces the probabilistic approach. The corresponding optimization scheme is a nested structure consisting of an inner optimization and outer optimization loop, as is illustrated in Figure \ref{fig:ProbMethodology}. The inner optimization serves as an optimizer for the beam weights by optimizing using a given percentile estimate. The outer optimization loop makes sure that the percentile estimate remains accurate during the optimization.

\subsubsection{The inner optimization}
\label{subsubsec:innerOptimization}
The inner optimization focuses on optimizing the beam weights for a given percentile estimate $d_i^{\alpha\%}(\bi{x})$. We choose to optimize the CTV underdosage probability by defining a quadratic underdose penalty as
\begin{equation} \label{eq:Methods_CTVobj}
    f_{CTV}^{\alpha,\gamma}(\bi{x}) = \frac{1}{N_{CTV}} \sum_{i \in CTV} w_i^{CTV} \bigl[ \gamma_i - d_i^{\alpha \%}(\bi{x}) \bigr]_+^2,
\end{equation}
where $w_i^{CTV}$ is the CTV voxel weight and $N_{CTV}$ is the number of CTV voxels. We use $[ h ]_+ = \max\{ 0, h \}$ to only penalize voxels if $h \geq 0$, meaning that only underdosed CTV voxels are penalized ($d_i^{\alpha \%}(\bi{x}) < \gamma_i$).

In order to use an analytical gradient and Hessian of the objective, we rewrite $d_i^{\alpha \%}(\bi{x})$ using the expectation value $\mathbb{E}[d_i(\bi{x}, \bxi)]$ and standard deviation $SD[d_i(\bi{x}, \bxi)]$ of the voxel dose \citep{Fabiano2022}, as
\begin{align}
d_i^{\alpha \%}(\bi{x}) & = \mathbb{E}[d_i(\bi{x},\bxi)] - \delta_i^{\alpha \%}(\bi{x}) SD[d_i(\bi{x},\bxi)] \nonumber \\ 
& = \mathbb{E}[d_i(\bi{x},\bxi)] - \delta_i^{\alpha \%}(\bi{x})  \Bigl[\mathbb{E}[d_i^2(\bi{x},\bxi)] - \mathbb{E}^2[d_i(\bi{x},\bxi)] \Bigr]^{1/2}, \label{eq:Methods_objectiveFunctions_percentileEstimate1}
\end{align}
where $\delta_i^{\alpha \%}(\bi{x}) \in \mathbb{R}$ (also referred to as $\delta$-factor) defines the number of standard deviations the percentile $d_i^{\alpha \%}(\bi{x})$ is displaced from the expectation value $\mathbb{E}[d_i(\bi{x}, \bxi)]$. The $\delta$-factor is determined by using the accurate percentile $d_i^{\alpha \%}(\bi{x})$ obtained from the PCE as
\begin{equation} \label{eq:Methods_deltaAlpha}
    \delta_i^{\alpha \%}(\bi{x}) = \frac{ \mathbb{E}[d_i(\bi{x}, \bxi)] - d_i^{\alpha \%}(\bi{x}) }{SD[d_i(\bi{x}, \bxi)]}.
\end{equation}
Similarly, quadratic overdose penalties for CTV ($d_i^{\beta} > \epsilon_i$) and OAR $d_i^{\nu} > \mu_i$) voxels are defined as
\begin{align}
    f_{CTV}^{\beta,\epsilon}(\bi{x}) & = \frac{1}{N_{CTV}} \sum_{i \in CTV} w_i^{CTV} \bigl[ d_i^{\beta \%}(\bi{x}) - \epsilon_i \bigr]_+^2,  \label{eq:Methods_CTVobj_overdosage} \\
    f_{OAR}^{\nu,\mu}(\bi{x}) & = \frac{1}{N_{OAR}} \sum_{i \in OAR} w_i^{OAR} \bigl[ d_i^{\nu \%}(\bi{x}) - \mu_i \bigr]_+^2, \label{eq:Methods_OARobj_overdosage}
\end{align}
where $w_i^{OAR}$ is the OAR voxel weight and $N_{OAR}$ is the number of OAR voxels. The $\delta$-factors are similarly calculated assuming that the $\beta^{th}$-percentile and $\nu^{th}$-percentiles are written as
\begin{align}
    d_i^{\beta \%}(\bi{x}) & = \mathbb{E}[d_i(\bi{x},\bxi)] + \delta_i^{\beta \%}(\bi{x}) SD[d_i(\bi{x},\bxi)], \label{eq:Methods_objectiveFunctions_percentileEstimate2} \\
    d_i^{\nu \%}(\bi{x}) & = \mathbb{E}[d_i(\bi{x},\bxi)] + \delta_i^{\nu \%}(\bi{x}) SD[d_i(\bi{x},\bxi)], \label{eq:Methods_objectiveFunctions_percentileEstimate3}
\end{align}
leading to the corresponding multiplicative factors $\delta_i^{\beta \%}(\bi{x}), \delta_i^{\nu \%}(\bi{x}) \in \mathbb{R}$. 

The objectives in Equation \ref{eq:Methods_CTVobj}, Equation \ref{eq:Methods_CTVobj_overdosage} and Equation \ref{eq:Methods_OARobj_overdosage} vanish for $d_i^{\alpha \%}(\bi{x}) > \gamma_i$, $d_i^{\beta \%}(\bi{x}) < \epsilon_i$ and $d_i^{\nu \%}(\bi{x}) < \mu_i$, respectively. To ensure that CTV and OAR voxels are always included in the optimization, we further optimize for the expected quadratic dose difference (we have a similar term for normal tissue) with a low weight, respectively as
\begin{align}
    f_{CTV}(\bi{x}) & = \frac{1}{N_{CTV}} \sum_{i \in CTV} w_i^{CTV} \mathbb{E}[(d_i(\bi{x},\bxi) - d_i^p)^2], \label{eq:fCTV} \\
    f_{OAR}(\bi{x}) & = \frac{1}{N_{OAR}} \sum_{i \in OAR} w_i^{OAR} \mathbb{E}[(d_i(\bi{x},\bxi))^2], \label{eq:fOAR_conformal} \\ 
    f_{Tissue}(\bi{x}) & =\frac{1}{N_{Tissue}} \sum_{i \in Tissue} w_i^{Tissue} \mathbb{E}[(d_i(\bi{x},\bxi))^2], \label{eq:fTissue_conformal}
\end{align}
where $d_i^p$ is the prescribed voxel dose, $N_{Tissue}$ is the number of tissue voxels and $w_i^{Tissue}$ is the tissue weight. The tissue objective in Equation \ref{eq:fTissue_conformal} aims to achieve dose conformity to the CTV. \\

The complete probabilistic (inner) optimization for a given set of $\delta$-factors and objective weights $\Pi = \{ \pi_{CTV}^{\alpha}, \pi_{CTV}^{\beta}, \pi_{OAR}^{\nu}, \pi_{CTV}^{low}, \pi_{OAR}^{low}, \pi_{Tissue} \}$ is given by
\begin{align}
    \min_{\bi{x}} \Bigl[ & \pi_{CTV}^{\alpha} f_{CTV}^{\alpha,\gamma}(\bi{x}) + \pi_{CTV}^{\beta} f_{CTV}^{\beta,\epsilon}(\bi{x}) + \pi_{OAR}^{\nu} f_{OAR}^{\nu,\mu}(\bi{x}) + \nonumber \\
    & \pi_{CTV}^{low} f_{CTV}(\bi{x}) + \pi_{OAR}^{low} f_{OAR}(\bi{x}) + \pi_{Tissue} f_{Tissue}(\bi{x}) \Bigr] \label{eq:objective} \\
    \textrm{s.t.} \qquad & d_i^{\alpha \%}(\bi{x}) = \mathbb{E}[d_i(\bi{x},\bxi)] - \delta_i^{\alpha \%}(\bi{x})  SD[d_i(\bi{x},\bxi)] \quad & \forall i \in CTV \label{eq:percentileApprox1} \\
    & d_i^{\beta \%}(\bi{x}) = \mathbb{E}[d_i(\bi{x},\bxi)] + \delta_i^{\beta \%}(\bi{x}) SD[d_i(\bi{x},\bxi)] & \forall i \in CTV  \label{eq:percentileApprox2} \\
    & d_i^{\nu \%}(\bi{x}) = \mathbb{E}[d_i(\bi{x},\bxi)] + \delta_i^{\nu \%}(\bi{x}) SD[d_i(\bi{x},\bxi)] & \forall i \in OAR \label{eq:percentileApprox3} \\
    & d_i(\bi{x},\bxi) = \sum_{j \in \mathbb{B}} D_{ij}(\bxi) x_j  \label{eq:voxelDose}, \qquad x_j \geq 0, & \forall j \in \mathbb{B}.
\end{align}

We solve the optimization using the interior-point method provided by \verb"fmincon" in Matlab \citep{matlab2024}, with an optimality tolerance of $10^{-8}$. To make sure the optimality tolerance is reached before the step- and function tolerance, we define the latter two to be $10^{-25}$. The analytical gradient and Hessian of the objective in Equation \ref{eq:objective} are derived in Appendix \ref{app:GradientAndHessian}.

\subsubsection{The outer optimization loop}
\label{subsubsec:outerOptimization}
For current iteration $k$, the inner optimization starts with initial guess $\bi{x}_{init}$ and results in beam weight $\bi{x}_{*}^k$, which in general is significantly different from $\bi{x}_{init}$. As a result, the PDF associated to each voxel dose (and thus the voxel dose percentiles) may have changed. We warm-start the next (inner) optimization (iteration $k+1$) using the previous initial ($\bi{x}^{k}$) and final ($\bi{x}_{*}^k$) beam weights as
\begin{equation}
    \bi{x}^{k+1} = \bi{x}^{k} + \kappa \cdot (\bi{x}_{*}^k - \bi{x}^{k}),
\end{equation}
where the damping factor $\kappa = 0.2$. Once the beam weights are updated, in the outer optimization loop the new $\delta$-factors are determined by Equation \ref{eq:percentileApprox1}, Equation \ref{eq:percentileApprox2} and Equation \ref{eq:percentileApprox3}, making use of the fast PCE sampling. The dampening of the beam weights implicitly dampens the $\delta$-factors as well (i.e., the percentiles that are optimized for), because the $\delta$-factors depend on the beam weights. \\

The outer optimization loop is terminated when the voxel dose percentiles corresponding to the damped pencil-beam weights converge for all CTV and OAR voxels. Since the voxel dose percentiles are determined by PCE sampling, a sampling noise is involved (which is propagated to the $\delta$-factors). As a result, the percentiles can only converge within a tolerance that is larger than the the sampling noise. We consider the percentiles to be converged if their trend does not change within a certain tolerance. To quantify this trend, we smooth the percentiles at iteration $k$ associated with the damped beam weight $\bi{x}^k$, using a moving average (MA) of window $\Delta W$ (denoted as $\text{MA}\Delta W$). For the lower percentile of the CTV (as in Equation \ref{eq:percentileApprox1}) this is done as
\begin{equation}
    d_i^{\alpha \%, \text{MA} \Delta W}(\bi{x}^{k}) = \frac{1}{\Delta W} \sum_{t=k-\Delta W+1}^{k} d_i^{\alpha \%}(\bi{x}^t), \qquad \forall i \in CTV,
\end{equation}
for $k \geq \Delta W$. We define the convergence criteria such that the relative change of $d_i^{\alpha \%, \text{MA} \Delta W}(k)$ within $\Delta k$ iterations is smaller than $\tau_{CTV,\alpha}$ for all CTV voxels, i.e., the convergence criterion for probability level $\alpha$ is given by
\begin{equation} \label{eq:Methods_convergenceCriteria_CTV}
    \left| \frac{d_i^{\alpha \%, \text{MA} \Delta W}(k) - d_i^{\alpha \%, \text{MA} \Delta W}(k-\Delta k)}{d_i^{\alpha \%, \text{MA} \Delta W}(k)} \right| < \tau_{CTV,\alpha}, \qquad \forall i \in CTV,
\end{equation}
for $k \geq \Delta W + \Delta k$. The same convergence criteria are applied to $d_i^{\beta \%, \text{MA}\Delta W}$ and $d_i^{\nu \%, \text{MA}\Delta W}$ for all voxels in the structure, with corresponding convergence tolerances of $\tau_{CTV,\beta}$ and $\tau_{OAR,\nu}$.

\subsubsection{Initialization of the probabilistic optimization}
\label{subsubsec:initialization}
As will be shown in Section \ref{subsubsec:VanHerk}, the probabilistic optimization is probabilistically equivalent to a PTV-based optimization (with a spherically symmetric dose distribution and static dose cloud approximation), therefore serving as a good initial estimate to warm-start the probabilistic optimization. The PTV is defined by extending the CTV by a PTV-margin $M_{PTV}$ isotropically. The PTV-optimization is initialized by using uniform beam weights (all 0.01) and minimizes for the quadratic difference of the nominal voxel dose $d_i^{nom}(\bi{x})$ and the prescribed dose $d_i^p$ as $\sum_{i \in \mathbb{V}} w_i (d_i^{nom}(\bi{x}) - d_i^p)^2$. The resulting output beam weights $\bi{x}_{PTV}$ are used in the first $k=1$ iteration of the probabilistic optimization, i.e., $\bi{x}_{init} = \bi{x}_{PTV}$. 

Before starting the inner optimization, we specify the voxel dose thresholds $\gamma_i$, $\epsilon_i$ and $\mu_i$ with the corresponding desired probability levels $\alpha$, $\beta$ and $\nu$. Then, we determine the voxel dose percentiles by PCE, for the CTV and OAR voxels that correspond to the current beam weights (so for $k=1$ that is $\bi{x}_{PTV}$). After the percentiles are converted to the $\delta$-factors by Equation \ref{eq:percentileApprox1}, Equation \ref{eq:percentileApprox2} and Equation \ref{eq:percentileApprox3}, the probabilistic optimization is started.

\subsection{Composite-wise robust optimization as comparison} \label{subsec:Methods_robustOptimization}
Various types of robust treatment planning exist \citep{UnkelbachPaganetti2018}, but in this work we restrict ourselves to comparing to a composite-wise worst-case robust approach. We robustly optimize for the worst-case scenario within scenario set $S$ for CTV and OAR with a nominal tissue objective, as
\begin{align} % Robust: QDD_robustCTVandSpine_compWise 
\min_{\bi{x}} \Bigl[ \max_{s \in S} \{ & \omega_{CTV} f_{CTV}(d(\bi{x},s)) + \omega_{OAR} f_{OAR}(d(\bi{x},s)) + \omega_{OAR}^{max} f_{maxOAR}(d(\bi{x},s)) \} \label{eq:robOpt_general} \\
+ \, &\omega_{CTV}^{nom} f_{CTV}^{nom}(d(\bi{x})) + \omega_{Tissue} f_{Tissue}^{nom}(d(\bi{x})) \Bigr] \nonumber  \\ 
\textrm{s.t.} \qquad &\bi{x} \geq 0, \nonumber \\
 &f_{CTV}(d(\bi{x},s)) = \frac{1}{N_{CTV}} \sum_{i \in CTV} w_i^{CTV} \cdot (d_i(\bi{x},s) - d_i^p)^2, \nonumber \\
 &f_{OAR}(d(\bi{x},s)) = \frac{1}{N_{OAR}} \sum_{i \in OAR} w_i^{OAR} \cdot (d_i(\bi{x},s) - d_i^p)^2, \nonumber \\
 &f_{maxOAR}(d(\bi{x},s)) = \frac{1}{N_{OAR} }\sum_{i \in OAR} w_i^{OAR} \cdot (d_i(\bi{x},s) - d_i^{maxOAR})_+^2, \nonumber \\
 &f_{Tissue}^{nom}(d(\bi{x})) = \frac{1}{N_{Tissue} }\sum_{i \in Tissue} w_i^{Tissue} \cdot (d_i^{nom}(\bi{x}) - d_i^p)^2, \nonumber \\
 &f_{CTV}^{nom}(d(\bi{x})) = \frac{1}{N_{CTV} }\sum_{i \in CTV} w_i^{CTV} \cdot (d_i^{nom}(\bi{x}) - d_i^p)^2, \nonumber
\end{align}
where the set of objective weights is given by $\Omega=\{\omega_{CTV}, \omega_{OAR}, \omega_{OAR}^{max}, \omega_{CTV}^{nom}, \omega_{Tissue}\}$. Large OAR voxel doses are penalized by a piecewise quadratic dose difference between the scenario voxel dose $d_i(\bi{x},s)$ and the voxel dose threshold $d_i^{maxOAR}$. A nominal CTV objective with corresponding objective weight $\omega_{CTV}^{nom}$ is used for some geometries to increase the importance of the nominal scenario.

\begin{figure}[b]
    \centering
    \includegraphics[width=0.4\textwidth]{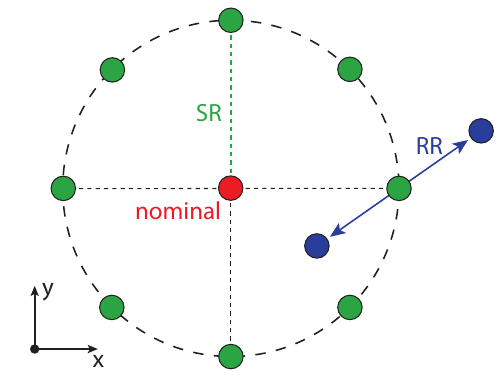}
    \caption{The robust error scenario set used during optimization. For every setup error with given setup robustness (SR) a range error with given range robustness (RR) is included. The nominal scenario (red) is always included.}
    \label{fig:robustScenarios}
\end{figure}

Scenario set $S$ is commonly defined as illustrated in Figure \ref{fig:robustScenarios}, consistent with approaches implemented in commercial treatment planning systems, such as by RayStation \citep{RayStation_refManual}. For setup errors in $X$ and $Y$, scenarios are included at primary axes corresponding to the used setup robustness (SR). Scenarios in the $(\pm 1, \pm 1)$ direction of the XY-plane are included and lie on the circle defined by the SR. For every setup scenario, range error scenarios are included with a fixed range robustness (RR). The nominal scenario is always included. \\

\subsection{Phantom geometries} 
\label{subsec:Methods_PhantomGeometries}

\begin{figure}[]
    \centering
    \includegraphics[width=\textwidth]{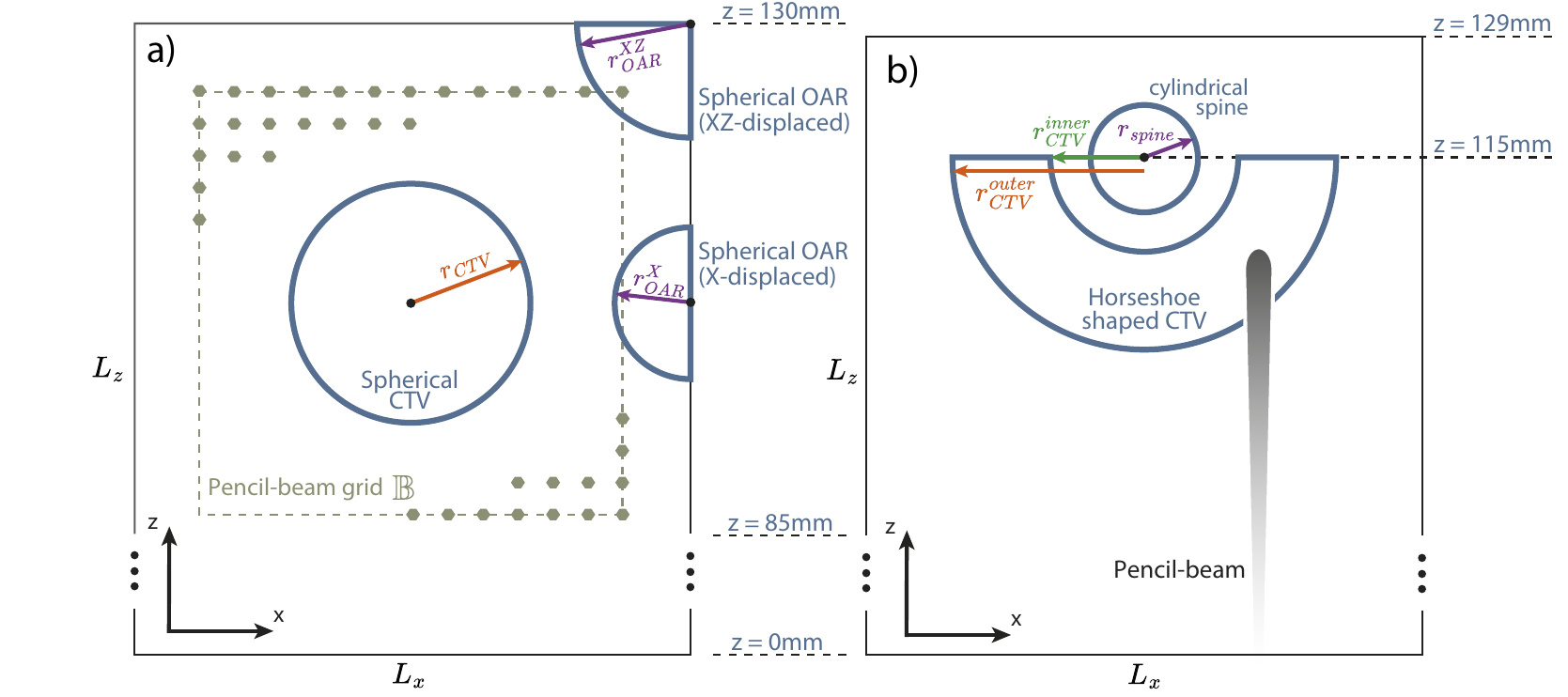}
    \caption{Illustration of the three-dimensional homogeneous (water) phantom geometries used in this work, for the a) spherical and b) spinal case in the XZ-plane. Dimensions are $(L_x,L_y,L_z) = \left(\qtylist[list-final-separator={, }]{45; 45 ; 130}{\mm}\right)$ and $\left(\qtylist[list-final-separator={, }]{70; 30 ; 129}{\mm}\right)$ for the respective cases. Voxel grids are defined for $z > \qty{85}{\mm}$, with respectively \numproduct{1 x 1 x 1} \unit{\milli\meter^3} and \numproduct{2 x 2 x 2} \unit{\milli\meter^3} voxels. Pencil-beam $j \in \mathbb{B}$ is directed at the pencil-beam spot in the grid and travels in the positive $z$-direction.}
    \label{fig:geometries}
\end{figure}

Figure \ref{fig:geometries} shows the three-dimensional homogeneous (water) phantom geometry that is used in this work for the spherical (left) and spinal (right) case in the XZ-plane. The spherical geometry has dimensions $(L_x,L_y,L_z) = \left(\qtylist[list-final-separator={, }]{45; 45; 130}{\mm}\right)$ and consists of \numproduct{1 x 1 x 1} \unit{\milli\meter^3} voxels. As the ROI in the $z$-direction is limited from $L_z = \qtyrange{85}{130}{\milli\metre}$, the number of considered voxels is $N_v^{spheres} = 91.125$. The spherical CTV has radius $r_{CTV} = \qty{9}{mm}$. As a first case, only the spherical CTV is considered (referred to as CTV-only), i.e., no OARs are included so that the remaining volume is normal tissue. For the CTV+OAR case, we include a single OAR to the geometry. We distinguish between the XZ-displaced OAR (radius $r_{OAR}^{XZ} = \qty{9}{mm}$) and X-displaced OAR (radius $r_{OAR}^{X} = \qty{5}{mm}$), which centers are respectively located at $(x,y,z)_{XZ} = \left(\qtylist[list-final-separator={, }]{44.5; 22.5; 129.5}{\mm}\right)$ and $(x,y,z)_{X} = \left(\qtylist[list-final-separator={, }]{44.5; 22.5; 107.5}{\mm}\right)$. As only part of the OARs fall within the ROI, only a quarter of the XZ-displaced and half of the X-displaced OAR is included. The remaining part of the geometry is normal tissue. The CTV, OAR and tissue voxels form the ROI $\mathbb{V}$, i.e., $\mathbb{V} = CTV \cup OAR \cup Tissue$.

% 11550 total, 1204 CTV, 9956 tissue, spine 390
The spinal geometry has dimensions $(L_x,L_y,L_z) = \left(\qtylist[list-final-separator={, }]{70; 30; 129}{\mm}\right)$ consisting of \numproduct{2 x 2 x 2} \unit{\milli\meter^3} voxels, thus $N_v^{spinal} = 11.550$. It consists of a cylindrical spine that is surrounded by a horseshoe shaped CTV. The spine has radius $r_{Spine} = \qty{6}{mm}$ and is positioned parallel to the $y$-direction along the entire geometry at $(x,z) = (\qty{35}{\mm}, \qty{115}{\mm})$. Parallel to the spine lies the CTV (for $\qty{9}{mm} < y < \qty{21}{mm}$), extending from $r_{CTV}^{inner} = \qty{12}{mm}$ to $r_{CTV}^{outer} = \qty{24}{mm}$ in the XZ-plane for $z < \qty{115}{mm}$. The remaining part of the geometry is normal tissue. \\

In both cases, the CTV is irradiated by pencil-beams traveling in the positive $z$-direction. In the spherical geometry we use 13 pencil-beam spots in each direction (spaced $\qty{3}{\milli\meter}$ apart), so that the number of pencil-beam spots $N_b^{spherical} = 13^3 = 2197$. The center pencil-beam spot is located at the CTV center, such that the grid extends \qty{18}{mm} from the CTV center in all directions. In the spinal geometry, the number of pencil-beam spots in the respective directions are $(N_b^x,N_b^y,N_b^z) = \left(\numlist[list-final-separator={, }]{21; 9; 13} \right)$, spaced $\qty{3}{mm}$ apart ($N_b^{spinal} = 2457$). Dose dependencies in this work are obtained by analytical approximations of the Bragg curve \citep{Bortfeld1997}, where dose values below 0.01\% of the maximum in $D_{ij}$ are neglected. Each beam has a Gaussian-distributed lateral profile that is assumed to have an energy-independent initial width of $\sigma_{b} = \qty{3}{\mm}$, increasing in depth. 

\subsection{Probabilistic evaluation of treatment plans} \label{subsec:Methods_probEvaluation}

\subsubsection{Cost accuracy analysis of the Polynomial Chaos Expansions} \label{subsec:Methods_probEvaluation_doseInfluence}
To ensure that the dose approximation is sufficiently accurate in the relevant uncertainty domain during optimization, a cost-accuracy analysis is done for the PCE in Appendix \ref{app:accuracies_PCE}. For this purpose, a $\Gamma$-evaluation \citep{Biggs2022} is done (with distance-to-agreement \qty{0.1}{\gray} and 1\% dose difference criteria) for two different proton pencil-beams (and voxel doses $\geq \qty{0.1}{\gray}$), in 123 different error scenarios that lie within the 99\% confidence ellipsoid of the input phase space (taking into account all uncertain variables simultaneously). For every scenario in the 99\%-ellipsoid we check the accepted voxel fraction.

Moreover, the PCE accuracy is quantified by determining the dose difference between the PCE and the dose engine for both test pencil-beams. For all test scenarios within the 99\% confidence ellipsoid, we determine the minimum voxel dose difference among the 2\% of the voxels having the largest dose difference, which we denote by $\Delta D_{2\%}$. Then, we calculate the scenario fraction for which the $\Delta D_{2\%}$ is larger than a certain dose value. Moreover, we determine the voxel dose difference averaged over all test scenarios (denoted by $\Delta D$), and check what voxel fraction has $\Delta D$ larger than a certain dose value. \\

After the optimization is done, treatment plan quality is checked by constructing an independent PCE of the voxel dose, in line with previous work \citep{RojoSantiago2021,RojoSantiago2023,RojoSantiago2023_2,RojoSantiago2024,Oud2024}. As the voxel dose distribution for a single pencil-beam is different than for the final treatment plan, the necessary PCE accuracy is determined by a separate $\Gamma$-analysis (with distance-to-agreement \qty{0.1}{\gray} and 1\% dose difference criteria for voxel doses $\geq \qty{0.1}{\gray}$).

\subsubsection{Probabilistic evaluation metrics} \label{subsec:Methods_probEvaluation_probMetrics}

In the following we discuss the probabilistic evaluation metrics that are used to get insights in the probabilistic outcomes of the treatment plans \citep{Perk2016}. The well-known DVH can be probabilistically extended towards the DVH-distribution \citep{Trofimov2012}, where the plan robustness is captured by the width of the DVH bands, representing confidence intervals of the DVHs. For example, the 95\% confidence band is defined such that in 95\% of the error scenarios - or in other words with 95\% probability - the DVH-curves lie within the $2.5^{th}$ and $97.5^{th}$ percentile of the dose value.

Probabilities of voxel-wise under- and overdosage can be shown to understand to which extent the treatment plan reaches the probabilistic objectives. The probabilities are obtained by counting the fraction of error scenarios for which a voxel is below or above the desired threshold. \\

\begin{figure}[]
    \centering
    \includegraphics[width=0.9\textwidth]{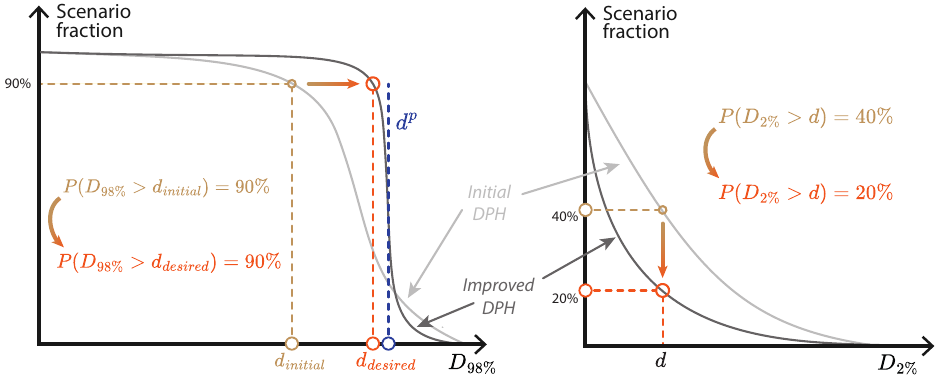}
    \caption{Illustration of dose population histograms for the CTV metric $D_{98\%}$ (left) and OAR metric $D_{2\%}$ (right), showing the fraction of error scenarios that exceed a given dose level (e.g., 90\% of scenarios yield $D_{98\%} \geq d_{desired}$, where $d_{desired}$ may for example be 95\% of prescription dose $d^p$).}
    \label{fig:DPHillustration}
\end{figure}

A more complete understanding can be obtained by the dose population histogram (DPH) \citep{vanHerk2000}, which is strongly related to the cumulative distribution function (CDF). It shows the (error) scenario fraction (i.e., probability) in which a dose metric of interest exceeds a given dose level. An illustrative example of the DPH is shown in Figure \ref{fig:DPHillustration} for the $D_{98\%}$ (left) and $D_{2\%}$ (right), but other dose metrics may be used. For DVH-metrics of the CTV (e.g., $D_{98\%}$) we aim to have steep DPH curves, ideally being a step function that starts at a scenario fraction of 100\% and drops to 0\% at prescription dose $d^p$. This ideal shape corresponds to a perfectly homogeneous dose distribution, because in all error scenarios the $D_{98\%}$ is exactly equal to $d^p$. In practice, some scenarios would result in the $D_{98\%} > d^p$ (as a result of partially overdosing the CTV), or would result in $D_{98\%} < d^p$ (as a result of partially underdosing the CTV). The corresponding DPH curve would be less steep and deviates from the ideal curve. In Figure \ref{fig:DPHillustration}, the initial DPH curve shows that 90\% of the error scenarios has a $D_{98\%}$ of at least $d_{initial}$. For the same fraction of error scenarios, the minimum $D_{98\%}$ can be increased (i.e., improved) towards $d_{desired} > d_{initial}$ (in practice, $d_{desired}$ is for example 95\% of the prescription dose), such that 90\% of the error scenarios has a $D_{98\%}$ of at least $d_{desired}$.

For OARs we aim to have DVH-metrics (e.g., $D_{2\%}$) that have low dose values in most scenarios, which correspond to DPH curves in the bottom left of the figure. By going from the initial to the improved DPH we improve the probability in which $D_{2\%} > d$.

\subsubsection{Probabilistic scaling to compare probabilistic and robust plans} \label{subsec:Methods_scaling}
The probabilistic plans (Section \ref{subsec:Methods_optimizationScheme}) are compared to robust optimizations (Section \ref{subsec:Methods_robustOptimization}). Robust plans are matched to the probabilistic plans by manually tuning the robust objective weights, until either 1) a \textit{similar CTV coverage}, or 2) a \textit{similar OAR dose} as in the probabilistic plan is achieved. The CTV coverage is defined as the $10^{th}$ percentile of the $D_{98\%}$ (as used by \citet{Tilly2019}, denoted as $D_{98\%}^{10\text{th}}$), which is the maximum $D_{98\%}$ value among the 10\% of scenarios with the lowest $D_{98\%}$ values. Alternatively, when the OAR dose is matched, the $90^{th}$ percentile of the $D_{2\%}$ is used (denoted as $D_{2\%}^{90\text{th}}$), which is the minimum $D_{2\%}$ value among the 10\% of scenarios with the highest $D_{2\%}$ values. Thus, once the probabilistic plan is done, either its $D_{98\%}^{10\text{th}}$ or its $D_{2\%}^{90\text{th}}$ is determined, and robust plans are made and similarly evaluated using PCE. 

As a next step, the probabilistic plan is scaled such that the $50^{th}$ percentile of $D_{50\%}$ (the median dose, an ICRU-recommended metric \citep{ICRU83}), matches the prescribed dose ($D_{50\%}^{50\text{th}} = 100\% \, d^p$). For the robust and probabilistic plans that are matched based on CTV coverage, the robust plan is scaled identically as the probabilistic plan, i.e., so that the $50^{th}$ percentile of the $D_{50\%}$ equals $100\% \,d^p$. For similar OAR dose, the robust plan is scaled to the probabilistic plan by matching the $D_{2\%}^{90\text{th}}$. 

\section{Results} \label{sec:Results}
In the following, results are first presented on the spherical CTV-only case (Section \ref{subsec:Results_sphericalCTVonly}) and compared with the Van Herk margin recipe. The spherical CTV+OAR case (Section \ref{subsec:Results_CTVandOAR}) and the spinal case (Section \ref{subsec:Results_spinal}) are shown as well. The optimization parameters for each geometry case are discussed in Appendix \ref{app:probRobOpt} and are summarized in Table \ref{tab:probOptParameters} and Table \ref{tab:robOptParameters} for the probabilistic and robust optimizations. 

Expectation values in Equation \ref{eq:Methods_objectiveFunctions_percentileEstimate1}, Equation \ref{eq:Methods_objectiveFunctions_percentileEstimate2} and Equation \ref{eq:Methods_objectiveFunctions_percentileEstimate3} ($\mathbb{E}[D_{ij}(\bxi)]$ and $\mathbb{E}[D_{ij}(\bxi)D_{ij'}(\bxi)]$) are calculated for CTV and OAR by Gauss-Hermite cubature (so without PCE) using 105 dose calculations. The PCE of the dose-influence matrix is constructed using 1637 dose calculations. Details on their cost-accuracy analyses are shown in Appendix \ref{app:accuracies_exp}.

\begin{table}[b]
\caption{\label{tab:robustness_comparison_XY}Comparison of the CTV coverage in the robust plans for a setup robustness (SR) ranging from \qty{4}{\milli\meter} till \qty{7}{\milli\meter} versus the probabilistic plan in the CTV-only \textit{setupXY} case.}
\begin{center}
\begin{tabular}{@{}lcccc}
\toprule                             
Treatment Plan & $D_{98\%}^{2\text{nd}}$ (\unit{\gray}) & $D_{98\%}^{5\text{th}}$ (\unit{\gray}) & $D_{98\%}^{10\text{th}}$ (\unit{\gray}) & $P(D_{98\%} \geq 0.95 \cdot d^p)$ \\[0.5ex]
\toprule
\text{SR of \qty{4}{\milli\meter} } & 35.95 & 41.36 & 46.30 & 55\% \cr
\text{SR of \qty{5}{\milli\meter} } & 43.98 & 48.70 & 52.55 & 75\% \cr 
\text{SR of \qty{6}{\milli\meter} } & 49.36 & 53.18 & 55.96 & 87\% \cr 
\text{SR of \qty{7}{\milli\meter} } & 53.70 & 56.42 & 57.98 & 94\% \cr 
\text{Probabilistic plan}& 50.19 & 53.59 & 55.98 & 85\% \cr 
\bottomrule
\end{tabular}
\end{center}

\end{table}

\begin{table}[]
\caption{\label{tab:robustness_comparison_RRSR}Comparison of the CTV coverage of the robust plan for a setup robustness (SR) of \qty{6}{\milli\meter} and range robustness (RR) of 4\% and 5\%, versus the probabilistic plan in the CTV only case.} 

\begin{center}
\begin{tabular}{@{}lcccc}
\toprule                             
Treatment Plan&$D_{98\%}^{2\text{nd}}$ (\unit{\gray}) & $D_{98\%}^{5\text{th}}$ (\unit{\gray}) & $D_{98\%}^{10\text{th}}$ (\unit{\gray}) & $P(D_{98\%} \geq 0.95 \cdot d^p)$ \\[0.5ex]
\toprule  
\text{SR/RR: \qty{6}{mm}/4\% } & 42.94 & 48.93 & 53.01 & 77\% \cr
\text{SR/RR: \qty{6}{mm}/5\% } & 45.86 & 50.84  & 54.31 & 81\% \cr 
\text{Probabilistic plan} & 46.58 & 50.92 & 54.03 & 79\% \cr 
\bottomrule
\end{tabular}
\end{center}

\end{table}

\subsection{Probabilistic and robust plans for the spherical CTV-only case}
\label{subsec:Results_sphericalCTVonly}

For the spherical CTV-only case we consider two combinations of systematic uncertainties: 1) setup errors in $X$ and $Y$ (i.e., \textit{setupXY}) and 2) setup errors in $X$ and $Y$ together with range errors (i.e., \textit{setupXYrange}). 

\subsubsection{Probabilistic and robust optimization}
\label{subsec:Results_CTVonlyProb}
The probabilistic optimization minimizes for CTV underdosage ($P(d_i \leq 0.95 \cdot d_i^p) \leq 10\%$) and overdosage ($P(d_i \geq 1.07 \cdot d_i^p) \leq 10\%$) probabilities, at the same time pushing the expected dose in the CTV to $d_i^p = \qty{60}{\gray}$. As a comparison to the probabilistic plan, we perform a robust optimization for the CTV and use a nominal objective for tissue, with the same objective weights as in the probabilistic optimization. \\

Robust plans with setup errors ranging from 4 till 7 mm were made and compared to the probabilistic plan by matching their CTV coverage probability ($D_{98\%}^{10\text{th}}$). Additionally, we checked the $D_{98\%}^{2\text{nd}}$ and $D_{98\%}^{5\text{th}}$, and determined the probability of $D_{98\%}$ exceeding $95\%$ of the prescribed dose $d^p$, i.e., $P(D_{98\%} \geq 0.95\cdot d^p)$. As Table \ref{tab:robustness_comparison_XY} shows, the required setup robustness (SR) to achieve similar CTV coverage ($D_{98\%}^{10\text{th}}$) as in the probabilistic \textit{setupXY} plan, is $\text{SR} = \qty{6}{\milli\meter}$. Figure \ref{fig:CTVonly_XY_XYr_RobProb} (top) compares the dose distributions for the robust ($\text{SR} = \qty{6}{\milli\meter}$) and probabilistic \textit{setupXY} plans for the XY-plane ($z = \qty{117.5}{\milli\meter}$). Additionally, cross sections through the CTV center for both plans along the $x$-axis are shown.

As the setup robustness was tuned to the $D_{98\%}^{10\text{th}}$, the dose extension beyond the CTV is very similar for the probabilistic and robust plans. Moreover, both plans are very conformal to the CTV, because of the way the uncertainty set is defined. According to \citet{RayStation_refManual}, the uncertainty set for $\text{SR} = \qty{6}{\milli\meter}$ includes the nominal scenario and eight others located on a circle of radius SR: $\left( x,y \right) = \left( 0, 0\right)$, $\left( x,y \right) = \left( \pm \text{SR}, 0\right)$, $\left( x,y \right) = \left( 0,\pm \text{SR} \right)$ and $\left( x,y \right) = \left( \pm \text{SR}/\sqrt{2},\pm \text{SR}/\sqrt{2} \right)$. No intermediate scenarios are included.

For $\text{SR} = \qty{6}{\milli\meter}$, we proceed to determine the necessary range robustness (RR) to achieve comparable CTV coverage (where we use range errors additional to setup errors in the $X$ and $Y$-directions). The results are listed in Table \ref{tab:robustness_comparison_RRSR}, showing that an RR of 5\% gives the closest match of CTV coverage to the probabilistic plan. The robust ($\text{SR/RR: \qty{6}{mm}/5\%}$) and probabilistic \textit{setupXYrange} plans are compared in Figure \ref{fig:CTVonly_XY_XYr_RobProb} (bottom) for the XZ-plane ($y = \qty{22.5}{\milli\meter}$). Cross sections through the CTV center along the diagonal ($z = x + \qty{85}{\milli\meter}$) for $y = \qty{22.5}{\milli\meter}$ are shown as well.

Compared to the probabilistic plan, the robust plan shows a larger (more conservative) dose expansion along the diagonals of the XZ-plane. This is a consequence of the way the discrete error scenarios are constructed (see Figure \ref{fig:robustScenarios}): scenarios that include a range-shift error are positioned farther from the nominal case than those without a shift. Probabilistic optimization, however, takes into account that large-shift scenarios are less probable to occur (compared to other scenarios in the set), resulting in a more conformal margin. In the XY-plane, where all scenarios in the set are equidistant from the nominal scenario (with distance SR), the dose distribution ends up as conformal as in the probabilistic plan.

\subsubsection{Verification against the Van Herk margin recipe}
\label{subsubsec:VanHerk}

\begin{figure}[]
    \centering
    
    \begin{subfigure}[b]{0.36\textwidth}
        \centering
        \includegraphics[width=\textwidth]{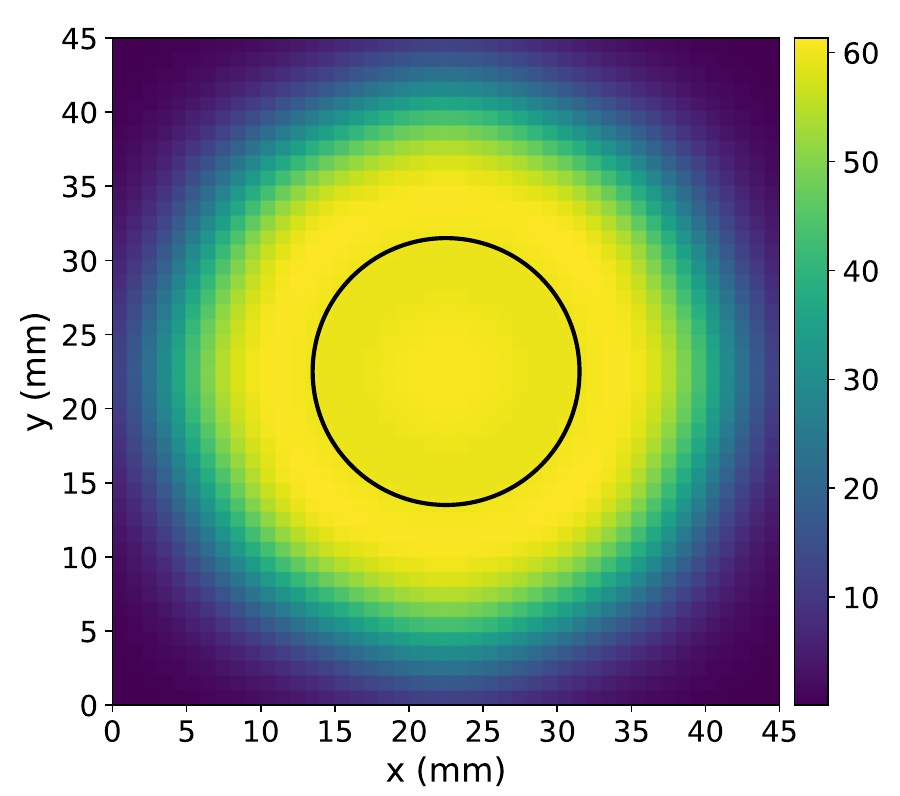}
        \caption{}
        \label{fig:CTVonly_XY_Rob}
    \end{subfigure}
    \hfill
    \begin{subfigure}[b]{0.36\textwidth}
        \centering
        \includegraphics[width=\textwidth]{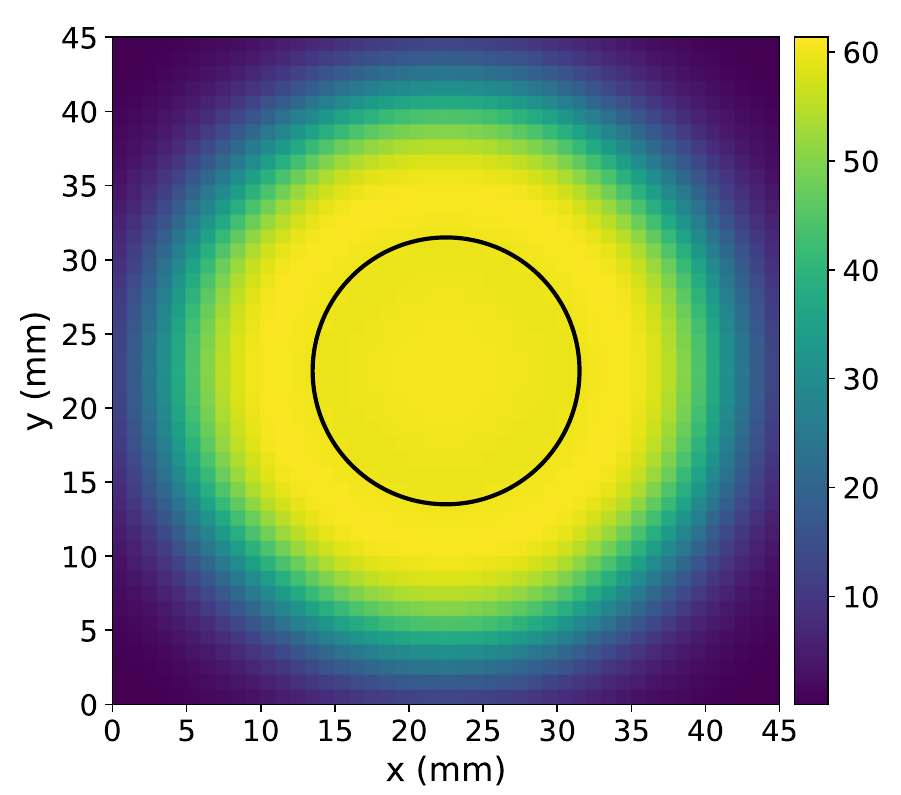}
        \caption{}
        \label{fig:CTVonly_XY_Prob}
    \end{subfigure}
    \hfill
    \begin{subfigure}[b]{0.25\textwidth}
        \centering
        \includegraphics[width=\textwidth]{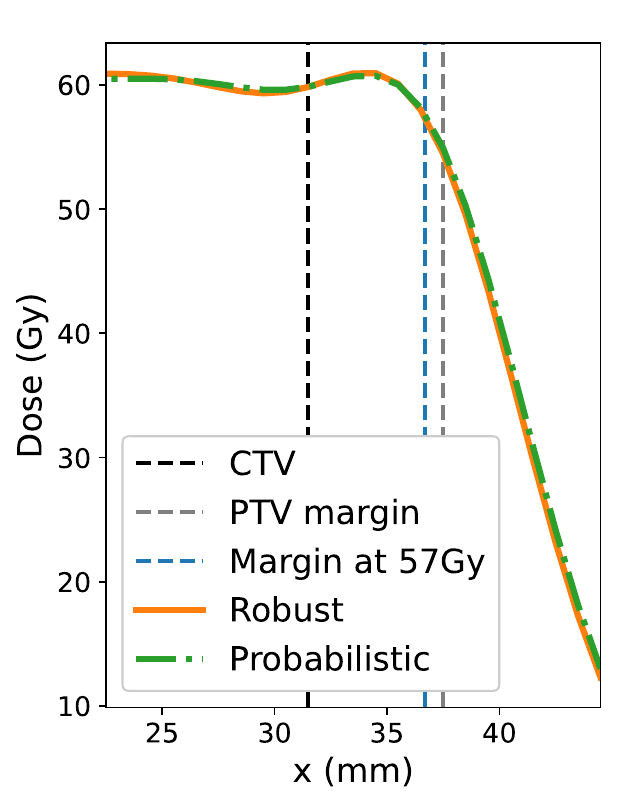}
        \caption{}
        \label{fig:CTVonly_XY_crossSection}
    \end{subfigure}

    \begin{subfigure}[b]{0.36\textwidth}
        \centering
        \includegraphics[width=\textwidth]{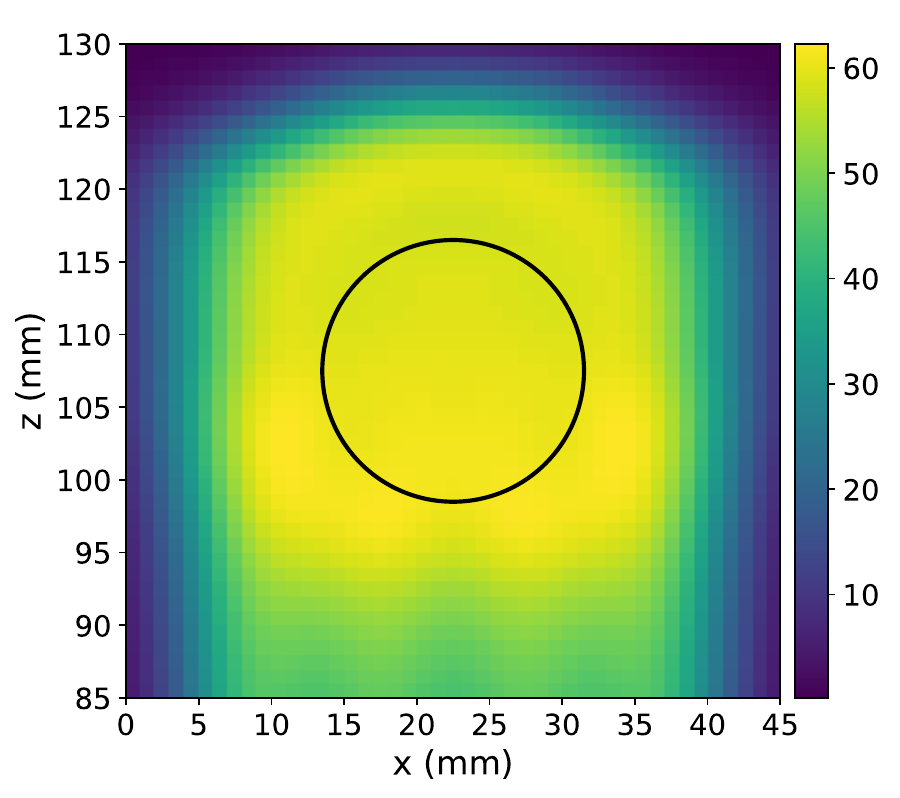}
        \caption{}
        \label{fig:CTVonly_XYr_Rob}
    \end{subfigure} 
    %\hfill
    \begin{subfigure}[b]{0.36\textwidth}
        \centering
        \includegraphics[width=\textwidth]{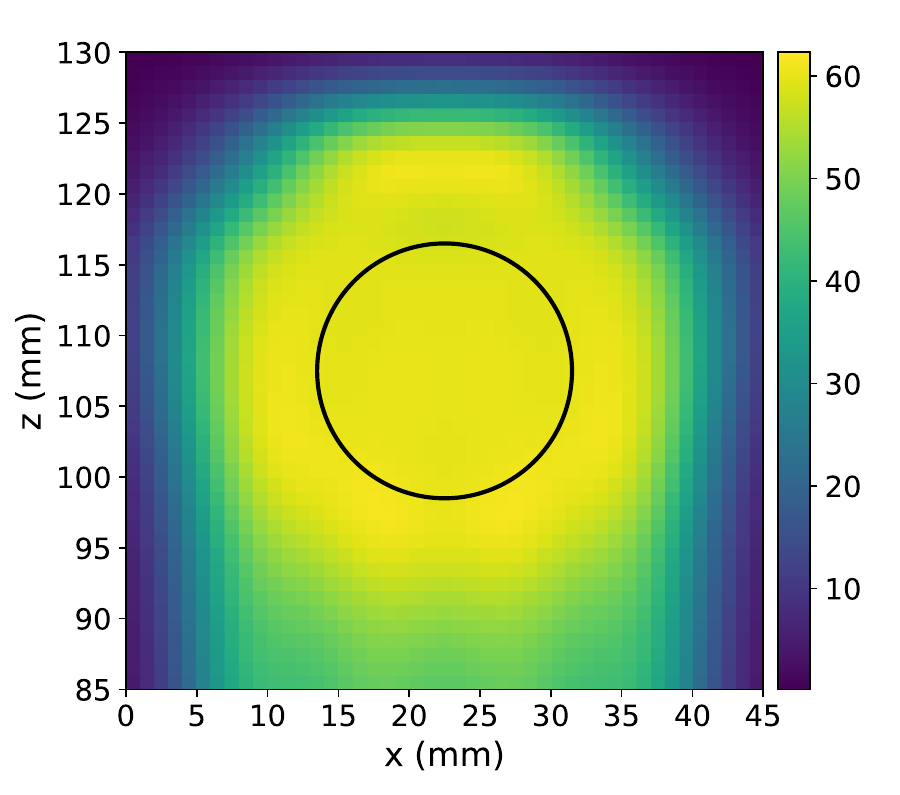}
        \caption{}
        \label{fig:CTVonly_XYr_Prob}
    \end{subfigure}
    %\hfill
    \begin{subfigure}[b]{0.25\textwidth}
        \centering
        \includegraphics[width=\textwidth]{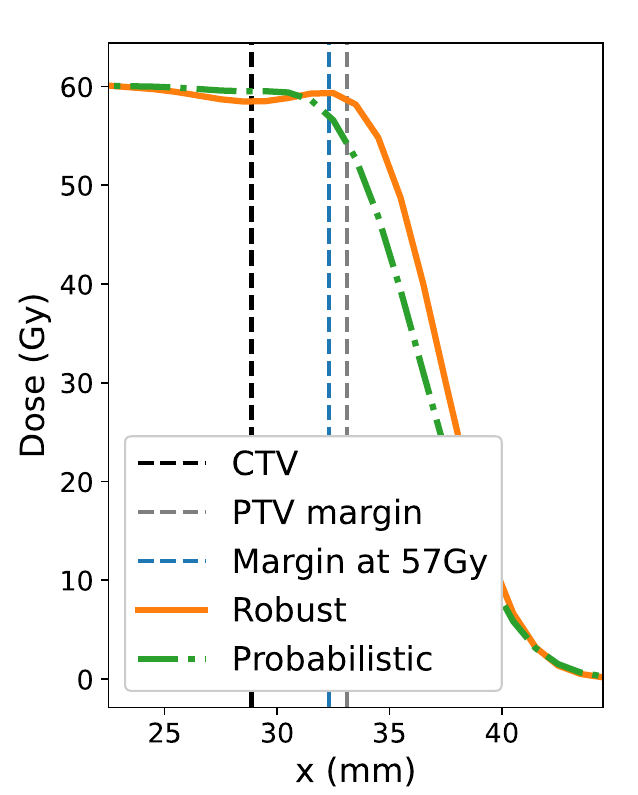}
        \caption{}
        \label{fig:CTVonly_XYr_crossSections}
    \end{subfigure} 
    \caption{Nominal dose distributions in the XY-plane through the CTV center for the (top) \textit{setupXY} and (bottom) \textit{setupXYrange} case: (left) robust plan, (middle) probabilistic plan, and (right) their cross-sections along the $x$-axis through the CTV center.}
    \label{fig:CTVonly_XY_XYr_RobProb}
\end{figure}

The margin recipes by \citet{vanHerk2000} for the PTV are based on a spherical CTV with ideal dose conformation (i.e., spherical symmetry in the dose) and static dose cloud approximation (i.e., invariance of the dose distribution shape under uncertainties). As only systematic errors are considered in this work, we compare to the Van Herk margin recipe without considering random errors. As the CTV-only probabilistic plan for setup errors in $X$ and $Y$ (Figure \ref{fig:CTVonly_XY_XYr_RobProb}, top) obeys these criteria (it is a homogeneous water phantom), we use this plan for the comparison.

The Van Herk margin recipe is based on the fact that a certain patient population (e.g., 90\% of the patients) should receive at least a minimum dose threshold (e.g., $0.95 \cdot d^p \, (= \qty{57}{\gray})$). For the probabilistic plan in Figure \ref{fig:CTVonly_XY_Prob}, we determined that the patient population (i.e., scenario fraction) in which the minimum CTV dose exceeds $0.95 \cdot d^p$ is 64.7\%. The resulting (2D) Van Herk margin\footnote{For two-dimensional systematic setup errors ($\Sigma_{X} = \qty{3}{\milli\meter}$, $\Sigma_{Y} = \qty{3}{\milli\meter}$), the Van Herk margin $M_{PTV}$ is calculated as $0.647 = 1 - \exp[-(M_{PTV}/\Sigma)^2/2]$, where $\Sigma = \sqrt{\Sigma_{X}^2 + \Sigma_{Y}^2} \approx \SI{4.24}{mm}$.} (using 64.7\% patient population and $0.95 \cdot d^p$ dose level) is thus $M_{PTV} = 1.44 \Sigma \approx \SI{6}{mm}$ and should correspond to the \qty{57}{\gray} dose level. This aligns well with the SR that leads to equivalent target coverage (Table \ref{tab:robustness_comparison_XY}).

In Figure \ref{fig:CTVonly_XY_crossSection}, besides the CTV margin (dashed black), the Van Herk margin (dashed gray) is shown with the (probabilistic) margin corresponding to the \qty{57}{\gray} dose level (dashed blue). In fact, the margin corresponding to the \qty{57}{\gray} dose level in the probabilistic plan is slightly smaller than \qty{6}{\milli\meter}. This is because the probabilistic plan was not optimized for the minimum CTV dose, but for the $10^{th}$-percentile of the CTV voxel dose. As the latter is a less conservative objective, a smaller (but comparable) margin is associated to it. This shows that the probabilistically optimized margin for ideal dose conformity is comparable to the Van Herk margin (for systematic errors).

\begin{table}[b!]
\caption{\label{tab:DVHmetrics_CTVandOAR_XZ}Statistical DVH-metrics to compare the CTV coverage ($D_{98\%}^{10\text{th}}$) and OAR overdosage ($D_{2\%}^{90\text{th}}$) of the robust and probabilistic plans for the XZ-displaced spherical CTV+OAR case. Metrics corresponding to the robust plans after scaling the beam weights are shown in brackets. The objective weights used in the robust plans are shown in the corresponding rows as $\{ \omega_{CTV}, \omega_{OAR}, \omega_{OAR}^{max}, \omega_{Tissue} \}$. }

\begin{center}
\begin{tabular}{@{}*{4}{l}}
\toprule
 & $D_{98\%}^{10\text{th}}$ (\unit{\gray}) & $D_{50\%}^{50\text{th}}$ (\unit{\gray}) & $D_{2\%}^{90\text{th}}$ (\unit{\gray}) \\[0.5ex]
\toprule
Robust $\{120, 1, 1, 160\}$ &  52.4 (52.3)
   & 60.1 (60.0) & 40.8 (40.8) \cr 
Robust ($\{100, 10, 10, 100\}$) & 45.4 (44.6)  &  60.2 (59.1) & 27.6 (27.1) \cr
Probabilistic &  53.3 (53.3)
   & 60.0 (60.0) & 27.0 (27.1) \cr 
\bottomrule
\end{tabular}
\end{center}
\end{table}

\begin{figure}[h!]
    \centering

    \begin{subfigure}[b]{0.32\textwidth}
        \includegraphics[width=\linewidth]{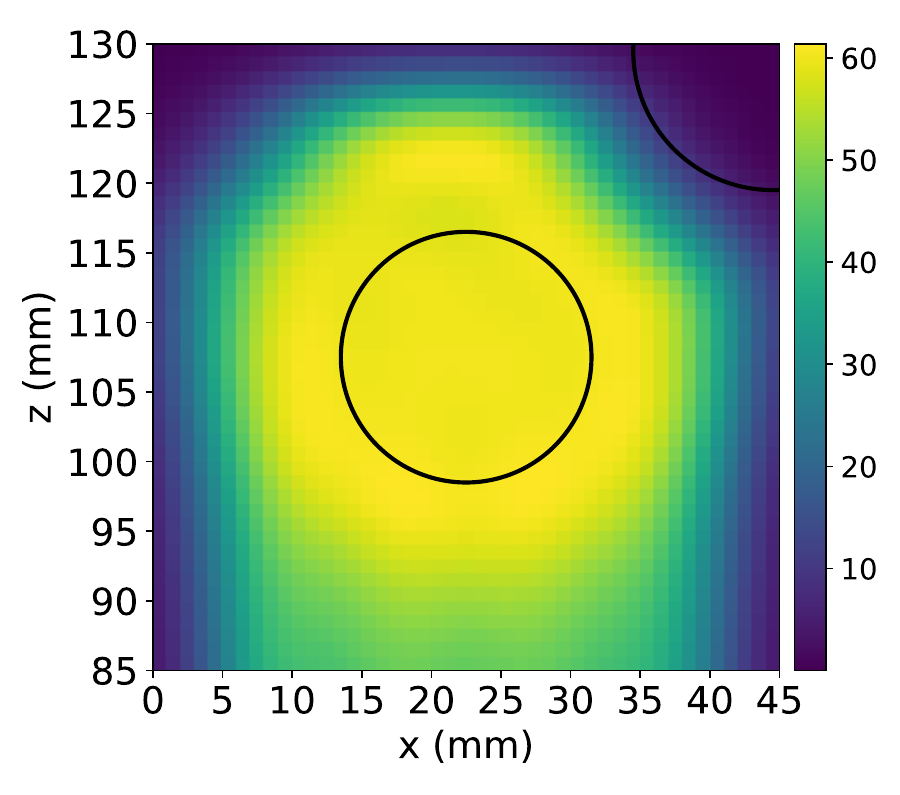}
        \caption{}
    \end{subfigure}
    \hfill
    \begin{subfigure}[b]{0.32\textwidth}
        \includegraphics[width=\linewidth]{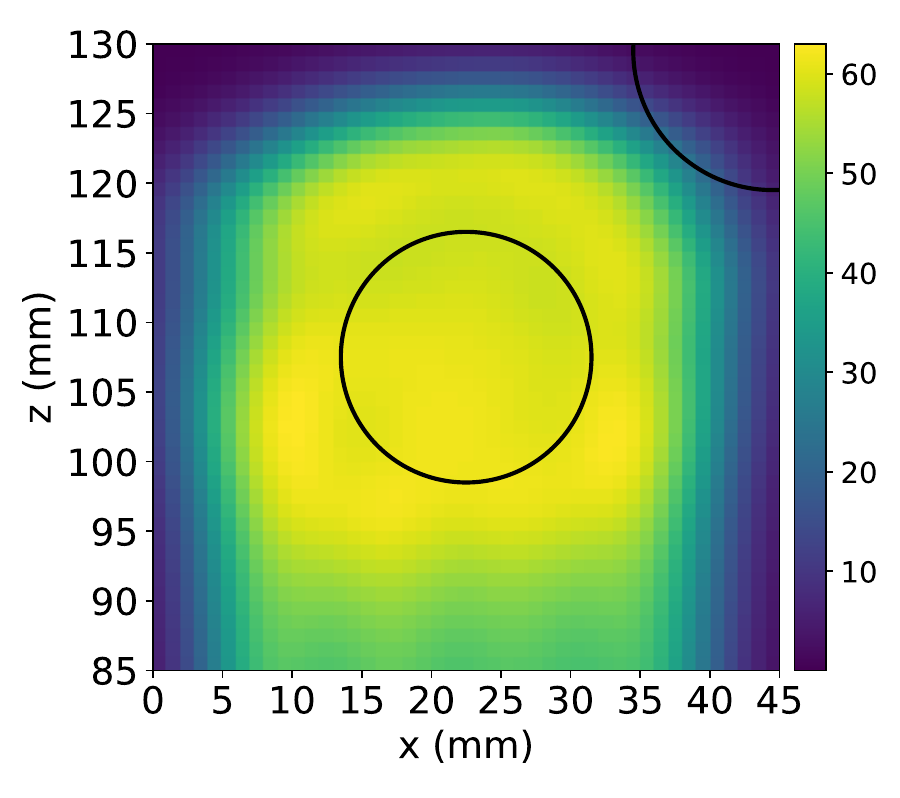}
        \caption{}
    \end{subfigure}
    \hfill
    \begin{subfigure}[b]{0.32\textwidth}
        \includegraphics[width=\linewidth]{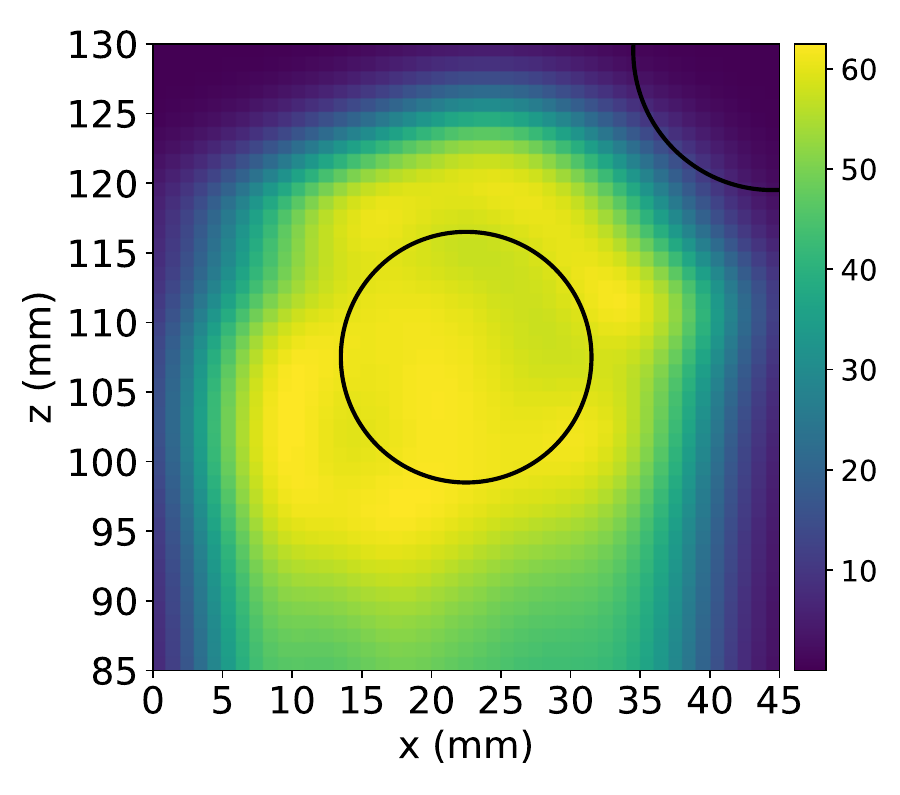}
        \caption{}
    \end{subfigure}

    %\vspace{0.5cm}

    \begin{subfigure}[b]{0.32\textwidth}
        \includegraphics[width=\linewidth]{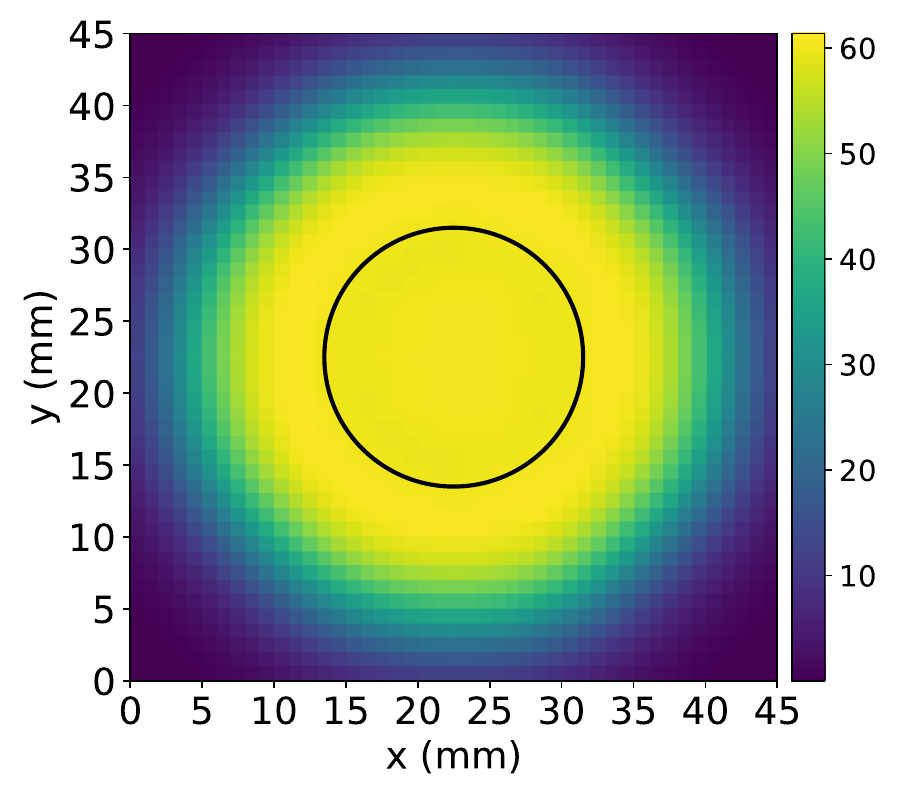}
        \caption{}
    \end{subfigure}
    \hfill
    \begin{subfigure}[b]{0.32\textwidth}
        \includegraphics[width=\linewidth]{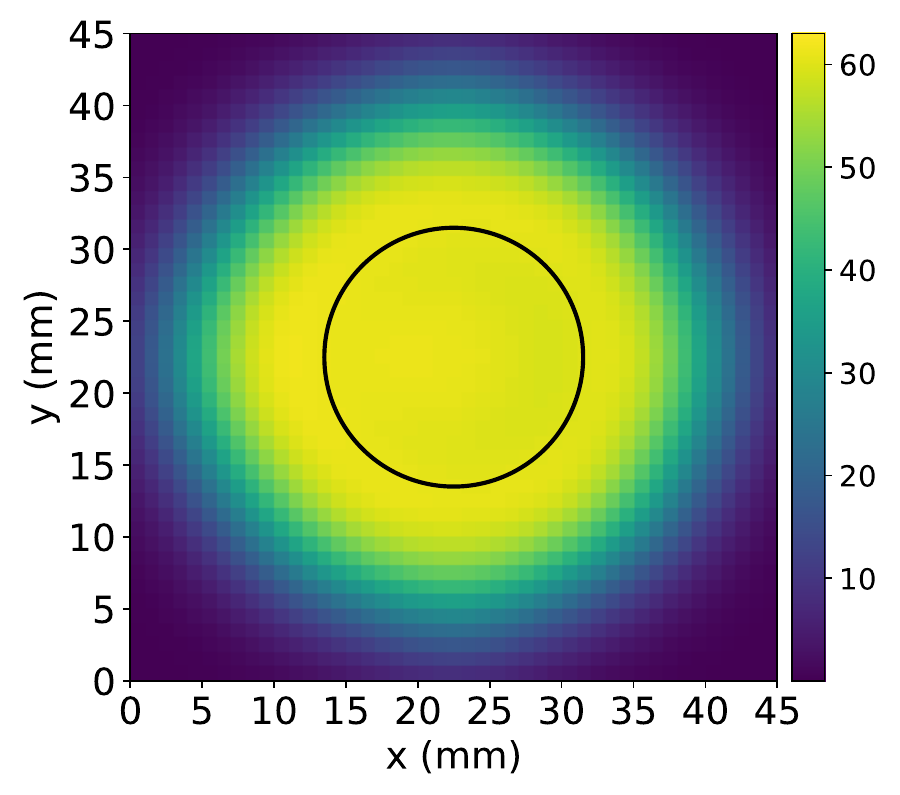}
        \caption{}
    \end{subfigure}
    \hfill
    \begin{subfigure}[b]{0.32\textwidth}
        \includegraphics[width=\linewidth]{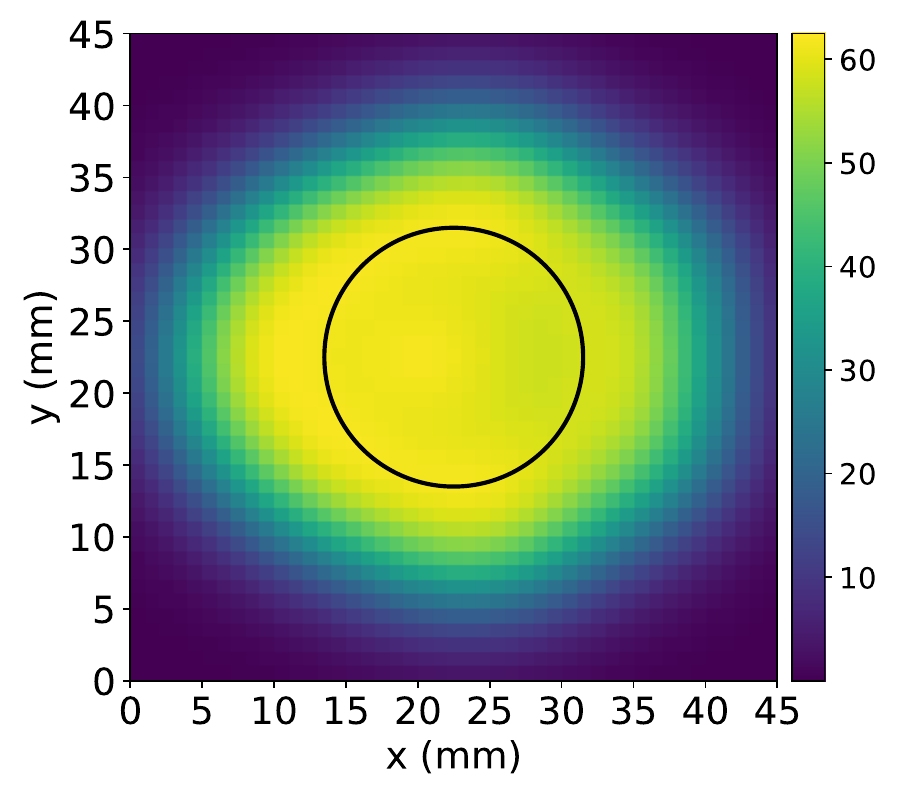}
        \caption{}
    \end{subfigure}

    \caption{Comparison of nominal XZ-displaced dose distributions for the (left) probabilistic plan, with the robust plans that are matched based on the (middle) $D_{50\%}^{50\text{th}}$ and (right) $D_{2\%}^{90\text{th}}$ metrics. Both the (bottom) XY-plane and (top) XZ-plane through the CTV center are shown.}
    \label{fig:sphericalCTVandOAR_XZshift}
\end{figure}

\subsection{The spherical CTV and OAR case}
\label{subsec:Results_CTVandOAR}

% We show the XZ for 50thD50 and 90thD2. We show X-shift in appendix.
We perform similar probabilistic optimizations as in Section \ref{subsec:Results_sphericalCTVonly}, but here we additionally optimize probabilistically for a spherical OAR, either for the XZ-displaced case or for the X-displaced case. Accordingly, besides optimizing for CTV under- and overdosage probability, the OAR overdosage probability is limited as $P(d_i \geq \qty{30}{\gray}) \leq 10\%$. In the following, results for the XZ-displaced case are shown. Results of the X-displaced case are shown in Appendix \ref{app:additionalComparisons_spherical}. Robust plans are obtained by tuning its objective weights, to achieve 1) a similar CTV coverage ($D_{98\%}^{10\text{th}}$) and 2) a similar OAR dose ($D_{2\%}^{90\text{th}}$) as in the probabilistic plan. The robust plan with objective weights of $\{ \omega_{CTV}, \omega_{OAR}, \omega_{OAR}^{max}, \omega_{Tissue} \} = \{120, 1, 1, 160\}$ and $\{100, 10, 10, 100\}$ give similar CTV coverage and OAR dose, respectively. The resulting probabilistic DVH-metrics before and after scaling (within brackets) are shown in Table \ref{tab:DVHmetrics_CTVandOAR_XZ}. \\

The dose distributions for the XZ-displaced case are shown in Figure \ref{fig:sphericalCTVandOAR_XZshift}, where the probabilistic plan (left) is shown together with the robust plans, scaled by matching CTV coverage (middle) and OAR dose (right). We show the XZ-slice (top) and XY-slice (bottom) through the CTV center. The dose distribution of the probabilistic plan shows a slight reduction in the CTV margin on the OAR-side, at the same time being conformal to the other parts of the CTV. Compared to the probabilistic plan, the robust plan with similar CTV coverage (middle row) is less conformal in general and has larger margins on the OAR-side, leading to higher OAR doses. 

\begin{figure}[h!]
    \centering

    \begin{subfigure}[b]{0.32\textwidth}
        \includegraphics[width=\linewidth]{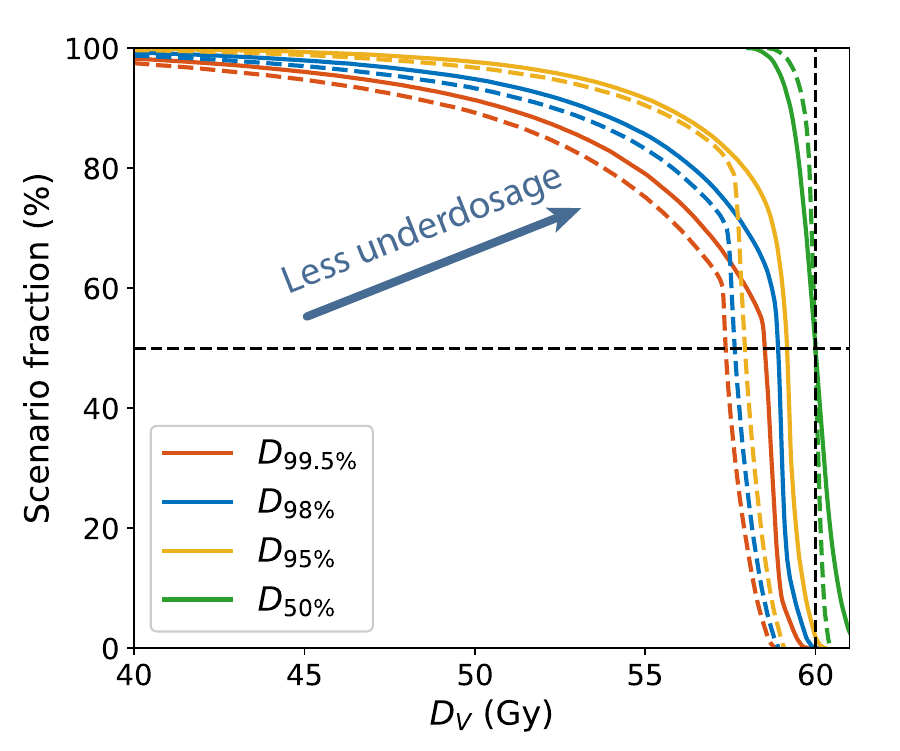}
        \caption{}
        \label{fig:CDFcomparisonSpheres_allDVHmetrics_CTV_50thD50}
    \end{subfigure}
    %\hfill
    \begin{subfigure}[b]{0.32\textwidth}
        \includegraphics[width=\linewidth]{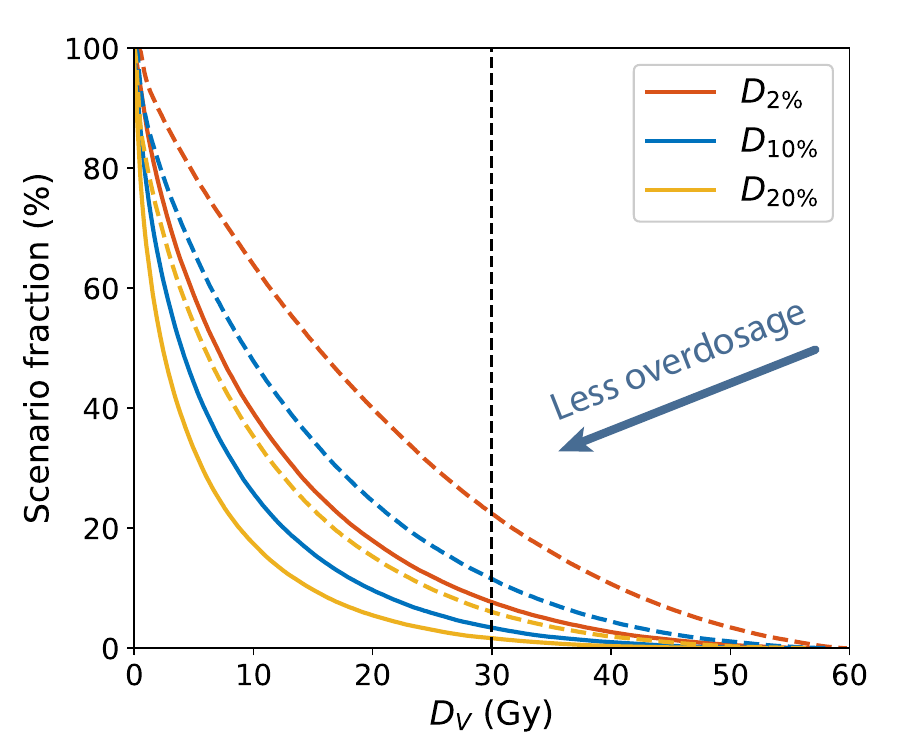}
        \caption{}
        \label{fig:CDFcomparisonSpheres_allDVHmetrics_OAR_50thD50}
    \end{subfigure}

    %\vspace{0.5cm}

    \begin{subfigure}[b]{0.32\textwidth}
        \includegraphics[width=\linewidth]{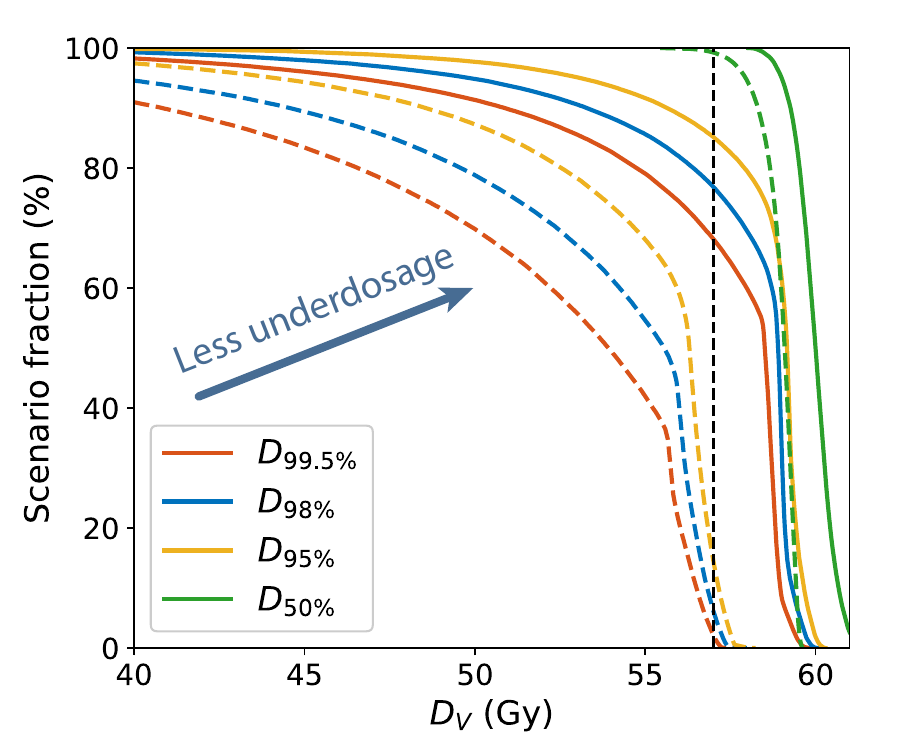}
        \caption{}
        \label{fig:CDFcomparisonSpheres_allDVHmetrics_CTV_90thD2}
    \end{subfigure}
    %\hfill
    \begin{subfigure}[b]{0.32\textwidth}
        \includegraphics[width=\linewidth]{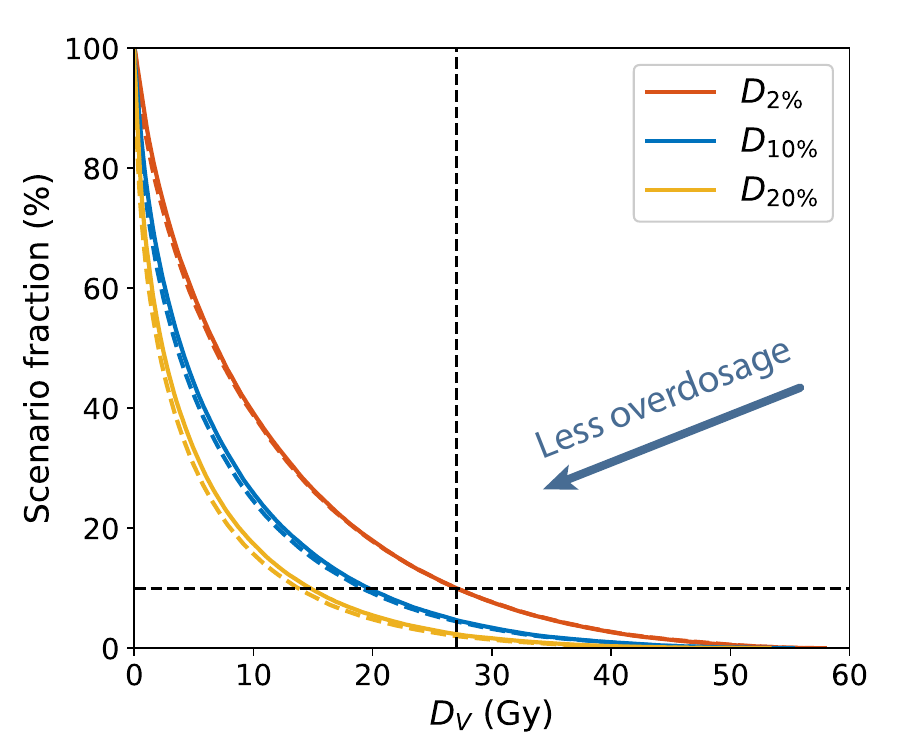}
        \caption{}
        \label{fig:CDFcomparisonSpheres_allDVHmetrics_OAR_90thD2}
    \end{subfigure}

    \caption{Dose population histograms of various DVH metrics ($D_{V}$, e.g., $V=98\%$) comparing XZ-displaced probabilistic (solid) and robust (dashed) plans in the CTV (left) and OAR (right). Plans matched on CTV coverage (top) have similar CTV coverage (a) and reduced OAR overdosage probability (b), whereas matching on (bottom) $D_{2\%}^{90\text{th}}$ reduces CTV underdosage probability (c) with similar OAR dose (d).}
    \label{fig:CDFcomparisonSpheres_allDVHmetrics}
\end{figure}

By increasing the relative OAR objective weight (i.e., its importance in the optimization) with respect to the $D_{50\%}^{50\text{th}}$ comparison, the margin at the OAR-side is reduced, leading to a similar OAR dose ($D_{2\%}^{90\text{th}}$). At the same time, this leads to reduced CTV conformity in both the XY- and XZ-plane. Since the OAR has an increased importance, shifts into the OAR (i.e., diagonal shifts) become worst-case scenarios more frequently during the optimization. This has two main effects. Firstly, the CTV margin in the $y$-direction is reduced because $y$-shifts are less likely to be worst-case scenarios (diagonal shifts lead to more OAR overdosage than $y$-shifts). In turn, this leads to a significant drop in CTV coverage. Secondly, a dose extension appears on the opposite side of the OAR (around $\left( x,z \right) = \left( \qty{12.5}{\milli\meter}, \qty{97.5}{\milli\meter} \right)$), ensuring that CTV coverage remains in cases where high-dose regions are shifted into the OAR. \\

% DPH figures
The CTV coverage of the plans is compared in Figure \ref{fig:CDFcomparisonSpheres_allDVHmetrics}, where the DPH of various DVH-metrics is compared between the probabilistic plan and either the (top) $D_{50\%}^{50\text{th}}$ or the (bottom) $D_{2\%}^{90\text{th}}$ scaled robust plan. For the former, CTV coverage (especially the high scenario fraction region) is similar and significant differences in the near-maximum OAR DVH-metrics can be seen. Specifically, in the probabilistic plan the $D_{2\%}$ is larger than \qty{30}{\gray} in about 7.5\% of the scenarios whereas this probability is increased to about 22.5\% in the scaled robust case.

\begin{figure}[t!]
    \centering

    \begin{subfigure}[b]{0.32\textwidth}
        \includegraphics[width=\linewidth]{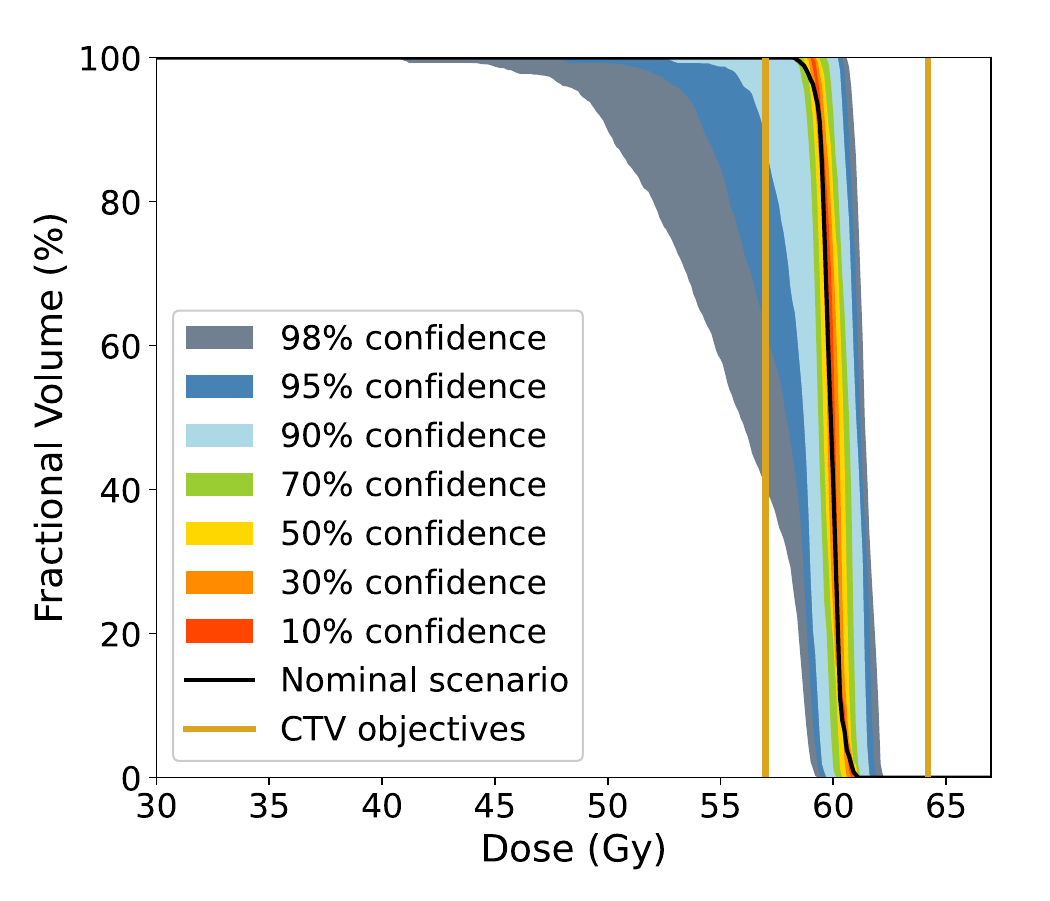}
        \caption{}
    \end{subfigure}
    %\hfill
    \begin{subfigure}[b]{0.32\textwidth}
        \includegraphics[width=\linewidth]{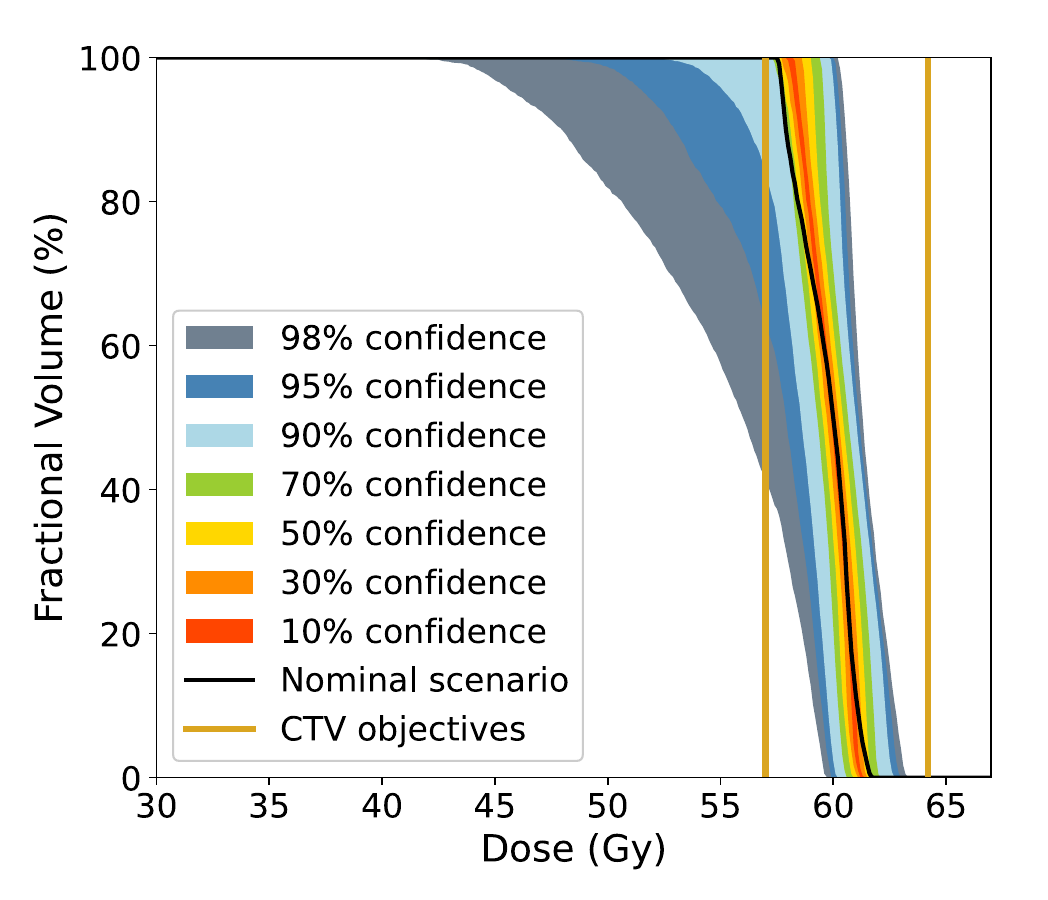}
        \caption{}
    \end{subfigure}
    %\hfill
    \begin{subfigure}[b]{0.32\textwidth}
        \includegraphics[width=\linewidth]{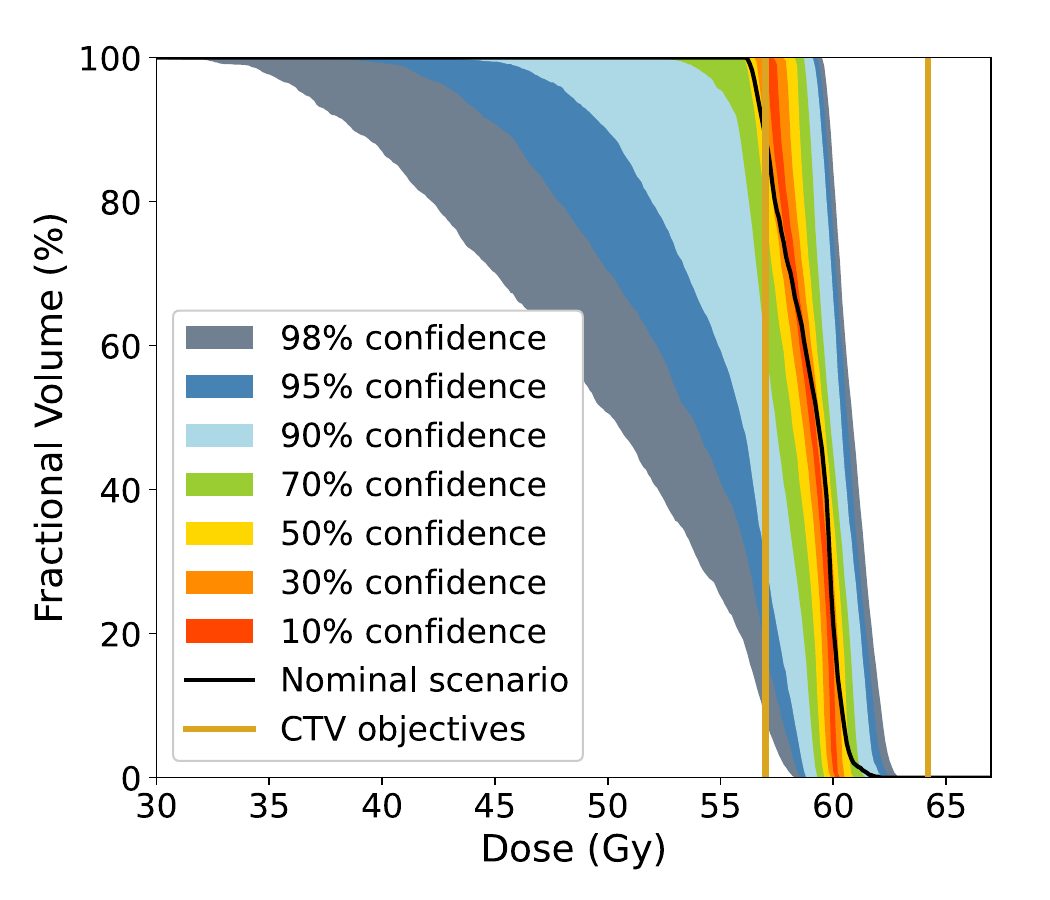}
        \caption{}
    \end{subfigure}

    %\vspace{0.5cm}

    \begin{subfigure}[b]{0.32\textwidth}
        \includegraphics[width=\linewidth]{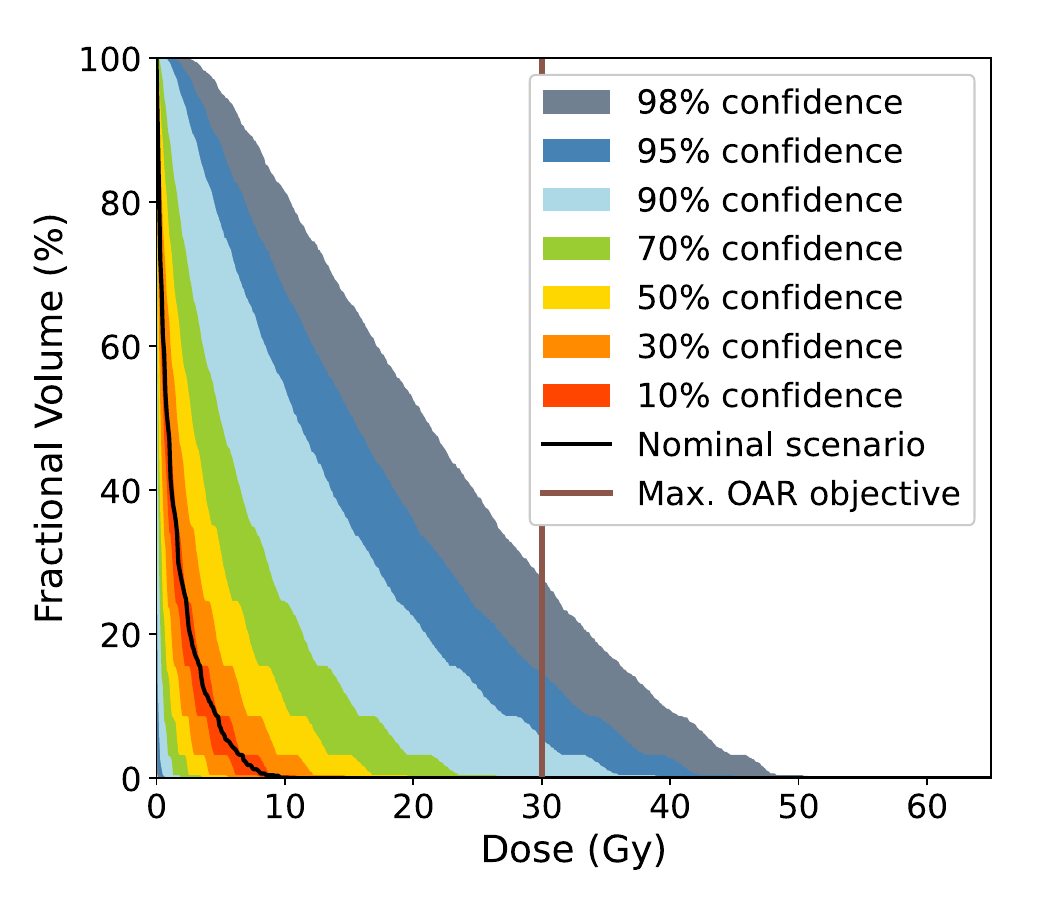}
        \caption{}
    \end{subfigure}
    %\hfill
    \begin{subfigure}[b]{0.32\textwidth}
        \includegraphics[width=\linewidth]{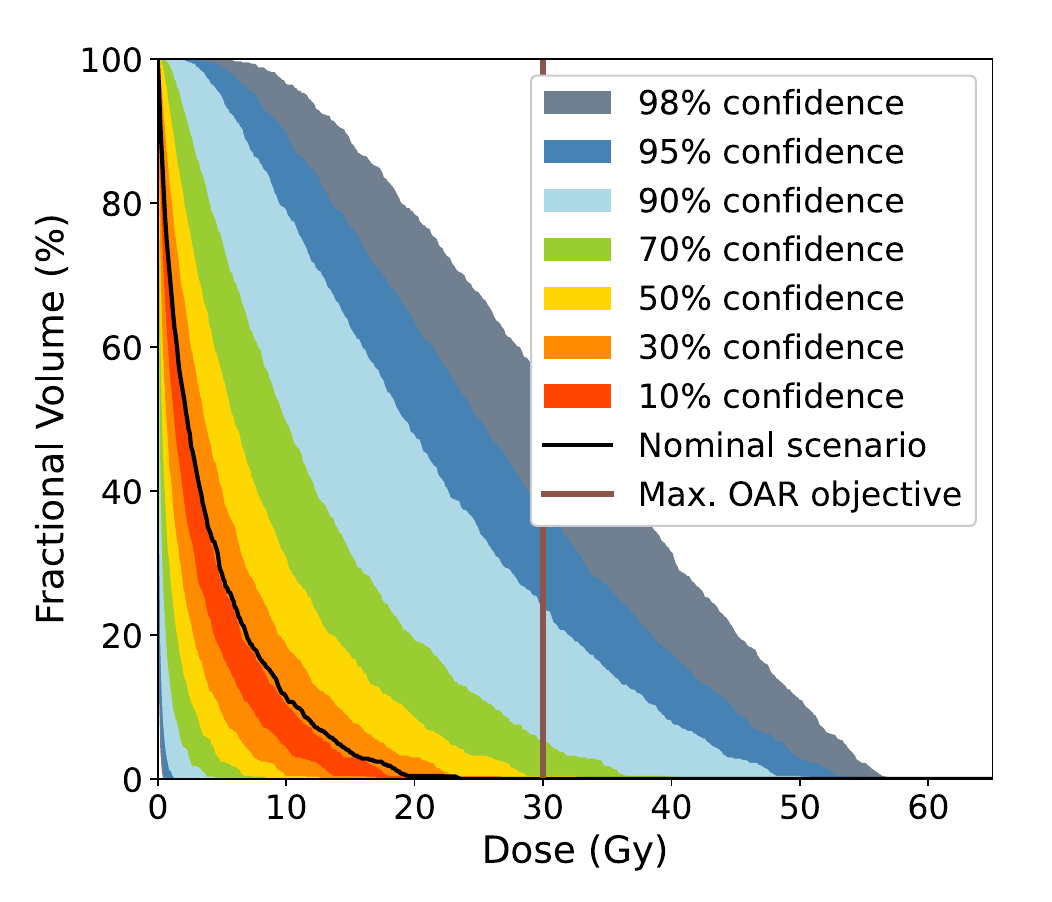}
        \caption{}
    \end{subfigure}
    %\hfill
    \begin{subfigure}[b]{0.32\textwidth}
        \includegraphics[width=\linewidth]{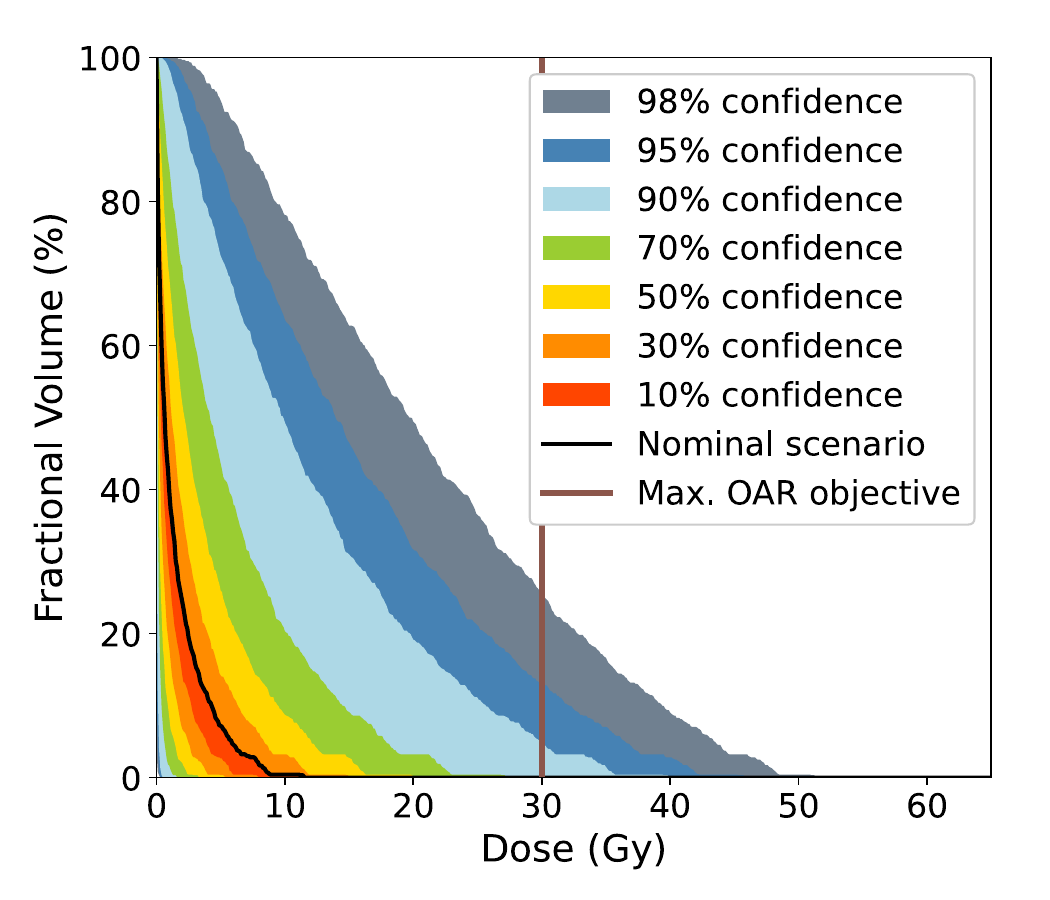}
        \caption{}
    \end{subfigure}

    \caption{Dose volume histogram distributions for the CTV (top) and OAR (bottom), comparing the XZ-displaced probabilistic (left) and robust plans that are matched based on the (middle) $D_{50\%}^{50\text{th}}$ and (right) $D_{2\%}^{90\text{th}}$ metrics. The nominal scenario (black) is shown together with various confidence bands.}
    \label{fig:CTVandOAR_DVHbands}
\end{figure}

The $D_{2\%}^{90\text{th}}$ scaled robust plan has similar near-maximum OAR DVH-metrics, whereas significant difference in the CTV coverage is visible. In the scaled robust plan, $P(D_{98\%} > \qty{57}{\gray}) \approx 6\%$, whereas this probability increases to approximately 77\% in the robust case (i.e., about 70\% increase). 

% XZ shift (scaled 50thD50)
%               Probabilistic       robust
% D2 at 30Gy:   7.7%                22.4

% XZ shift (scaled 90thD2)
%               Probabilistic       robust
% D98 at 57Gy:  76.9%               6.2

In Figure \ref{fig:CTVandOAR_DVHbands}, we compare the DVH bands of CTV (top) and OAR (bottom) between the probabilistic (left) and robust plans that are matched based on the (middle) $D_{50\%}^{50\text{th}}$ and (right) $D_{2\%}^{90\text{th}}$ metrics. The probabilistic plan shows more CTV homogeneity in the presence of uncertainty (as well as in the nominal plan) and shows a smaller spread in the DVH distributions of the CTV, compared to both robust plans. The scaled $D_{2\%}^{90\text{th}}$ robust plan was matched to the probabilistic plan by OAR dose, resulting in similar DVH-distributions for the OAR. Compared to the probabilistic plan, the DVH-distributions of the $D_{50\%}^{50\text{th}}$ scaled robust plan are wider (i.e., less robust) and shifted towards larger dose (i.e., increased OAR overdosage probability). \\

\subsection{The spinal case}
\label{subsec:Results_spinal}

\begin{table}[b!]
\caption{\label{tab:DVHmetrics_spinal}Statistical DVH-metrics to compare the CTV coverage and spinal overdosage of the robust and probabilistic spinal plans. Metrics corresponding to the robust plans after scaling the beam weights are shown in brackets.} 
\begin{center}
\begin{tabular}{@{}*{4}{l}}
\toprule                             
 & $D_{98\%}^{10\text{th}}$ (\unit{\gray}) & $D_{50\%}^{50\text{th}}$ (\unit{\gray}) & $D_{2\%}^{90\text{th}}$ (\unit{\gray}) \\[0.5ex]
\toprule 
Robust (equivalent of $\nu = 90\%$) & 33.1 (33.1)  &  60.0 (60.0) & 54.2 (54.3) \cr % with robust weights: 2 1 1 1 4
Probabilistic ($\nu = 90\%$) &  32.0 (31.6)   & 60.7 (60.0) & 54.8 (54.1) \cr 
\midrule
Robust (equivalent of $\nu = 95\%$)  &  30.4 (30.4) &  59.9 (60.0)  & 52.9 (52.9) \cr  % with robust weights: 3 2 2 2 6
Probabilistic ($\nu = 95\%$)   &  31.0 (30.6) &  60.8 (60.0)  & 52.7 (52.1) \cr 
\midrule
Robust (equivalent of $\nu = 98\%$)    &  27.5 (27.6) & 59.9 (60.0)  & 51.2 (51.3) \cr % with robust weights: 11 10 10 10 22
Probabilistic ($\nu = 98\%$) & 27.9 (27.5) &  60.9 (60.0)  & 48.7 (48.0)     \cr   
\bottomrule
\end{tabular}
\end{center}
\end{table}

We perform 3 optimizations for the spinal geometry as depicted as in Figure \ref{fig:geometries}, where the OAR overdosage ($>\qty{54}{\gray}$) is prioritized over CTV coverage, with the 3 cases corresponding to optimizing for different probability levels for the OAR ($\nu = 90\%$, $\nu = 95\%$ and $\nu = 98\%$). Robust plans are optimized by matching their CTV coverage (and scaling the $D_{50\%}^{50\text{th}}$) to the probabilistic plan. The statistical DVH-metrics that are used for scaling are shown in Table \ref{tab:DVHmetrics_spinal}.  Dose distributions of the $\nu = 90\%$ case are shown in this esection, while the $\nu = 95\%$ and $\nu = 98\%$ cases are similar and are presented in Appendix \ref{app:additionalComparisons_spinal}.

\begin{figure}[t!]
    \centering

    % First row
    \begin{subfigure}{0.45\textwidth}
        \centering
        \includegraphics[width=\linewidth]{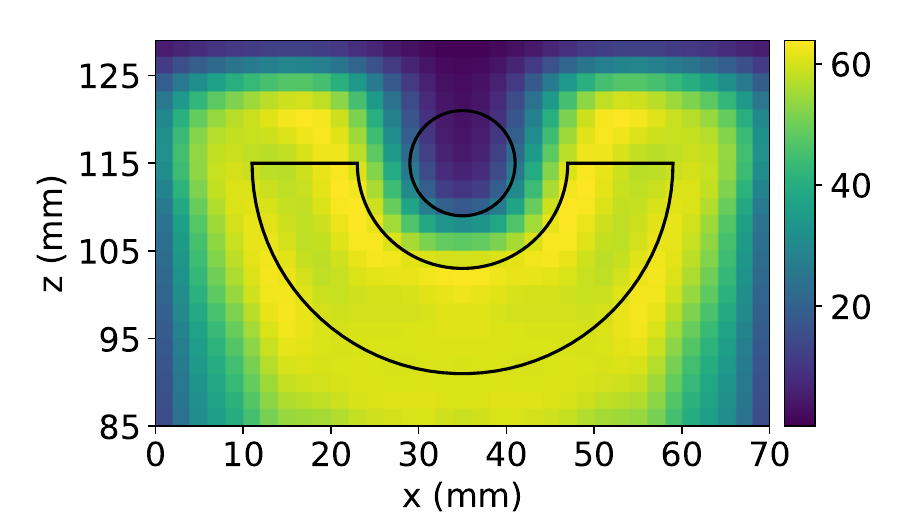}
        \caption{}
    \end{subfigure}
    %\hfill
    \begin{subfigure}{0.45\textwidth}
        \centering
        \includegraphics[width=\linewidth]{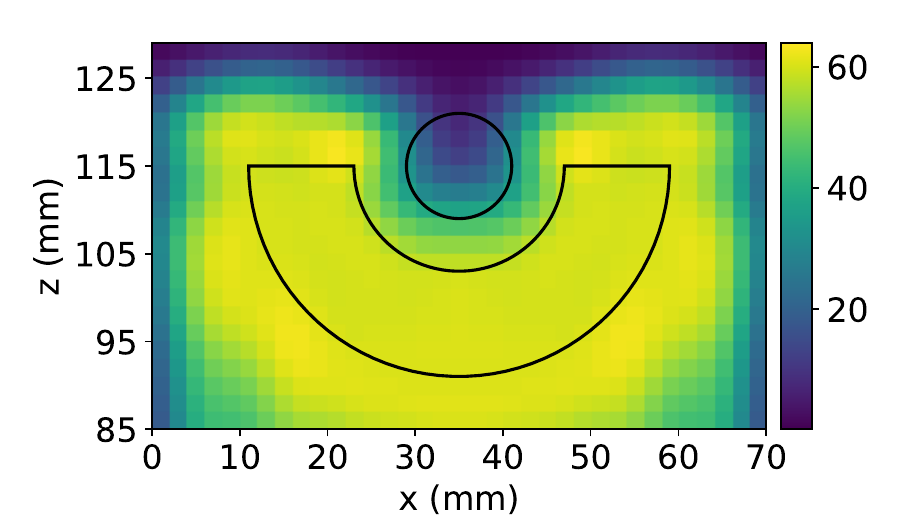}
        \caption{}
    \end{subfigure}

    %\vspace{1em}

    % Second row
    \begin{subfigure}{0.45\textwidth}
        \centering
        \includegraphics[width=\linewidth]{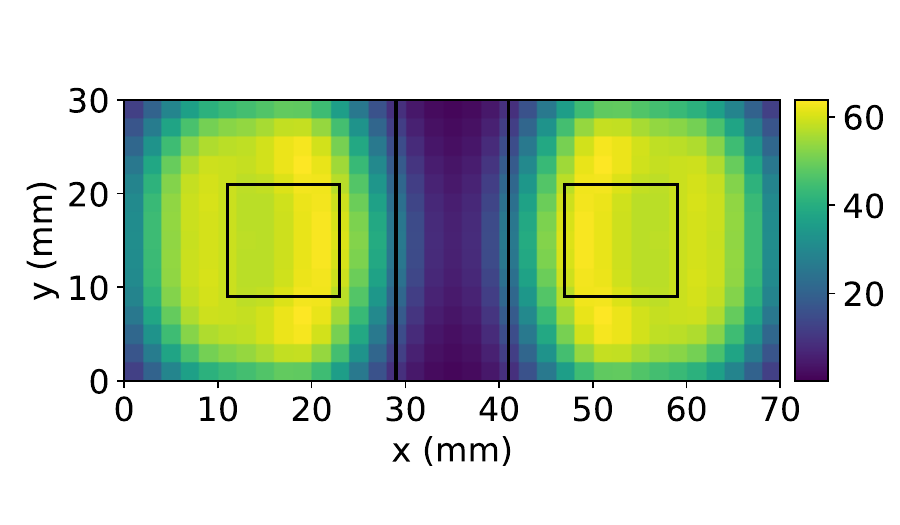}
        \caption{}
    \end{subfigure}
    %\hfill
    \begin{subfigure}{0.45\textwidth}
        \centering
        \includegraphics[width=\linewidth]{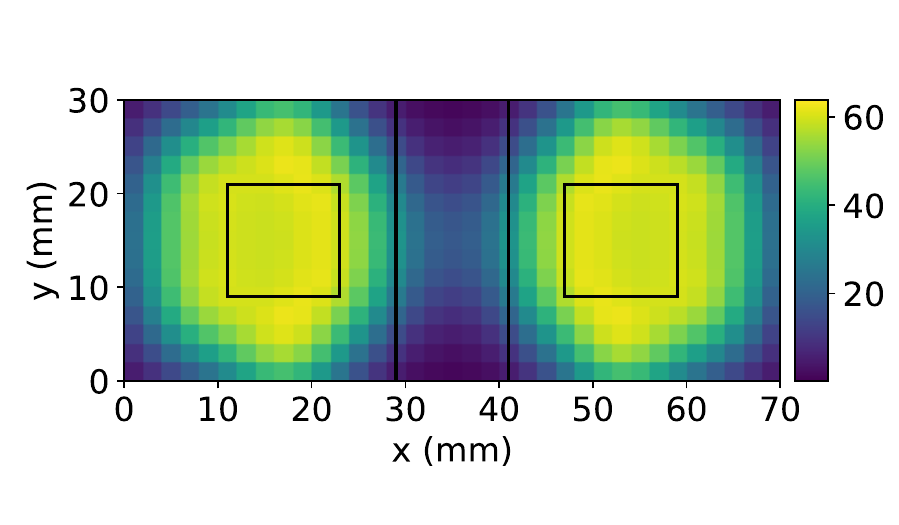}
        \caption{}
    \end{subfigure}

    \caption{Comparison of nominal dose distributions for the (left) probabilistic and (right) robust plan in the spinal case ($\nu = 90\%$). Both the (bottom) XY-plane and (top) XZ-plane through the CTV center are shown.}
    \label{fig:nominalDose_spinal_nu90}
\end{figure}

\begin{figure}[b!]
    \centering
    \begin{subfigure}{0.4\textwidth}
        \centering
        \includegraphics[width=\linewidth]{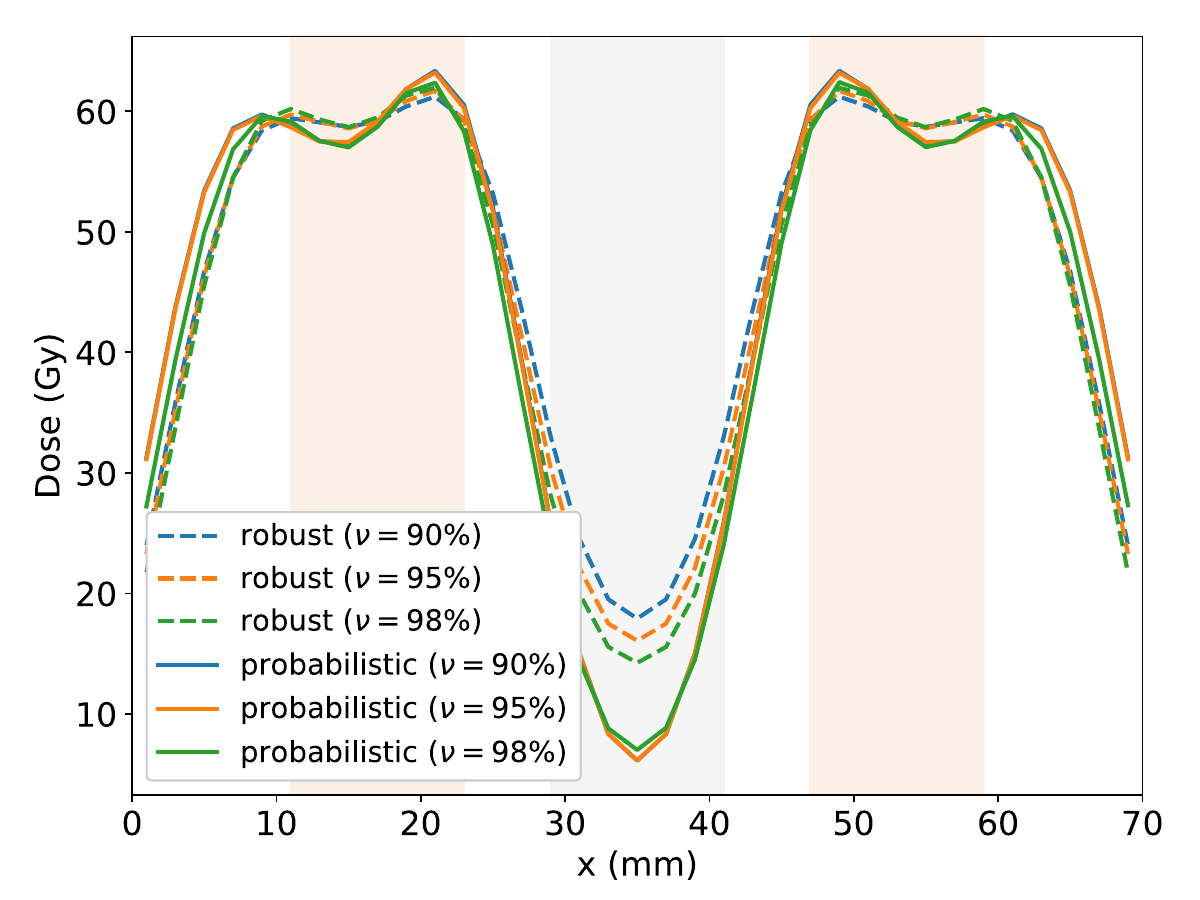}
        \caption{}
        \label{fig:left}
    \end{subfigure}
    %\hfill
    \begin{subfigure}{0.4\textwidth}
        \centering
        \includegraphics[width=\linewidth]{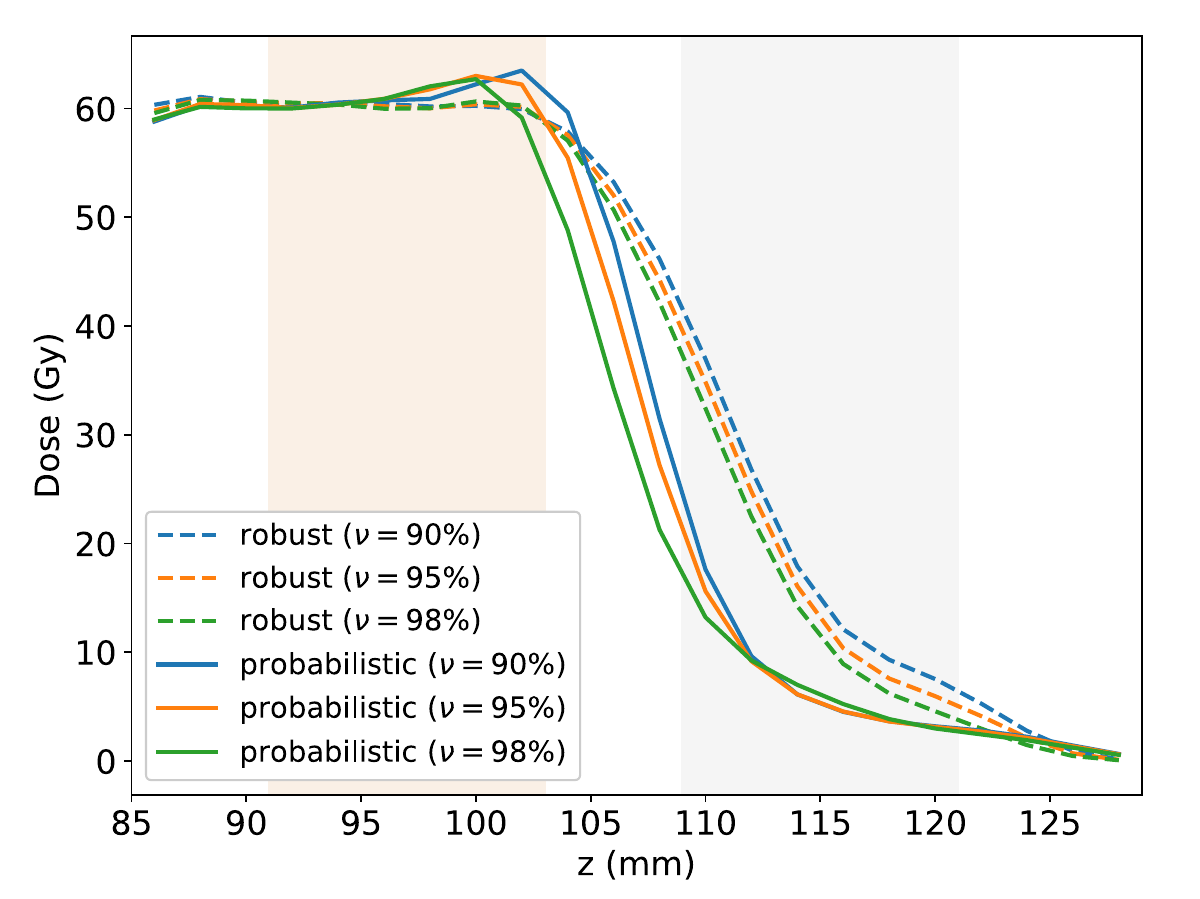}
        \caption{}
        \label{fig:right}
    \end{subfigure}
    
    \caption{Cross sections through the spine center along the $X$ and $Z$ axes for the probabilistic ($\nu = 90\%$, $\nu = 95\%$ and $\nu = 98\%$) and robustly matched plans.}
    \label{fig:nominalDose_spinal_crossSections}
\end{figure}

The resulting dose distributions for $\nu = 90\%$ are shown in Figure \ref{fig:nominalDose_spinal_nu90}, where the XZ-slice (top) and XY-slice (bottom) through the spine center are shown, respectively. The probabilistic plans show improved CTV conformity compared to the robust plans, the difference being especially significant in the XY-plane. A dose build-up occurs at the inner CTV edges (on the spinal side) in order to reduce the spinal dose (at the same time preserving CTV coverage). This effect is seen more clearly in Figure \ref{fig:nominalDose_spinal_crossSections}, where cross sections (for $\nu = 90\%$) along the $X$- and $Z$-axis passing through the spine center are shown, with the cross sections for the $\nu = 95\%$ and $\nu = 98\%$ plan comparisons for completeness. Compared to the robust plans, all probabilistic plans lead to reduction of spinal dose, at the expense of having more inhomogeneous CTV dose. As expected, the dose margin in the probabilistic plan is reduced if spinal overdosage is allowed in less error scenarios (i.e., increased $\nu$). This is especially visible in the range direction (along the $z$-axis). \\

% These are figures scaled such that they can be more easily compared with the robust overdosage/underdosage plots.
\begin{figure}[t]
    \centering

    % First row
    \begin{subfigure}{0.32\textwidth}
        \centering
        \includegraphics[width=\linewidth]{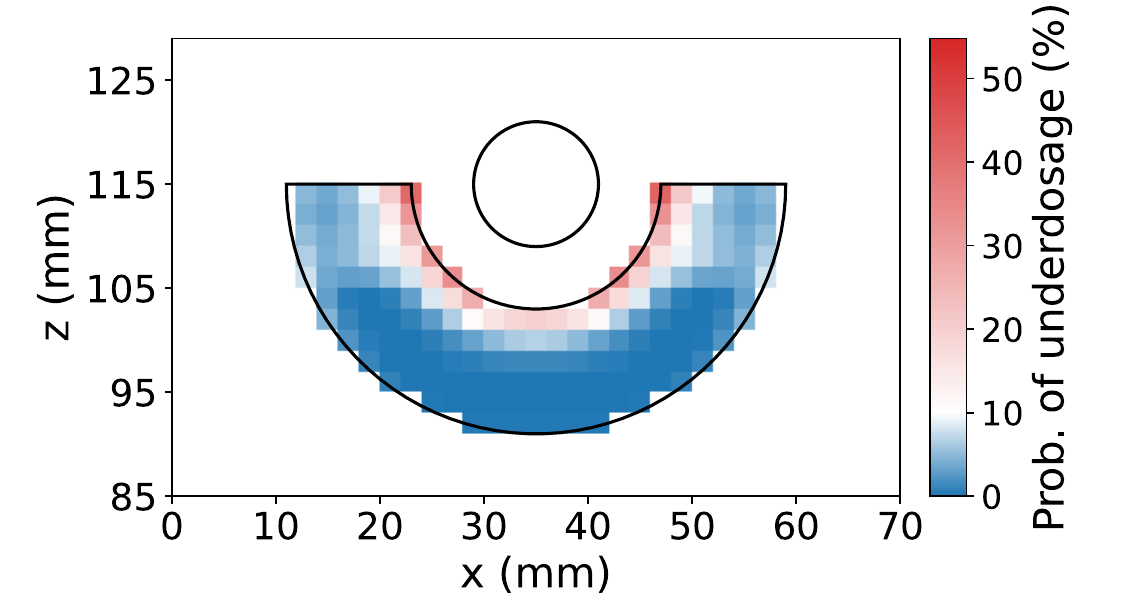}
        \caption{}
    \end{subfigure}
    %\hfill
    \begin{subfigure}{0.32\textwidth}
        \centering
        \includegraphics[width=\linewidth]{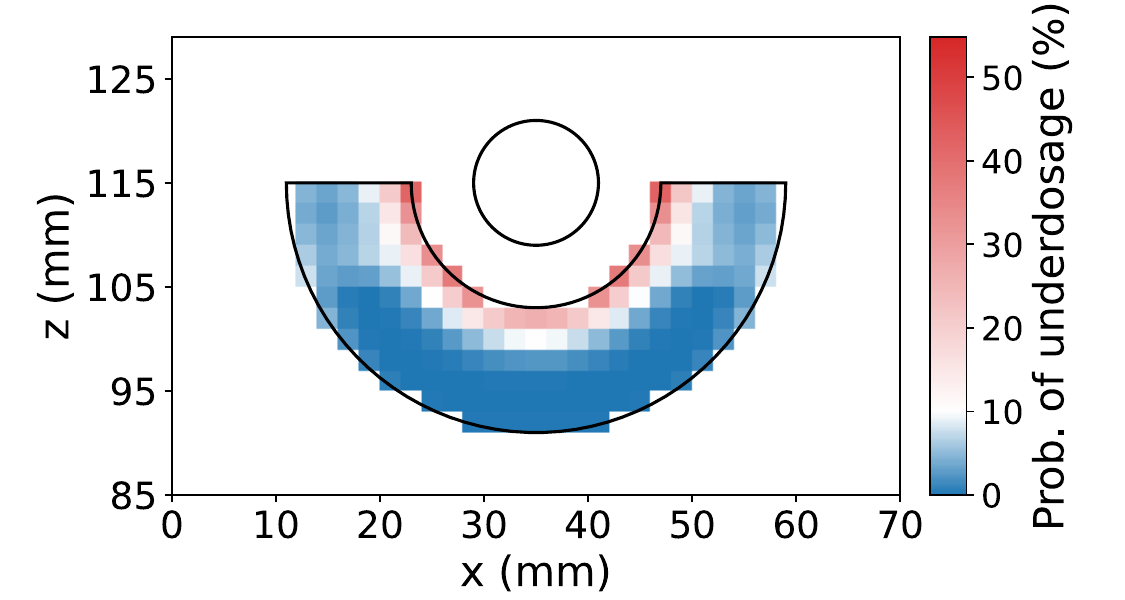}
        \caption{}
    \end{subfigure}
    %\hfill
    \begin{subfigure}{0.32\textwidth}
        \centering
        \includegraphics[width=\linewidth]{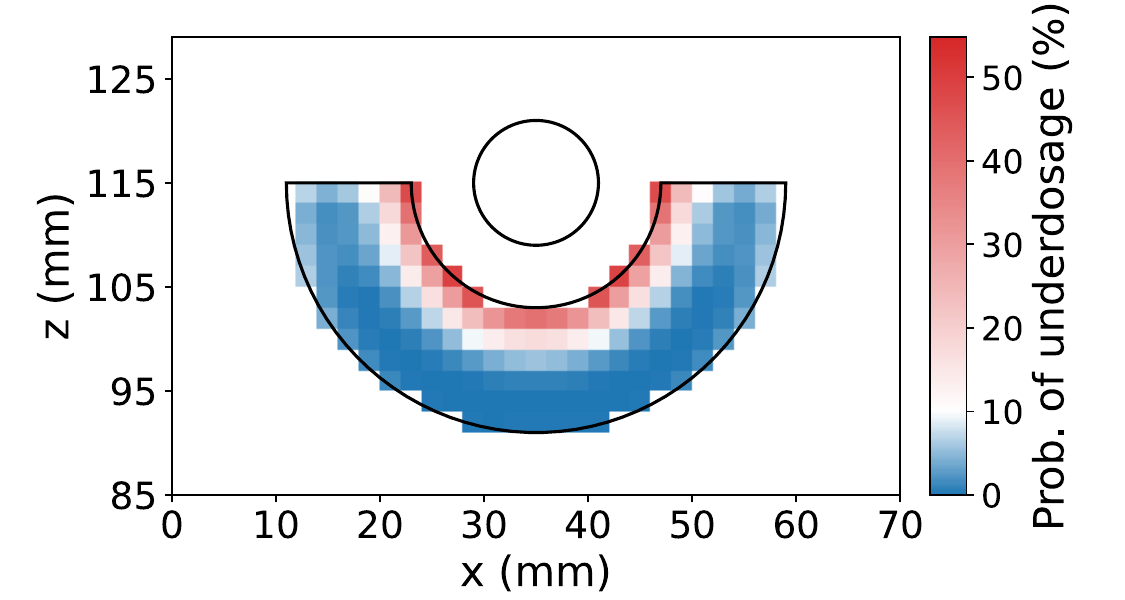}
        \caption{}
    \end{subfigure}

    %\vspace{1em}

    % Second row
    \begin{subfigure}{0.32\textwidth}
        \centering
        \includegraphics[width=\linewidth]{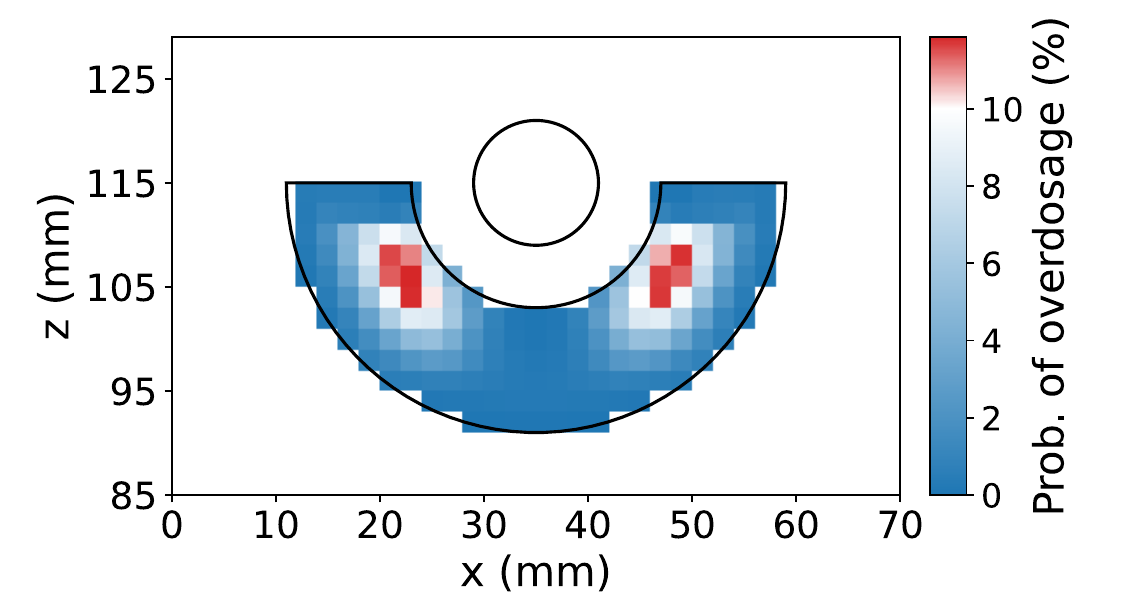}
        \caption{}
    \end{subfigure}
    %\hfill
    \begin{subfigure}{0.32\textwidth}
        \centering
        \includegraphics[width=\linewidth]{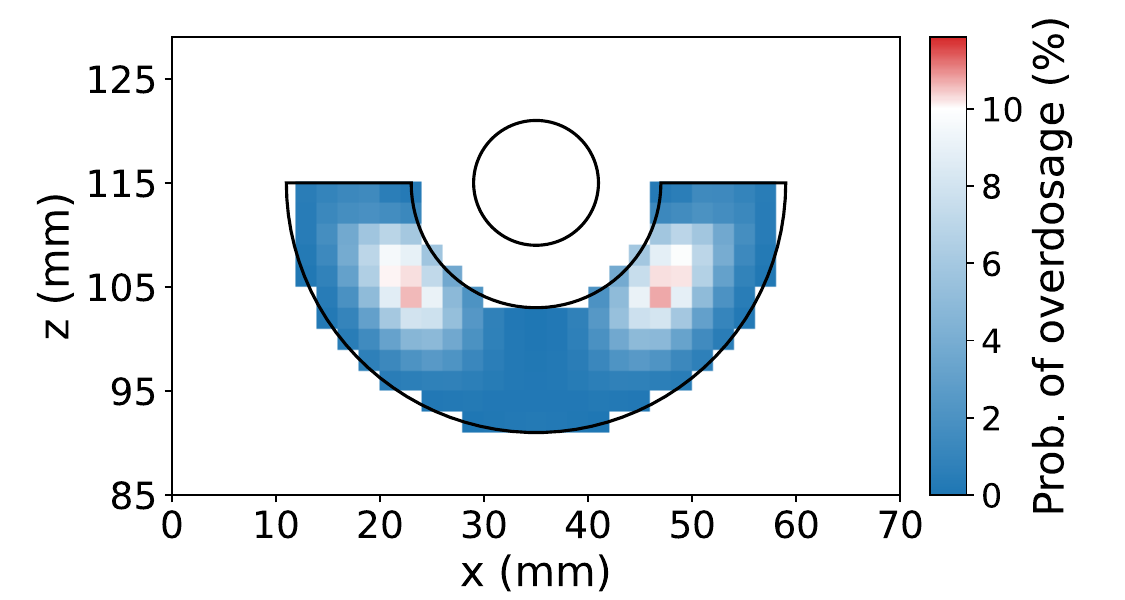}
        \caption{}
    \end{subfigure}
    %\hfill
    \begin{subfigure}{0.32\textwidth}
        \centering
        \includegraphics[width=\linewidth]{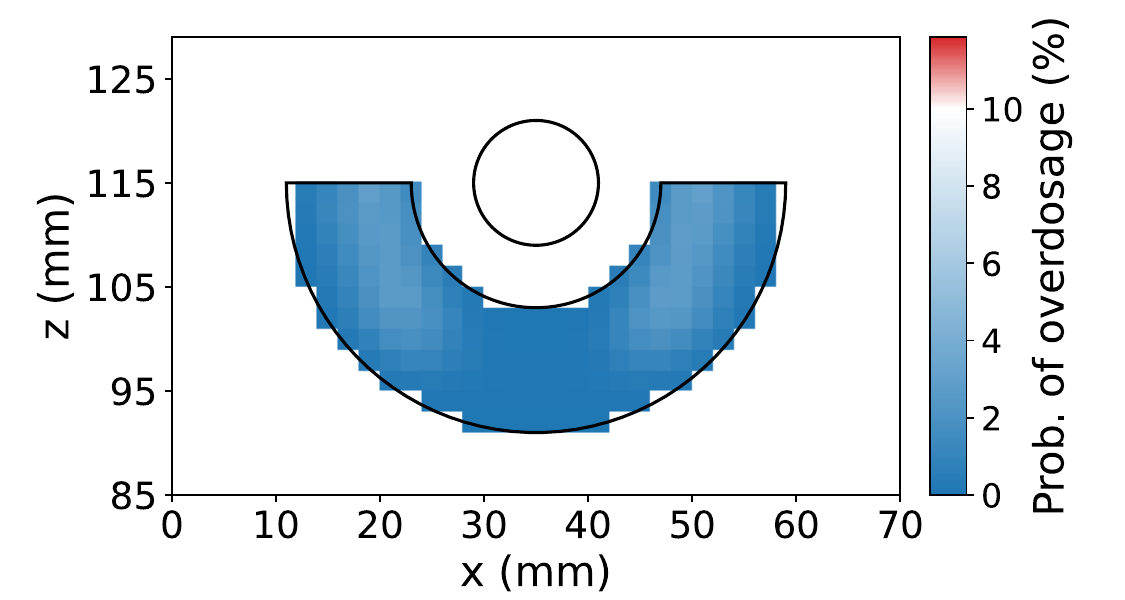}
        \caption{}
    \end{subfigure}

    %\vspace{1em}

    % Third row
    \begin{subfigure}{0.32\textwidth}
        \centering
        \includegraphics[width=\linewidth]{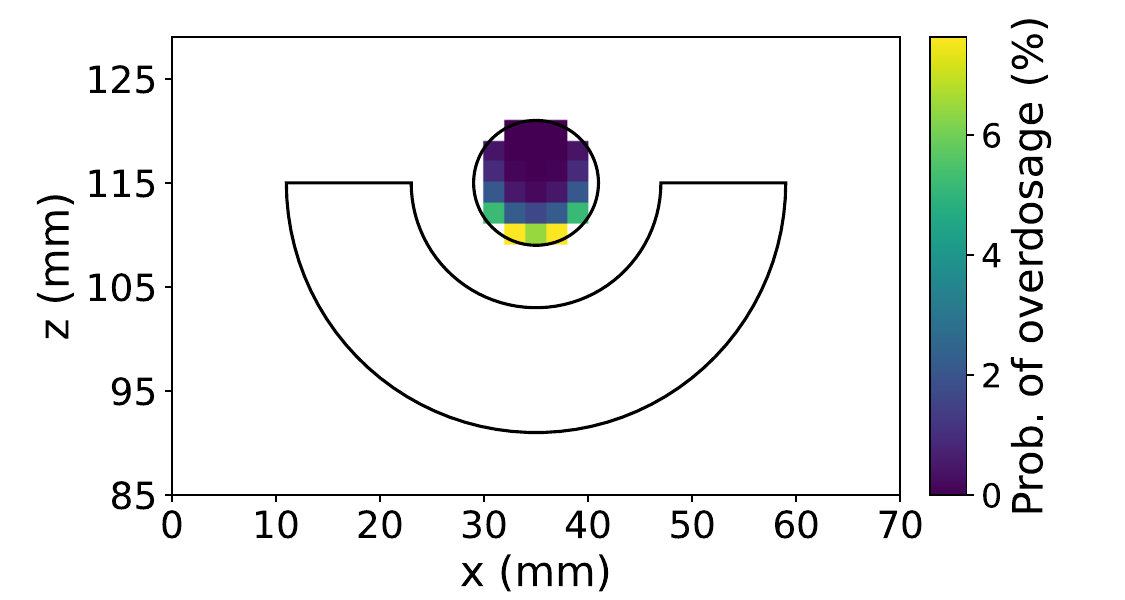}
        \caption{}
    \end{subfigure}
    %\hfill
    \begin{subfigure}{0.32\textwidth}
        \centering
        \includegraphics[width=\linewidth]{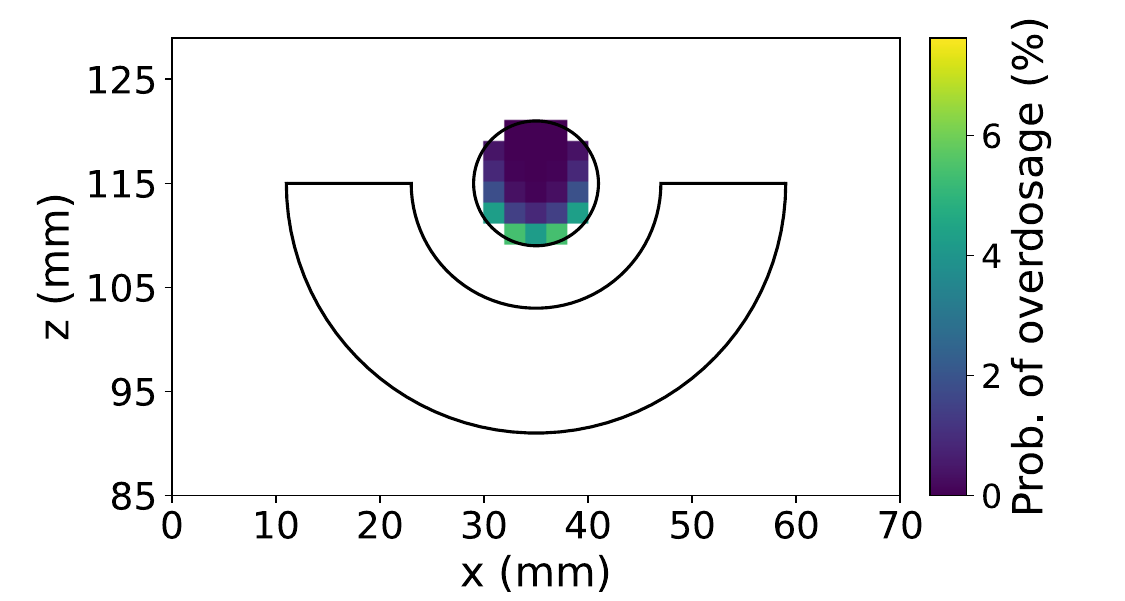}
        \caption{}
    \end{subfigure}
    %\hfill
    \begin{subfigure}{0.32\textwidth}
        \centering
        \includegraphics[width=\linewidth]{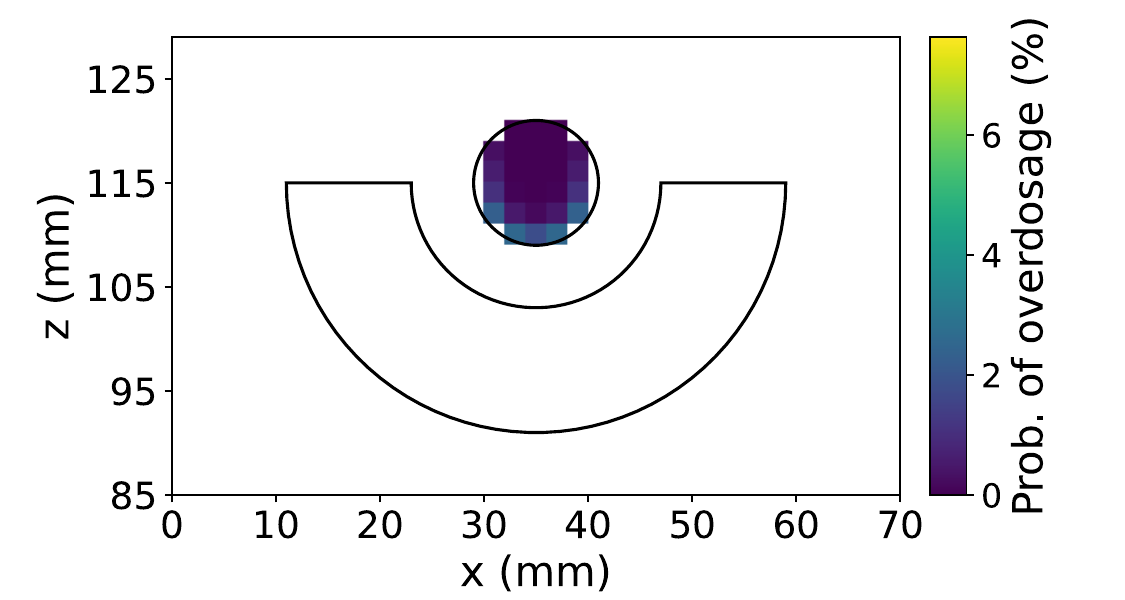}
        \caption{}
    \end{subfigure}

    \caption{The probability of CTV underdosage (top), CTV overdosage (middle) and spine overdosage (bottom) probability, for the (left to right) $\nu = 90\%$, $\nu = 95\%$ and $\nu = 98\%$ cases. Probabilistic CTV objectives are reached if the probability of under- and overdosage is below 10\% (blue).}
    \label{fig:spinal_probUnderOverDosage}
\end{figure}

\begin{figure}[h!]
    \centering
    \begin{subfigure}{0.35\textwidth}
        \centering
        \includegraphics[width=\linewidth]{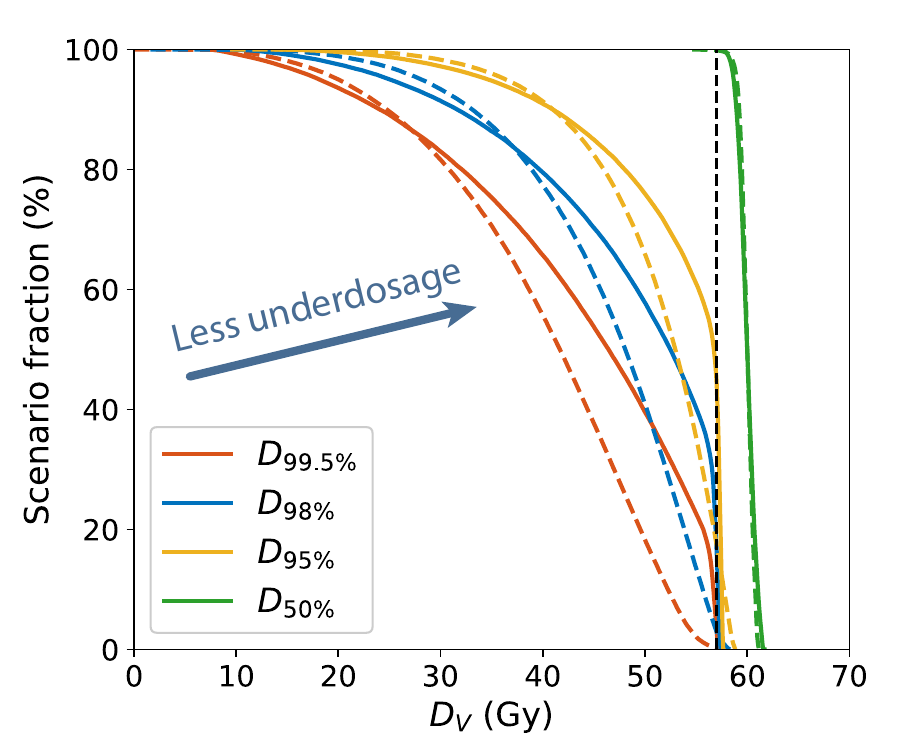}
        \caption{}
        \label{fig:spinal_CDF_nu90_CTV}
    \end{subfigure}
    %\hfill
    \begin{subfigure}{0.35\textwidth}
        \centering
        \includegraphics[width=\linewidth]{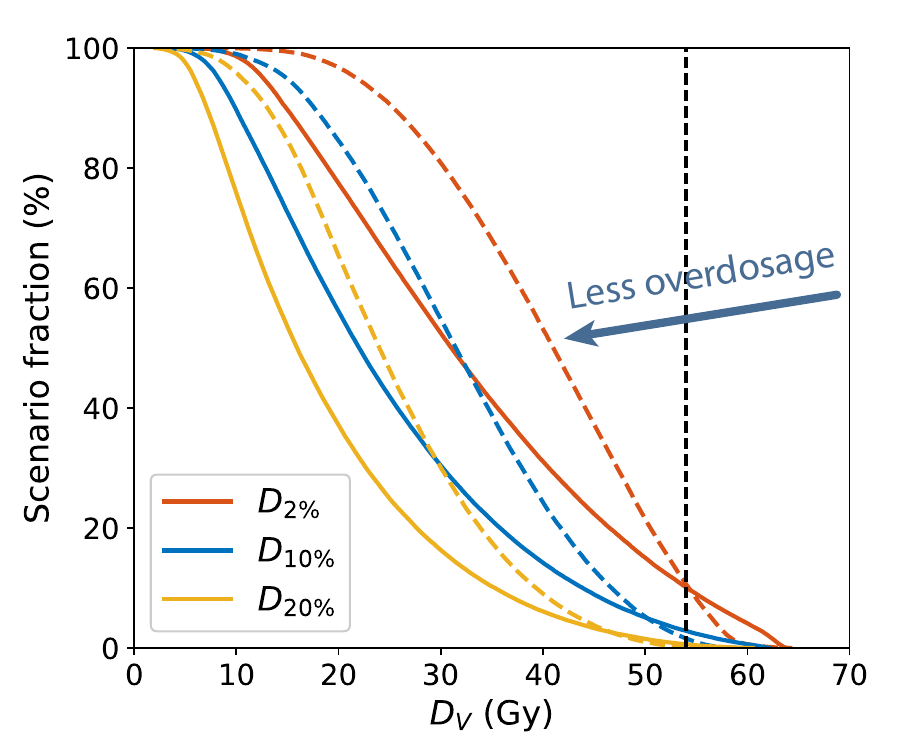}
        \caption{}
        \label{fig:spinal_CDF_nu90_OAR}
    \end{subfigure}
    
    \caption{Dose population histograms of various DVH metrics ($D_{V}$, e.g., $V=98\%$) comparing spinal probabilistic (solid) and robust (dashed) plans for $\nu = 90\%$ in the CTV (a) and spine (b), scaled by $D_{50\%}^{50\text{th}}$. The probabilistic plan shows reduced spinal overdosage probability and reduced CTV underdosage probability in the same plan.}
    \label{fig:spinal_DPH_nu90}
\end{figure}

\begin{figure}[b!]
    \centering

    \begin{subfigure}[b]{0.35\textwidth}
        \includegraphics[width=\linewidth]{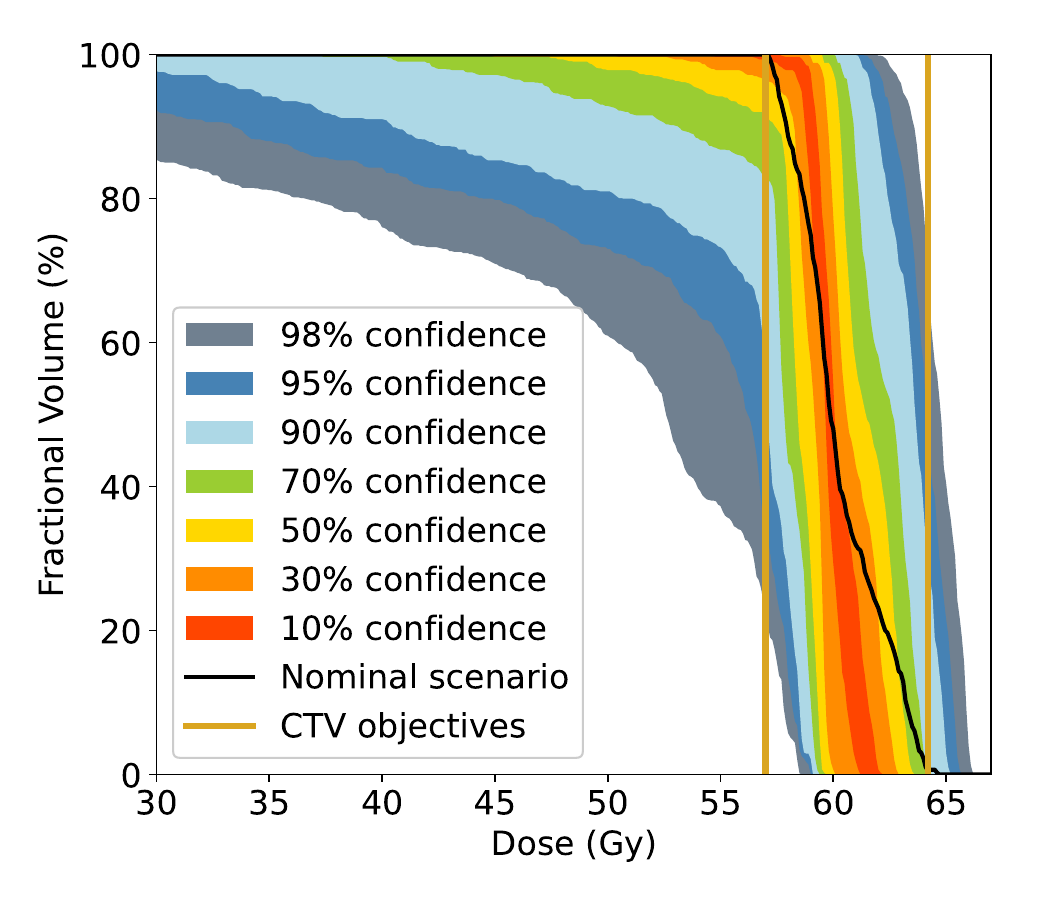}
        \caption{}
    \end{subfigure}
    %\hfill
    \begin{subfigure}[b]{0.35\textwidth}
        \includegraphics[width=\linewidth]{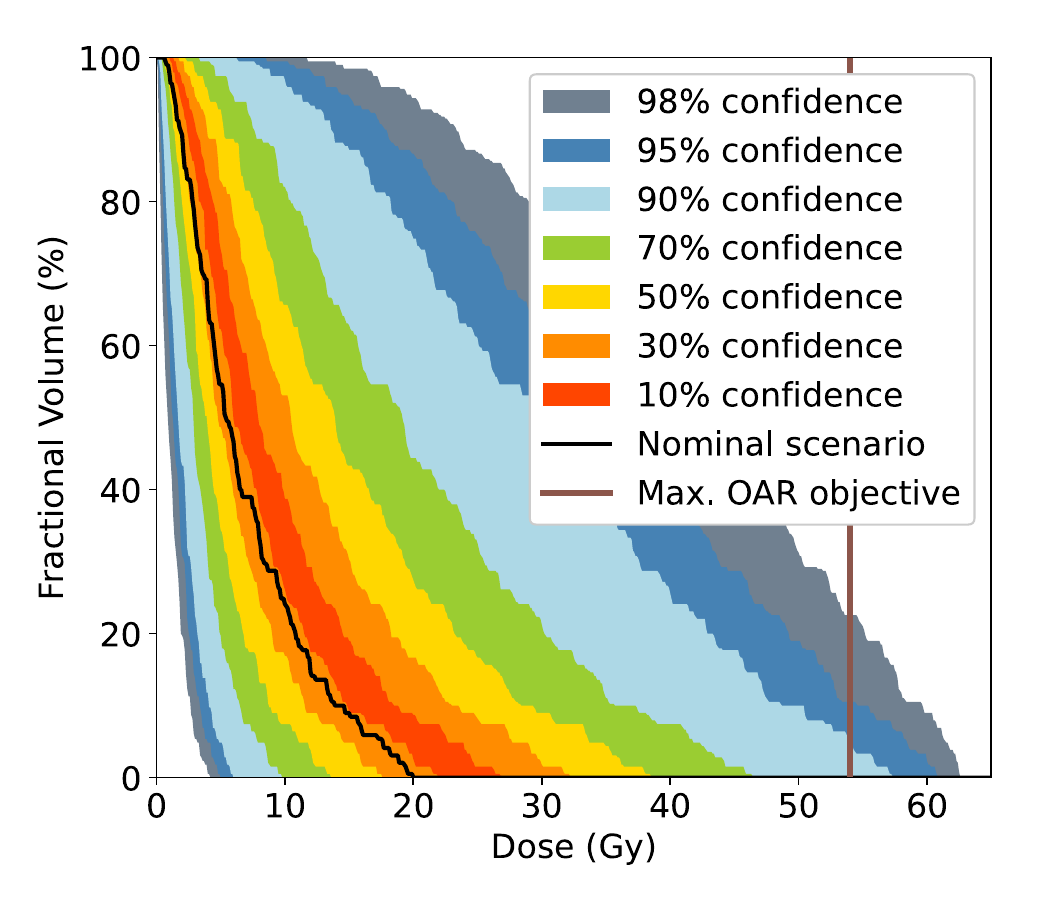}
        \caption{}
    \end{subfigure}
    
    %\vspace{0.5cm}

    \begin{subfigure}[b]{0.35\textwidth}
        \includegraphics[width=\linewidth]{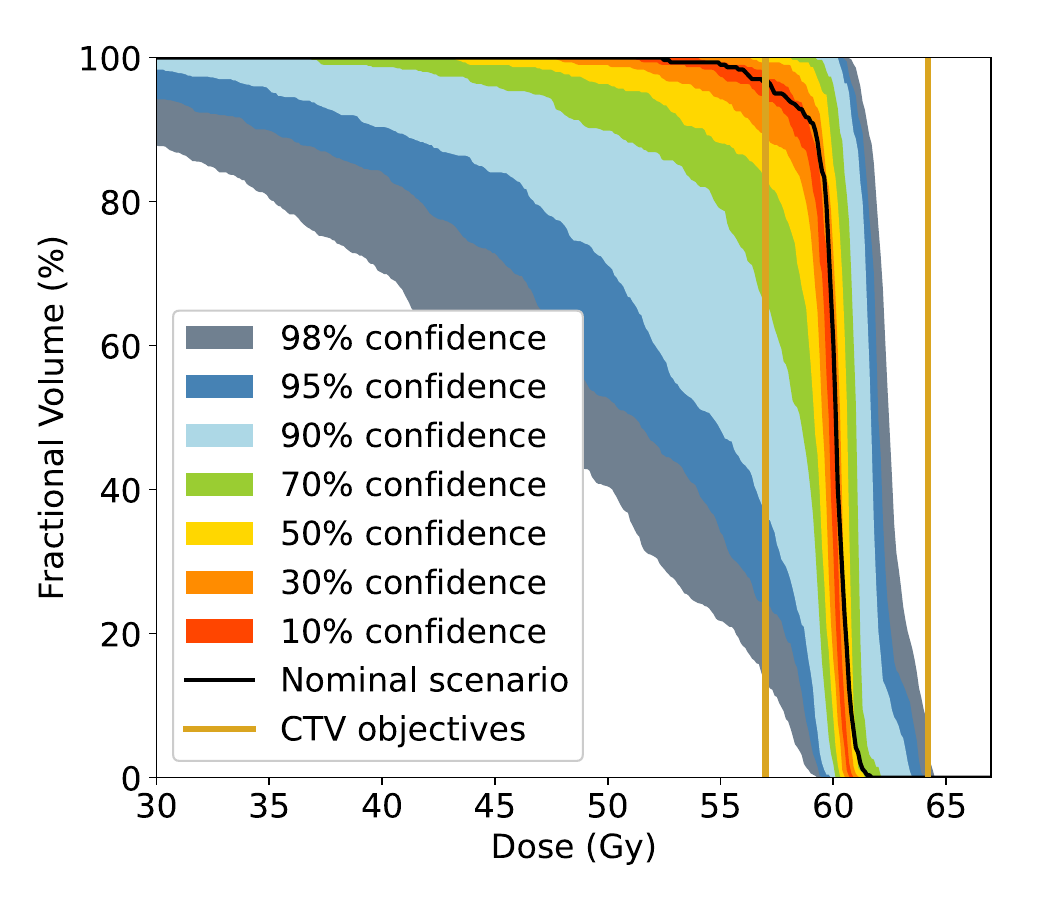}
        \caption{}
    \end{subfigure}
    %\hfill
    \begin{subfigure}[b]{0.35\textwidth}
        \includegraphics[width=\linewidth]{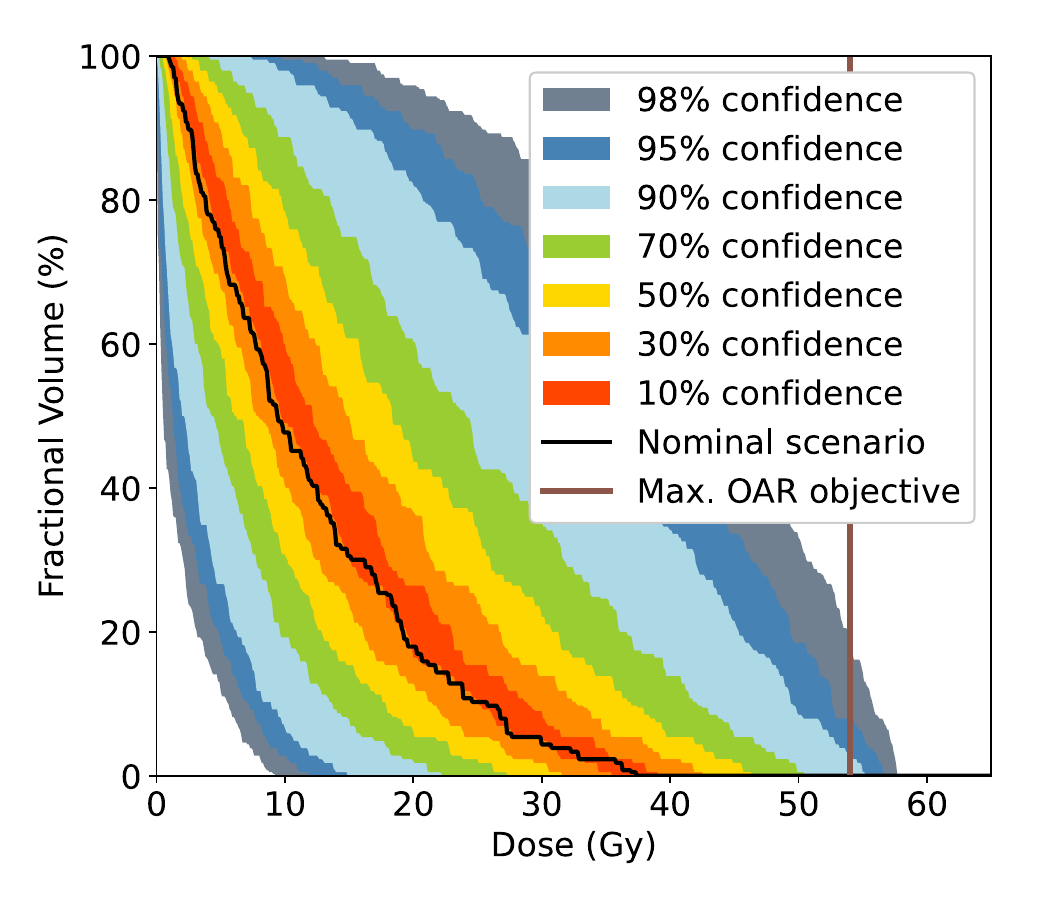}
        \caption{}
    \end{subfigure}

    \caption{Dose volume histogram distributions for the CTV (left) and spine (right) for the $\nu = 90\%$ plans. Probabilistic (top) and robust (bottom) plans are compared, showing the nominal scenario (black) with various confidence bands.}
    \label{fig:spinal_DVH_probAndRob}
\end{figure}

% prob under overdosage (comparison prob10, prob5, prob2).
To understand to which extent the probabilistic objectives are reached in the $\nu = 90\%$, $\nu = 95\%$ and $\nu = 98\%$ case, we show the probability of underdosing and overdosing CTV and spinal voxels in Figure \ref{fig:spinal_probUnderOverDosage} (in the XZ-plane). The probabilities of CTV underdosage, CTV overdosage and spinal overdosage are shown from top to bottom, where the $\nu = 90\%$, $\nu = 95\%$ and $\nu = 98\%$ cases are shown from left to right. CTV voxels that do not reach the probabilistic objectives are red (i.e., are under- or overdosed in more than 10\% of the scenarios). For the $\nu = 90\%$ plan, spinal overdosage probabilities are below 10\% (i.e., the probability that was optimized for) for all voxels. Probability levels of overdosage in the spinal edge voxels reach approximately 5\% and 2\% for the $\nu = 95\%$ and $\nu = 98\%$ plans, respectively. This shows that the probabilistic methodology allows for tuning desired probability levels. \\

% DPH figures
A comparison of the DVH-metrics is done in Figure \ref{fig:spinal_DPH_nu90}, where the DPH of various DVH-metrics is shown for the probabilistic and robust ($\nu = 90\%$) plans. As both plans are matched by the CTV coverage $D_{98\%}^{10\text{th}}$, both plans are similar the region of large scenario fraction (i.e., in the lower tails of the DVH-metrics). This means that in the worst error scenarios (where the CTV receives the lowest dose), the plan quality is similar. Also, the region of low scenario fraction of the spinal DVH-metrics (e.g., $D_{2\%}^{90\text{th}}$) is similar between both plans. All DPHs deviate significantly in the central region of the distributions. For the majority of the dose values, the probabilistic plan shows smaller probability of CTV underdosage (e.g., $P(D_{98\%} > \qty{57}{\gray})$ increased by about 14\%) and spinal overdosage (e.g., the $P(D_{2\%} > \qty{30}{\gray})$ is reduced by about 28.5\%) in the same plan. \\

DVH bands of the CTV and spine are compared between the probabilistic (top) and robust (bottom) plans in Figure \ref{fig:spinal_DVH_probAndRob} for the $\nu = 90\%$ case. The probabilistic CTV objectives at $0.95 \cdot d_p = \qty{57}{\gray}$ and $1.07 \cdot d_p = \qty{64.2}{\gray}$ are shown, together with the maximum OAR dose threshold at $\qty{54}{\gray}$. As the probabilistic plan focused on preventing spinal overdosage probability, the spinal DVH has lower dose values associated to the same fractional volume, but similar DVH spread. Only for approximately 5\% of the scenarios (corresponding to the upper bound of the 90\% confidence level), the robust plan reaches lower doses than in the probabilistic plan. Improvement in the spine is possible at the expense of a less homogeneous CTV, with a slight overdosage (as allowed for). Additionally, the DVH bands are wider, though they remain within the desired thresholds.

% spinal v=90
%               Probabilistic       robust
% D98 at 57Gy:  17.0%               3.0%    (increase by about 14%)
% D2 at 30Gy:   52.4%               80.9%   (decrease by about 28.5%)

\section{Discussion} \label{sec:Discussion}

In this work we present a proof-of-concept of a novel approach to probabilistic treatment planning, that allows for the precise tuning of voxel-wise under- and overdosage probabilities. 

Probabilistic planning — based on re-optimization after probabilistic evaluation — has shown potential to reduce inter-patient variation and improve trade-offs between target coverage and OAR sparing, as demonstrated by a study currently under review \citep{deJongJI_2025submission}. This approach differs from earlier work on probabilistic optimization, which primarily focused on optimizing for target coverage (e.g., $D_{98\%}$) objectives \citep{Gordon2010,Tilly2019} and constraints \citep{Mescher2017}, often using approximate DVH formulations. Our approach is equally applicable to such formulations, allowing objective functions of the type of Equation \ref{eq:Methods_CTVobj} to be reformulated accordingly. Since dose coverage optimization is inherently non-convex, \citet{Tilly2019} proposed a (convex) CVaR objective. By associating the $\delta$-factors to CVaR estimates rather than percentile estimates, our approach can likewise optimize the CVaR of the voxel dose. Also combinations (e.g., dose coverage objectives with CVaR) are possible. In future work, we can explore a broader range of probabilistic objectives, including commonly used radio-biological metrics, such as generalized equivalent uniform dose, biological effective dose and tumor control probability \citep{vanHaveren2019}.

Percentile optimization has been approximated in previous studies \citep{Sobotta2010,Chu2005,Fabiano2022} by assuming Gaussian-distributed responses, which does not allow for precise tuning towards desired probabilities or percentiles. Our approach directly applies to non-Gaussian response distributions, because PCE allows to model non-Gaussian responses and its efficient sampling allows to yield sufficient statistics for accurate percentile prediction. This was demonstrated in the spinal case, where the probability of spinal overdosage in the treatment plan matched the desired values or reached values below the threshold. 

In the current comparison we prioritized to limit spinal overdosage probabilities, resulting in large CTV underdosage probabilities. Still, CTV underdosage probabilities were larger in the robust case (for similar spine overdosage). The CTV coverage in these plans did not meet the clinical criterion that is often aimed for ($D_{98\%}^{10\text{th}} = 95\% d_i^p$), meaning that these plans are non-robust. This can be seen from Figure \ref{fig:spinal_DVH_probAndRob}, where the lower tails of the near-maximum DVH metrics of the CTV extend to dose values much smaller than was prescribed. Sufficient CTV coverage can be reached by using a stricter CTV underdosage objective, or by increasing the importance of the particular objective. A probabilistic optimization of the horseshoe-shaped CTV shows that the clinical criterion can be reached if there would be no spine (see Appendix \ref{app:additionalComparisons_spinal}). 

To enable a fair comparison, plans were scaled by the $D_{50\%}$. This ICRU-based metric describes the median CTV dose and is numerically robust. We want to emphasize that the exact choices and priorities chosen in treatment planning do not invalidate the approach. In fact, we observed that in some probabilistic optimizations, even one or two outer loop iterations were sufficient to achieve the desired under- and overdosage probabilities. Further research on clinical datasets is needed to explore how the probabilistic approach performs under realistic clinical trade-offs and priorities. \\

Although improved robust plans (e.g., in terms of conformity) can possibly be achieved by addition of other objectives or other types of robust optimization (e.g., objective-wise), the spherical and spinal probabilistic plans consistently show that OAR/spinal overdosage is reduced for identical CTV coverage. The reason for this is that, as opposed to probabilistic optimization, mini-max robust optimization relies on a chosen uncertainty set and treats every scenario within it as equally probable. 

The effect of using an uncertainty set in robust optimization can be clearly seen in the CTV-only \textit{setupXYrange} case (Figure \ref{fig:CTVonly_XY_XYr_RobProb}, bottom). The dose margin is extended into all directions where an error scenario is defined. As a result, the dose margins along the diagonals of the XZ-plane are more conservative (and thus less conformal) compared to the probabilistic plan. Similar results are seen in Figure \ref{fig:sphericalCTVandOAR_XZshift} for the $D_{2\%}^{90\text{th}}$ comparison, where the increase of the OAR objective weight leads to overcompensation of single shifts, at the same time giving less importance to other (potentially more important, higher probability) shifts. Besides a reduction in conformity, this leads to a smaller CTV margin in the $y$-direction, in turn resulting in lower CTV coverage. In the spherical XZ-displaced case, the probabilistic plan automatically leads to a conformal plan, because 1) there are no competing objectives and 2) it does not rely on single worst-case scenario that could overcompensate other scenarios. \\

\begin{table}
\caption{\label{tab:optTimes_Dij} Computation times of expectation values ($\mathbb{E}[D_{ij}(\bxi)]$, $\mathbb{E}[D_{ij}(\bxi)D_{ij'}(\bxi)]$ and $\sum_{i \in nonCTV} w_i \mathbb{E}[D_{ij}(\bxi)D_{ij'}(\bxi)]$) and PCE construction of the dose-influence matrix. } 
\begin{center}
\begin{tabular}{@{}*{4}{c}}
\toprule                             
 & spherical & spinal & Number dose calculations \cr
\midrule
Expectation values of $D_{ij}$ &  \qty{8435}{\second} & \qty{1412}{\second} & 105 \cr
PCE construction of $D_{ij}$ (CTV) &  \qty{6416}{\second} & \qty{1738}{\second} & 1637 \cr 
PCE construction of $D_{ij}$ (OAR)  &  \qty{2435}{\second} & \qty{1143}{\second} & 1637 \cr
\bottomrule
\end{tabular}
\end{center}
\end{table}

% Memory intensive
The current approach is memory intensive, because voxel-wise objectives are used. The memory expense is especially due to the voxel-wise term given by $\mathbb{E}[D_{ij}(\bxi)D_{ij'}(\bxi)]$, which arises in the gradient and Hessian of the variance (see Appendix \ref{app:GradientAndHessian}). This leads to the question of how this approach would scale to clinical patient cases, which involve significantly more voxels than the presented phantom cases. A straightforward improvement is to only apply probabilistic optimization to CTV edge voxels, or to randomly sampled voxels across the structure. Automatic methods can select voxel subsets using adaptive \citep{Martin2007} or deep-learning based \citep{Quarz2024} sampling.

The expectation value calculation in Equation \ref{eq:Methods_objectiveFunctions_percentileEstimate1}, Equation \ref{eq:Methods_objectiveFunctions_percentileEstimate2} and Equation \ref{eq:Methods_objectiveFunctions_percentileEstimate3} and PCE construction of the dose-influence matrix have been parallelized by 16 CPU-cores (2x Intel XEON E5-6248R 24C 3.0GHz) \citep{DHPC2024}. Computation times are reported in Table \ref{tab:optTimes_Dij}. Calculations for the spinal plans were significantly lower, because of the courser voxel grid. Long computation times for the expectation values are especially due to $\sum_{i \in nonCTV} w_i \mathbb{E}[D_{ij}(\bxi)D_{ij'}(\bxi)]$, which is computed for non-CTV voxels. This computational expense can be partly reduced by optimizing over tissue voxels without using expectation values. Computation speeds are expected to be improved by using GPU-cores. 

PCE construction times in this work can be treated as a conservative upper bound to what is clinically often regarded as sufficient, because stricter $\Gamma$-evaluation criteria are used here. Namely, a $\Gamma$-evaluation using \qty{3}{\milli\meter}/3\% instead of \qty{1}{\milli\meter}/\qty{0.1}{\gray} settings lead to 98\% of the voxels being accepted in all test scenarios for both pencil-beams.

\begin{table}[]
\caption{\label{tab:optTimes} Comparison of total probabilistic and robust optimization times. } 

\begin{center}
\begin{tabular}{@{}*{3}{l}}
\toprule
 & Probabilistic & Robust \cr 
\midrule
Spherical XZ-displaced &  \qty{3.7}{\hour} & \qty{4.0}{\hour} \cr 
Spherical X-displaced  &  \qty{6.8}{\hour} & \qty{4.4}{\hour} \cr
Spinal ($\nu = 90\%$)           &  \qty{7.6}{\hour} & \qty{9}{\minute} \cr 
Spinal ($\nu = 95\%$)           &  \qty{7.3}{\hour} & \qty{11}{\minute} \cr 

Spinal ($\nu = 98\%$)           &  \qty{11.4}{\hour} & \qty{20}{\minute} \cr
\bottomrule
\end{tabular}
\end{center}
\end{table}

Cumulative optimization times are listed in Table \ref{tab:optTimes}. Spherical X-displaced plans took longer to optimize than XZ-displaced plans, due to more conflicting probabilistic objectives (CTV and OAR are closer). For the same reason, even though the spinal geometry contains less voxels than the spherical geometry, the probabilistic spinal plans took longer to optimize than the spherical plans. Robust optimization times for spherical plans were similar to probabilistic ones, but spinal robust optimizations were much faster, likely because the choice of objective weights made the plans more challenging to optimize. Probabilistic optimization times could be improved by parallelization and by using different warm-start strategies. Large computation times are primarily due to the inner optimization, which is further analyzed in Appendix \ref{app:probOptTimes}. \\
% Spinal: number voxels total 11550 (CTV 1204, spine 390).
% Spherical: number voxels total 91125 (CTV 3071, OAR 1206).

\noindent Other suggestions to be included into future work are as follows:
\begin{enumerate}
    \renewcommand{\theenumi}{\roman{enumi}}  % affects numbering label
    \renewcommand{\labelenumi}{\theenumi.}  % adds period if desired
  
    \item Although the current work only focuses on probabilistic objectives, the approach can be extended to handle probabilistic constraints.
    \item The presented approach is not limited to proton therapy, but can likewise be applied to photon therapy, or radiation therapy in general. Although photons may be less sensitive to uncertainties than protons, photons - particularly volumetric modulated arc therapy - have shown inter-patient variation in PTV coverage \citep{RojoSantiago2023_2}. In cases involving complex anatomies, challenging trade-offs between target and OARs, or hypo-fractionated treatments, the probabilistic approach could potentially have great added value for photon therapy as well. Even when the expected improvements are less significant, the probabilistic approach can help in the interpretation of dosimetric outcomes.
    \item To extend the work to hypo-fractionated treatments, random errors should be included in the probabilistic approach. Possibly, fractionation schemes can be explicitly included into the optimization process by accounting for the number of fractions, similar to previous approaches for photon \citep{Unkelbach2004} and proton therapy \citep{Wahl2018}. 
\end{enumerate}

\section{Conclusions} \label{sec:Conclusions}

This work presents a new approach to probabilistic treatment planning, that is able to optimize for exact underdosage and overdosage probabilities of multiple structures, for individualized probability levels and dose thresholds. Systematic setup and range errors were considered, and for the former, probabilistic equivalence to the Van Herk margin recipe was demonstrated. Compared to composite-wise robust plans, the probabilistic plans achieve more OAR sparing with similar target coverage (or improved target coverage with similar OAR sparing) for all spherical and spinal comparisons. Probabilistic plans were found to be more conformal to the CTV, as probabilistic optimization accounts for the probability of different error scenarios rather than relying only on a predefined uncertainty set and optimizing with a single worst-case scenario. As the proposed method is sufficiently general to be extended to dose-coverage or CVaR optimization, this is an obvious follow-up. Besides that, following work should focus on improving computational efficiency through time and memory optimization techniques, so that clinical feasibility (i.e., the application to real clinical cases) can be demonstrated. 
\section*{Data availability statement}

The Open Source Generalized Polynomial Chaos Expansion Toolbox \citep{Perk2014} (\url{https://gitlab.com/zperko/opengpc/-/tree/4a2e11cc77617c705f8a555996aee3c7cd4a61db/}) is used to construct and evaluate Polynomial Chaos Expansions. The data that support the findings of this study are subject to contractual restrictions and therefore cannot be made publicly available upon publication. However, the data that support the findings of this study are available from the authors upon reasonable request.
\section*{Acknowledgements}

The authors would like to acknowledge that this work is funded by RaySearch Laboratories.
\section*{Conflicts of interest}

Zoltán Perkó is an associate professor at TU Delft and is employed as a Senior Applied Scientist at Radformation Inc.. His industry employment is unrelated to the submitted work. The remaining authors have no conflicts of interest to declare.
\section*{CRediT author and contributor statement}

\textbf{Jelte Rinus de Jong}: Conceptualization, Data Curation, Formal Analysis, Investigation, Methodology, Software, Validation, Visualization, Writing - Original Draft, Writing - Review and Editing \\
\textbf{Zoltán Perkó}: Conceptualization, Funding Acquisition, Methodology, Project Administration, Supervision, Resources, Writing - Review and Editing \\
\textbf{Danny Lathouwers}: Conceptualization, Methodology, Project Administration, Supervision, Resources, Writing - Review and Editing \\
\textbf{Mischa Hoogeman}: Conceptualization, Funding Acquisition, Writing - Review and Editing \\
\textbf{Steven Habraken}: Conceptualization, Writing - Review and Editing \\
\textbf{Sebastiaan Breedveld}: Conceptualization, Writing - Review and Editing

% Conceptualization, Data Curation, Formal Analysis, Funding Acquisition, Investigation, Methodology, Project Administration, Software, Supervision, Resources, Validation, Visualization, Writing - Original Draft, Writing - Review and Editing

\bibliography{myrefs}

\clearpage
\appendix

\section{Cost-accuracy analyses of the dose approximations}
\label{app:accuracies_PCE}

\subsection{Accuracy of the Polynomial Chaos Expansion for the dose-influence matrix}
\begin{figure}[b!]
    \centering
    \begin{subfigure}[b]{0.32\textwidth}
        \centering
        \includegraphics[width=\textwidth]{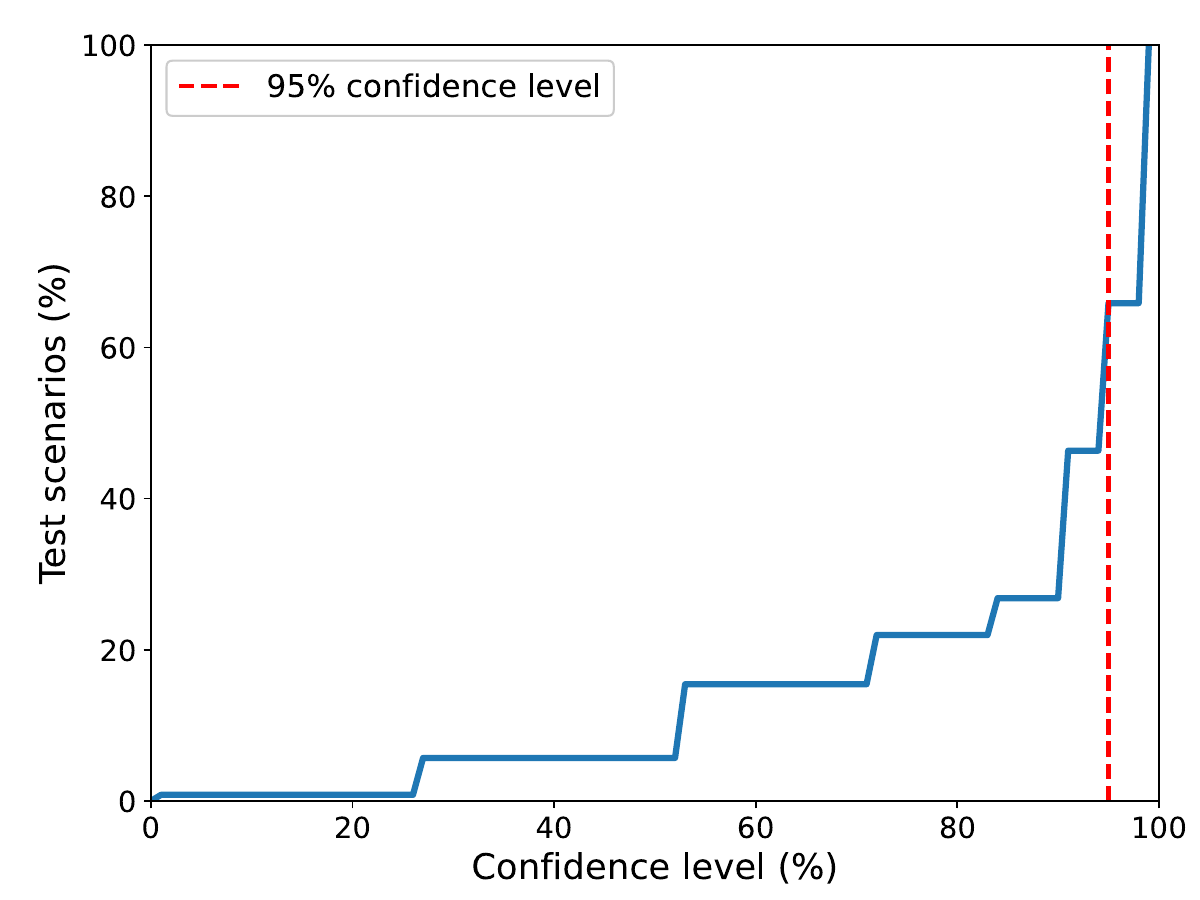}
        \caption{}
        \label{fig:gammaEvaluation_scenarios}
    \end{subfigure} % from scenarioVSconfidence.py
    %\hfill
    \begin{subfigure}[b]{0.32\textwidth}
        \centering
        \includegraphics[width=\textwidth]{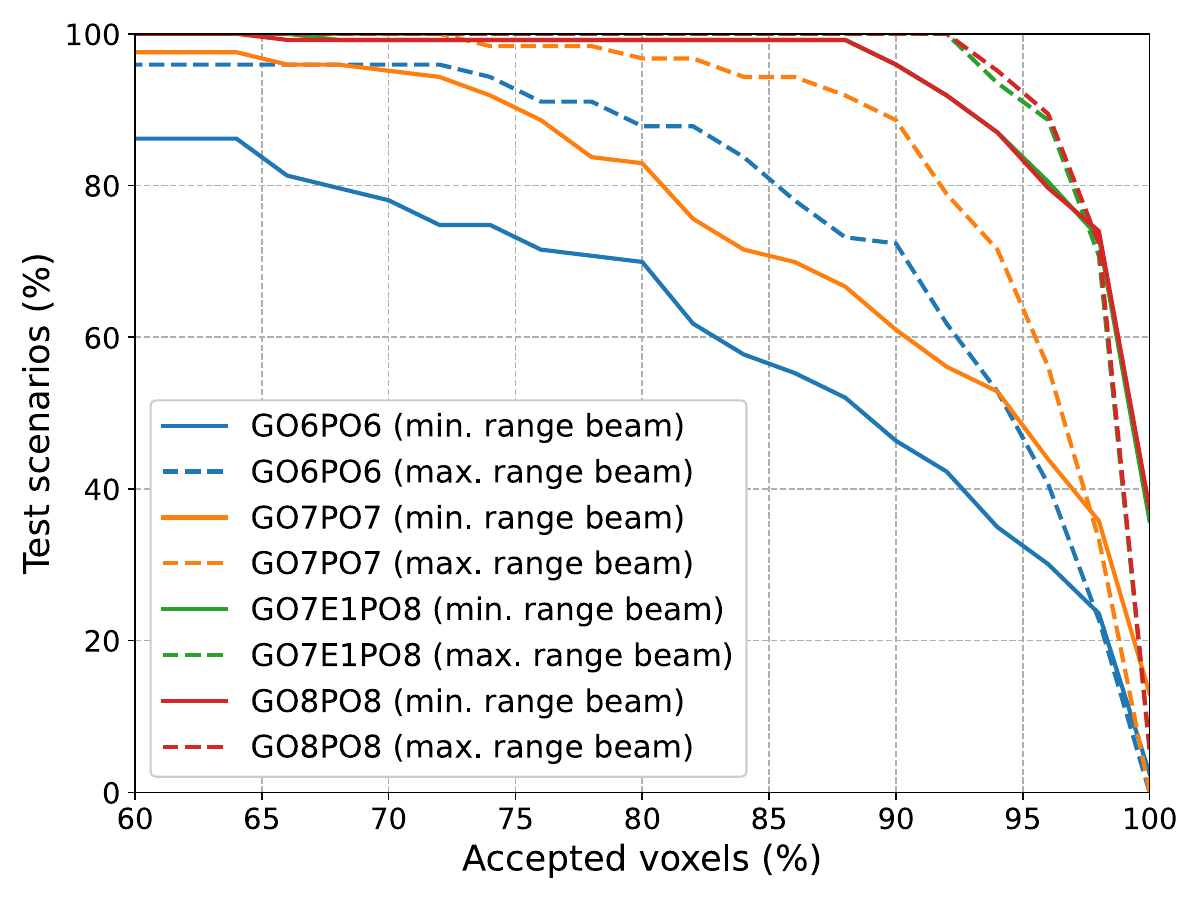}
        \caption{}
        \label{fig:gammaEvaluation_Dij}
    \end{subfigure} % from gammaEvaluation_Dij_compare.py
    %\hfill
    \begin{subfigure}[b]{0.32\textwidth}
        \centering
        \includegraphics[width=\textwidth]{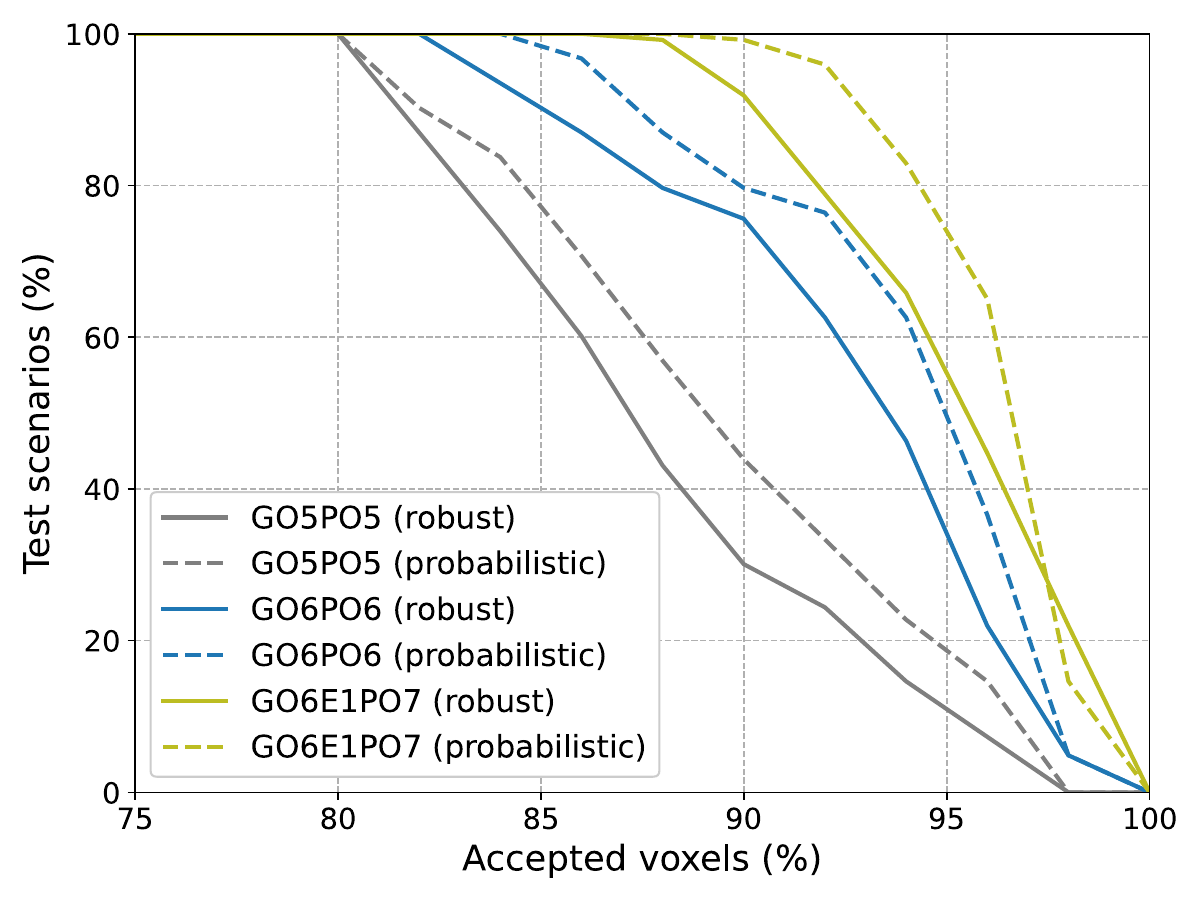}
        \caption{}
        \label{fig:gammaEvaluation_di}
    \end{subfigure} % from gammaEvaluation_di_compare.py
    \caption{A $\Gamma$-evaluation (\qty{0.1}{\gray}/1\%) is done for a) 123 test scenarios that are distributed within the 99\%-confidence ellipsoid. The PCE accuracy is shown for different polynomial and grid orders for the b) $D_{ij}$ for pencil-beam spots of minimum (test pencil-beam 1) and maximum range (test pencil-beam 2) and for the c) robust and probabilistic dose distributions for the CTV-only \textit{setupXYrange} case.}
    \label{fig:gammaEvaluation}
\end{figure}

To determine the grid and polynomials order that are needed to obtain a PCE of sufficient accuracy, we define a set of test scenarios. We uniformly distribute 7 test scenarios along each dimension such that the minimum and maximum scenarios coincide with the interval bounds of the truncated multivariate Gaussian distribution. By finding all combinations of the test scenarios along the axes, we obtain a rectangular grid consisting of 343 unique scenarios. Only the 123 scenarios that fall within the 99\% confidence ellipsoid of the input phase space (taking into account all uncertain variables simultaneously) are kept. Figure \ref{fig:gammaEvaluation_scenarios} shows that many test scenarios lie close to the 99\% confidence ellipsoid, meaning that a significant fraction of these test scenarios is unlikely to occur in reality (e.g., approximately 34\% of the test scenarios fall beyond the 95\% confidence level).

A $\Gamma$-evaluation \citep{Biggs2022} (\qty{0.1}{\gray}/1\%) was done (for voxel doses larger than $\qty{0.1}{Gy}$) to verify the PCE accuracy of Equation \ref{eq:PCE_Dij}, of which the results are shown in Figure \ref{fig:gammaEvaluation_Dij}. We analyse two different pencil-beams that have pencil-beam spots located at $(x,y,z) = \left(\qtylist[list-final-separator={, }]{4.5; 4.5; 89.5}{\mm}\right)$ and $(x,y,z) = \left(\qtylist[list-final-separator={, }]{40.5; 40.5; 125.5}{\mm}\right)$ (in the spherical geometry). We refer to these pencil-beam spots as \textit{Test pencil-beam 1} and \textit{Test pencil-beam 2}, corresponding to a minimum and maximum range, respectively. For grid order $GO = 7$ and polynomial order $PO = 7$ (denoted as GO7PO7), 90\% of the test scenarios pass the $\Gamma$-evaluation for 88\% of the voxels for test pencil-beam 2 (the maximum range pencil-beam), whereas test pencil-beam 1 (the minimum range pencil-beam) has lower accuracy. The GO8PO8 PCEs shows significantly improved accuracy, where 90\% of the test scenarios pass the $\Gamma$-evaluation for at least 96\% of the voxels. Moreover, about 66\% of the accepted scenarios has at least a 98\% accepted voxel fraction, which corresponds to the 95\% confidence ellipsoid. The same accuracy can be achieved when $GO = 7$ is increased only along the single dimensions by $lev_{extra} = 1$ (so that along the single dimensions $GO = 8$, denoted as GO7E8PO8). For this extended Smolyak sparse grid we only need $2 \cdot lev_{extra} \cdot N = 6$ additional calculations compared to $GO = 7$.

\begin{figure}[]
    \centering

    % Row 1
    \begin{subfigure}{0.35\textwidth}
        \includegraphics[width=\linewidth]{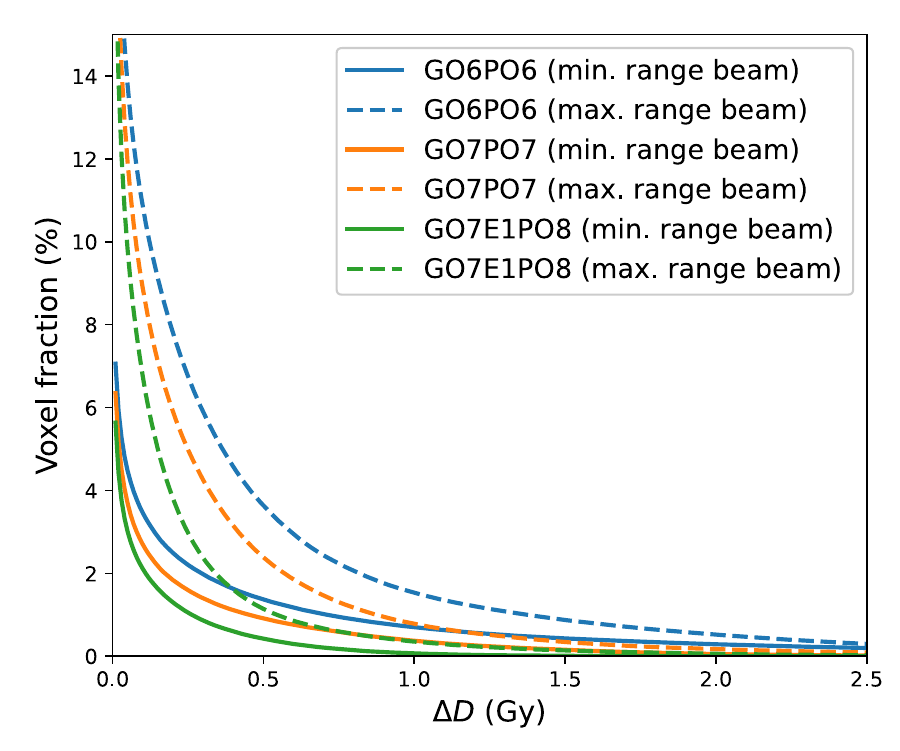}
        \caption{}
    \end{subfigure}
    %\hfill
    \begin{subfigure}{0.35\textwidth}
        \includegraphics[width=\linewidth]{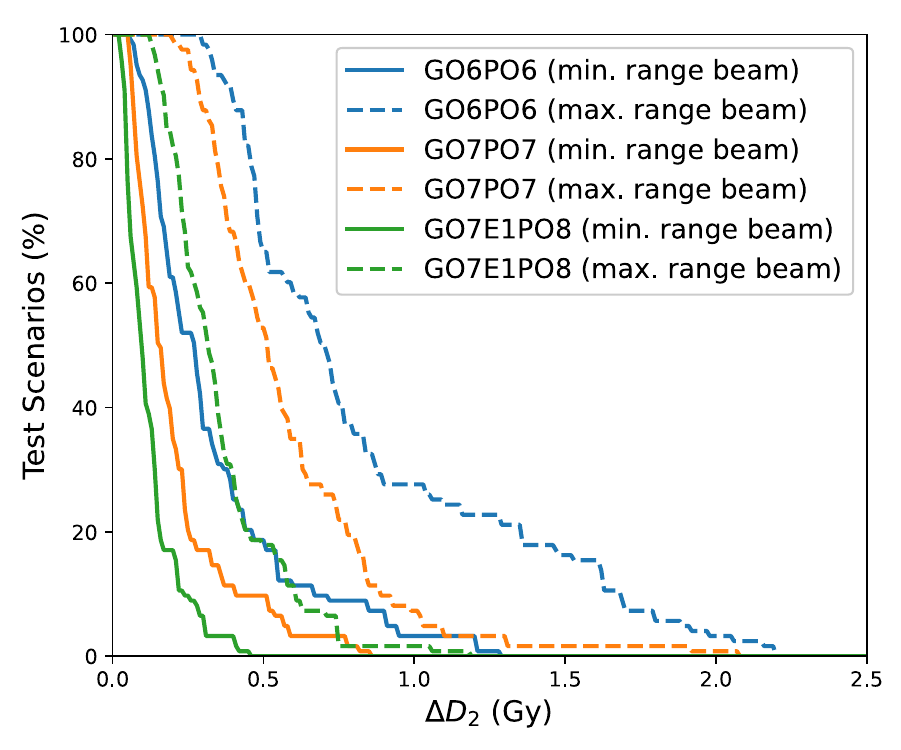}
        \caption{}
    \end{subfigure}

    % Row 2
    \begin{subfigure}{0.35\textwidth}
        \includegraphics[width=\linewidth]{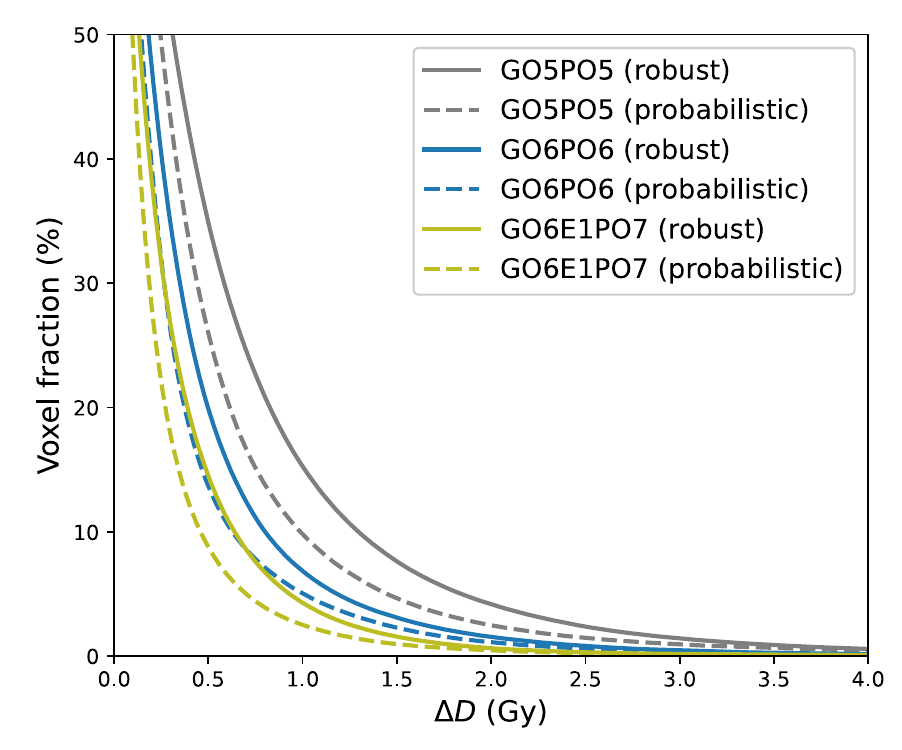}
        \caption{}
    \end{subfigure}
    %\hfill
    \begin{subfigure}{0.35\textwidth}
        \includegraphics[width=\linewidth]{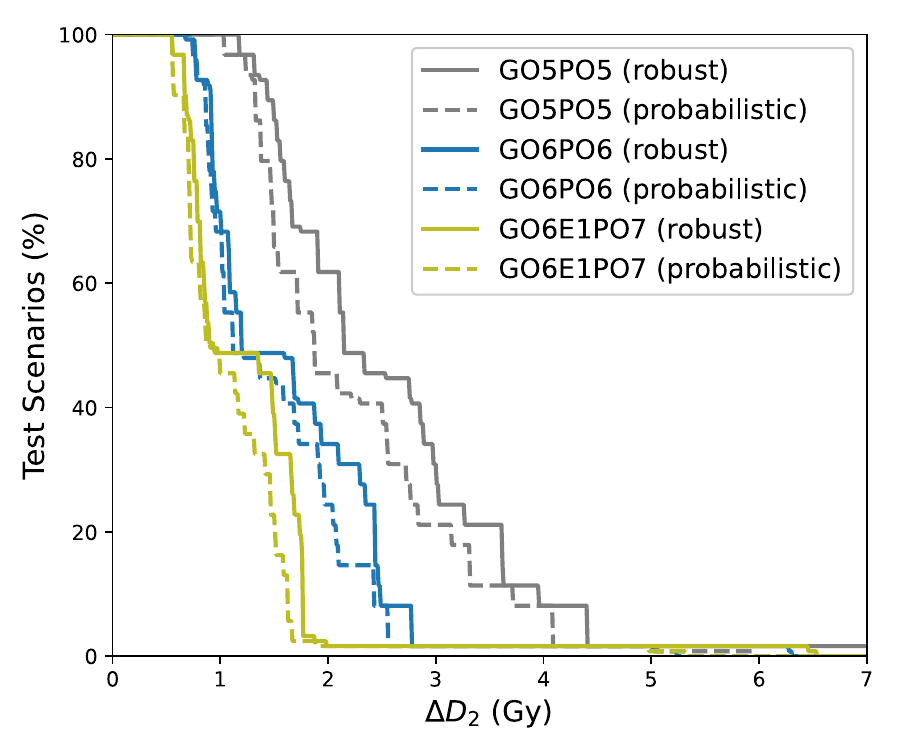}
        \caption{}
    \end{subfigure}
    \caption{The PCE accuracy analysis for the (top) $D_{ij}$ matrix and (bottom) voxel dose distributions, showing (left) the voxel fraction for which the mean dose difference over all scenarios ($\Delta D$) exceeds the dose value, and (right) the scenario fraction for which the minimum dose difference of the worst 2\% of the voxels ($\Delta D_{2\%}$) exceeds the dose value.}
    \label{fig:D2andMeanDose_DijandVoxelDose}
\end{figure}

Besides the $\Gamma$-evaluation, the PCE accuracy is quantified by determining the dose difference between the PCE and the dose engine for both test pencil-beams. For all test scenarios within the 99\% confidence ellipsoid, we determine the minimum voxel dose difference among the 2\% of the voxels having the largest dose difference, which we denote by $\Delta D_{2\%}$. Then, we calculate the scenario fraction for which the $\Delta D_{2\%}$ is larger than a certain dose value. Moreover, we determine the voxel dose difference averaged over all test scenarios (denoted by $\Delta D$), and check what voxel fraction has $\Delta D$ larger than a certain dose value. The results are shown in Figure \ref{fig:D2andMeanDose_DijandVoxelDose}. For GO7E8PO8, $\Delta D_{2\%}$ does not exceed \qty{1.2}{\gray} for both beams. Also, the mean dose difference over all scenarios is smaller than \qty{0.15}{\gray} in 95\% of the voxels. Based on this analysis, we choose to construct the $D_{ij}$ PCE using GO7E8PO8.

The accuracy of the entire $D_{ij}$ matrix was checked for GO7E8PO8, by obtaining the mean of the element-wise absolute difference (i.e., $\langle \Delta D_{ij} \rangle = \sum_{i,j} |D_{ij}^{true} - D_{ij}^{PCE}|_{>\qty{0.1}{\gray}} / N_{>\qty{0.1}{\gray}}$) for all test scenarios (for the $N_{>\qty{0.1}{\gray}}$ elements that have dose value larger than \qty{0.1}{\gray}), which was found to be \qty{0.013}{\gray} (range: \qty{2.5e-4}{\gray} till \qty{0.23}{\gray}).

\subsection{Accuracy of the Polynomial Chaos Expansion for the voxel dose distributions}
For plan evaluation, a PCE of the voxel dose is constructed independently. The required accuracy is determined by analyzing the spherical \textit{setupXYrange} CTV-only plan. Results for different grid and polynomial orders are shown in Figure \ref{fig:gammaEvaluation}c, where a $\Gamma$-analysis ($\qty{0.1}{\gray}/1\%$, for doses $\geq \qty{0.1}{\gray}$) is performed. For the GO6PO6 PCE, about 90\% of the voxels is accepted in 80\% of the scenarios. We can improve the accuracy significantly by the GO6E7PO7 PCE, where nearly all scenarios are accepted for 90\% of the voxels. Based on this analysis, we choose to construct the $d_i$ PCE using GO6E7PO7.

As shown in Figure \ref{fig:D2andMeanDose_DijandVoxelDose}, about 97.5\% of the test scenarios have $\Delta D_{2\%} < \qty{2}{\gray}$ for the GO6E7PO7 PCE. Also, about 66\% of the test scenarios (these scenario fall within the 95\% confidence ellipsoid) have $\Delta D_{2\%} < \qty{1.5}{\gray}$. The mean dose difference over all scenarios is smaller than about \qty{0.9}{\gray} in 95\% of the voxels. 

\begin{table}[h!]
\caption{\label{tab:accuracyDij_meanDiff}Mean differences of $\mathbb{E}[D_{ij}(\bxi)]$ matrix elements (only matrix elements are considered if the GO8 element has a dose larger than $\qty{0.1}{\gray}$) comparing different grid orders (GO3, GO4, GO5, GO6) with GO8 (considered true).} 
\begin{center}
\begin{tabular}{@{}*{2}{l}}
\toprule                     
Grid Order (GO) & Mean difference with GO8 (\unit{\gray}) \\[0.5ex]
\midrule
GO3 & $2.51 \times 10^{-2}$ \cr
GO4 & $2.74 \times 10^{-3}$ \cr
GO5 & $6.19 \times 10^{-4}$ \cr
GO6 & $4.10 \times 10^{-4}$ \cr
\bottomrule
\end{tabular}
\end{center}
\end{table}

\begin{figure}[h!]
    \centering

    \begin{subfigure}{0.35\textwidth}
        \centering
        \includegraphics[width=\linewidth]{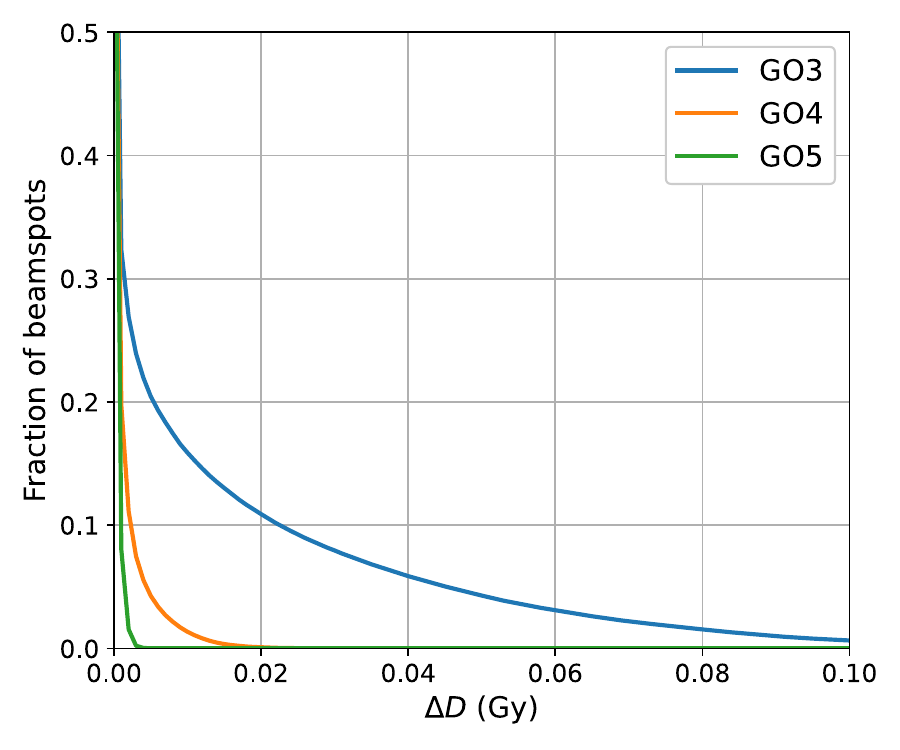}
        \caption{}
        \label{fig:accuracy_ExpDij_moreMeasures_deltaD}
    \end{subfigure}
    %\hfill
    \begin{subfigure}{0.35\textwidth}
        \centering
        \includegraphics[width=\linewidth]{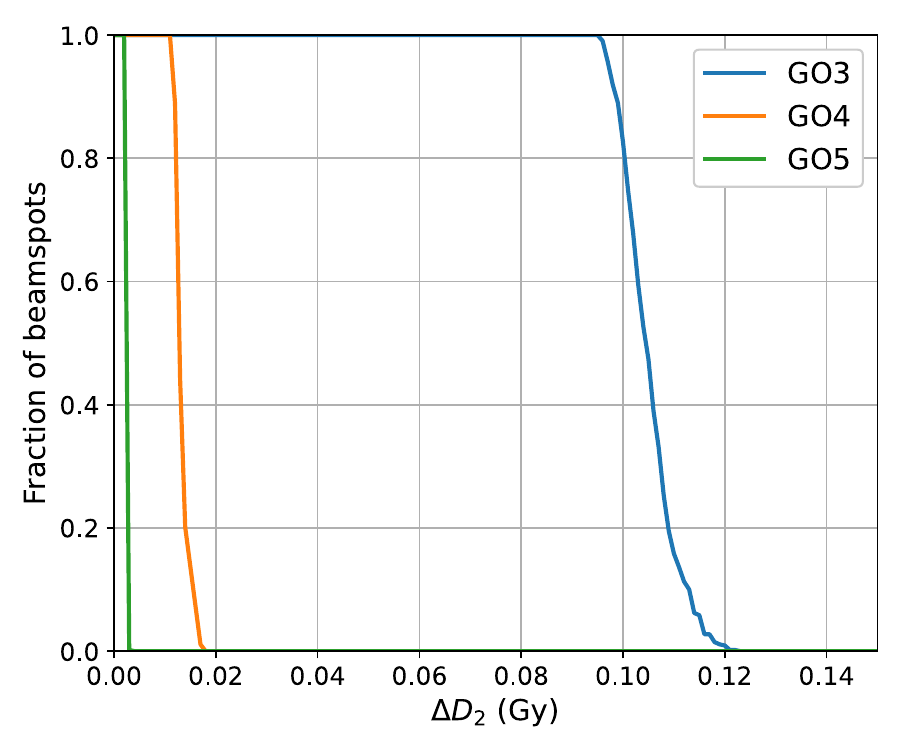}
        \caption{}
        \label{fig:accuracy_ExpDij_moreMeasures_deltaD2}
    \end{subfigure}

    \caption{An accuracy analysis of $\mathbb{E}[D_{ij}]_{GO}$, comparing different grid orders (GO3, GO4, GO5) with GO8 (considered true). The fraction of pencil-beam spots that have a) a $\Delta D_{2\%}$ and b) mean dose difference (over all voxels) exceeding a dose value is shown.}
    \label{fig:accuracy_ExpDij_moreMeasures}
\end{figure}

\subsection{Accuracy of the expectation values of the dose-influence matrices}
\label{app:accuracies_exp}

We aim to find the expected dose-influence matrix $\mathbb{E}[D_{ij}(\bxi)]$ associated to grid order $GO$ that is sufficiently close to the true $\mathbb{E}[D_{ij}(\bxi)]$ ($GO = 8$ is assumed to be true). Table \ref{tab:accuracyDij_meanDiff} shows the mean difference over all $D_{ij}$ elements larger than $\qty{0.1}{\gray}$, between $\mathbb{E}[D_{ij}(\bxi)]_{GO}$ and $\mathbb{E}[D_{ij}(\bxi)]_{GO=8}$ for grid orders $GO = 3$ till $GO = 6$. Figure \ref{fig:accuracy_ExpDij_moreMeasures_deltaD} shows the fraction of pencil-beam spots that has a mean dose difference (over the voxels) larger than a certain dose value. For example for $\mathbb{E}[D_{ij}(\bxi)]_{GO=4}$, all pencil-beam spots have a $\Delta D_{2\%}$ below \qty{0.018}{\gray} and the mean dose difference over all voxels is smaller than \qty{0.01}{\gray} for about 98.5\% of the pencil-beam spots. In Figure \ref{fig:accuracy_ExpDij_moreMeasures_deltaD2} we show the fraction of pencil-beam spots (each pencil-beam spot corresponds to a $\mathbb{E}[D_{ij}(\bxi)]$ column) that has a $\Delta D_{2\%}$ larger than a certain dose value. To be on the conservative side, we choose to use $\mathbb{E}[D_{ij}(\bxi)]_{GO=4}$ and $\mathbb{E}[D_{ij}(\bxi)D_{ij'}(\bxi)]_{GO=4}$ in this work.

\section{Probabilistic and robust optimizations details}
\label{app:probRobOpt}

This section presents a summary of the probabilistic (Section \ref{app:probOpt}) and robust (Section \ref{app:robOpt}) optimizations that are applied to all phantom geometries, as presented in Section \ref{subsec:Methods_PhantomGeometries}.

\subsection{Probabilistic optimizations}
\label{app:probOpt}

\begin{table}[b!]
\caption{\label{tab:probOptParameters} Probabilistic optimization parameters.} 

\begin{center}
\begin{tabular}{@{}*{6}{c}}
\toprule
 & \multicolumn{2}{c}{Spherical} & \multicolumn{3}{c}{Spinal} \\
\cmidrule(lr){2-3} \cmidrule(lr){4-6}
 & CTV-only & CTV and OAR & $\nu_i = 90\%$ & $\nu_i = 95\%$ & $\nu_i = 98\%$ \\[0.5ex]
\midrule
$w_i^{CTV}$ &  100 & 100     & 100 & 100 & 100 \cr
$w_i^{OAR}$ &  20 & 20     & 20 & 20 & 20 \cr
$w_i^{Tissue}$ &  1 & 1     & 1 & 1 & 1 \cr

$d_i^p [\SI{}{Gy}]$ &  60 & 60     & 60 & 60 & 60 \cr
$\alpha_i [\%]$ &  10 & 10     & 10 & 10 & 10 \cr
$\beta_i [\%]$ &  90 & 90     & 90 & 90 &  90\cr
$\nu_i [\%]$ &  --- & 90     & 90 & 95 & 98 \cr
$\gamma_i [\SI{}{Gy}]$  & 57 & 57     & 57 & 57 & 57 \cr
$\epsilon_i [\SI{}{Gy}]$ &  64.2  & 64.2     & 64.2 & 64.2 & 64.2 \cr
$\mu_i [\SI{}{Gy}]$ &  --- & 30     & 54 & 54 & 54 \cr
\midrule
$\pi_{CTV}^{\alpha}$ &  15 &  15     & 15  & 15 &  15 \cr     
$\pi_{CTV}^{\beta}$  &  15 &  15     & 15  & 15 &  15 \cr   
$\pi_{OAR}^{\nu}$    &  --- &  15     & 750 & 750 & 750 \cr 
$\pi_{CTV}^{low}$    &  1 &  1     & 5 & 5 & 5 \cr 
$\pi_{OAR}^{low}$    &  --- & 1     & 15 & 15 & 15 \cr 
$\pi_{Tissue}$      &  1 &  1       & 1 & 1 & 1 \cr 
\midrule
$\Delta W$   &  20 &  15     & 15 & 15 & 15 \cr 
$\Delta k$   &  10 &  5     & 5 & 5 & 5\cr
$\tau_{CTV, \alpha}$   &  $5 \cdot 10^{-4}$ &  $5 \cdot 10^{-4}$      & 0.01     & 0.01 & 0.005 \cr   
$\tau_{CTV, \beta}$    &  $5 \cdot 10^{-4}$ &  $5 \cdot 10^{-4}$      & $4 \cdot 10^{-3}$    & $4 \cdot 10^{-3}$ & $1 \cdot 10^{-3}$ \cr
$\tau_{OAR, \nu}$      &  --- & $7 \cdot 10^{-3}$      & 0.1  & 0.1 & 0.05 \cr
\bottomrule
\end{tabular}
\end{center}
\end{table}

Table \ref{tab:probOptParameters} summarizes the probabilistic optimization parameters that are used for each geometry. All variables are in accordance with Section \ref{subsubsec:innerOptimization}, where the general (inner) probabilistic optimization is shown in Equation \ref{eq:objective}. As an example, for the spherical CTV-only case, we aim to limit the CTV underdosage probability as $P(d_i(\bi{x},\bxi) \leq \gamma_i) \leq \alpha \% = P(d_i(\bi{x},\bxi) \leq \qty{57}{\gray}) \leq 10\%$ and CTV overdosage probability as $P(d_i(\bi{x},\bxi) \geq \epsilon) \leq (100 - \beta) \% = P(d_i(\bi{x},\bxi) \geq \qty{64.2}{\gray}) \leq 10\%$.

\begin{figure}[]
    \centering

    % Row 1
    \begin{subfigure}{0.32\textwidth}
        \centering
        \includegraphics[width=\linewidth]{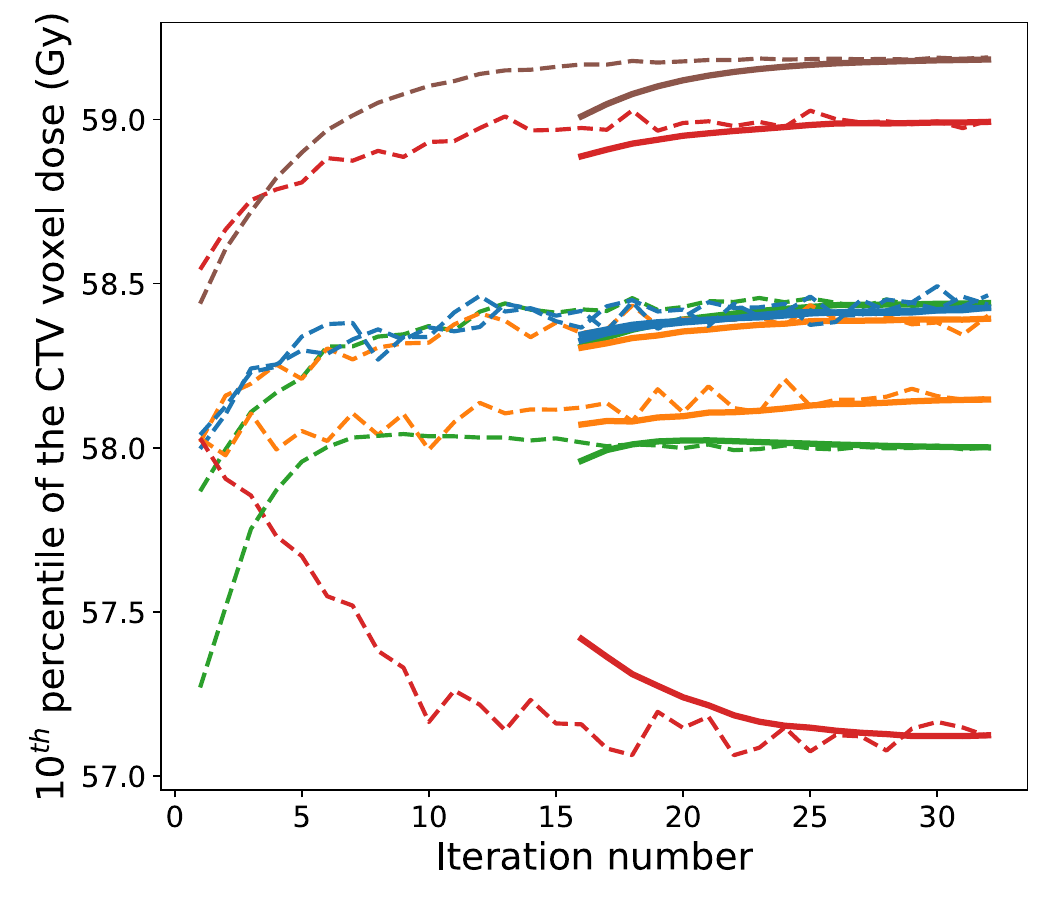}
        \caption{}
    \end{subfigure}
    %\hfill
    \begin{subfigure}{0.32\textwidth}
        \centering
        \includegraphics[width=\linewidth]{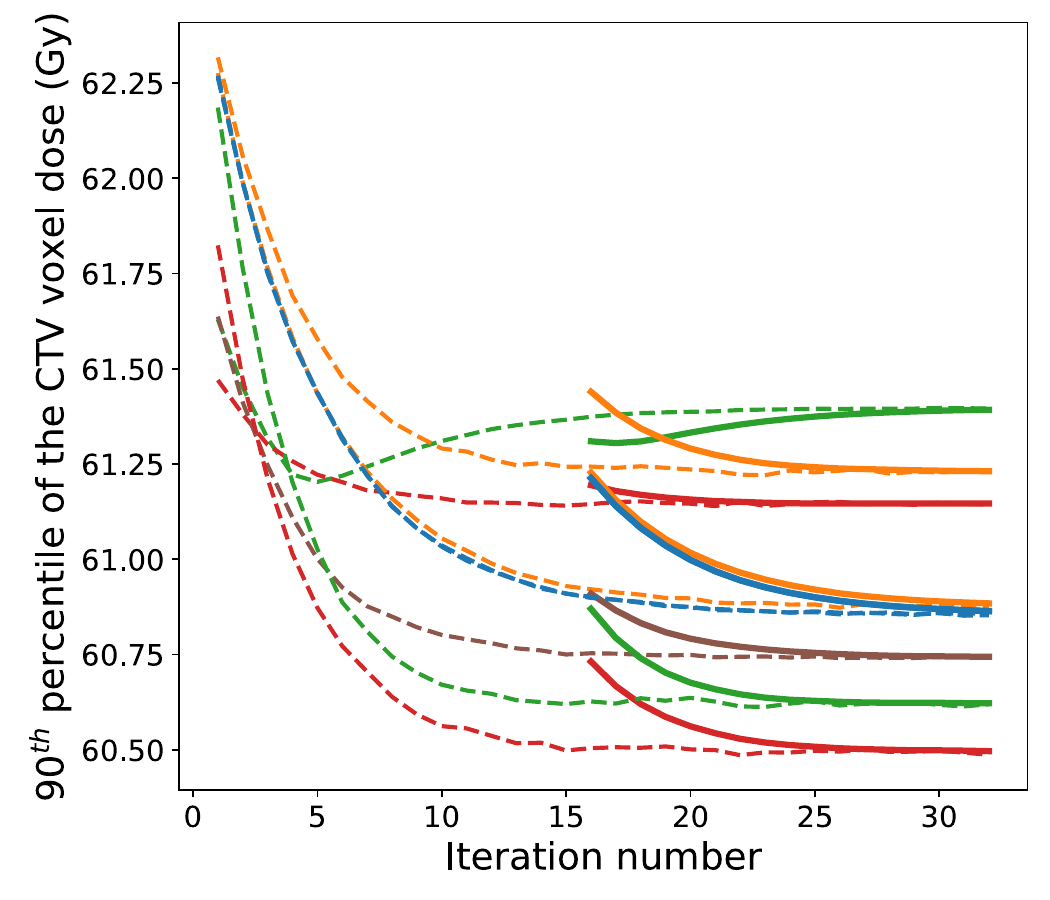}
        \caption{}
    \end{subfigure}
    %\hfill
    \begin{subfigure}{0.32\textwidth}
        \centering
        \includegraphics[width=\linewidth]{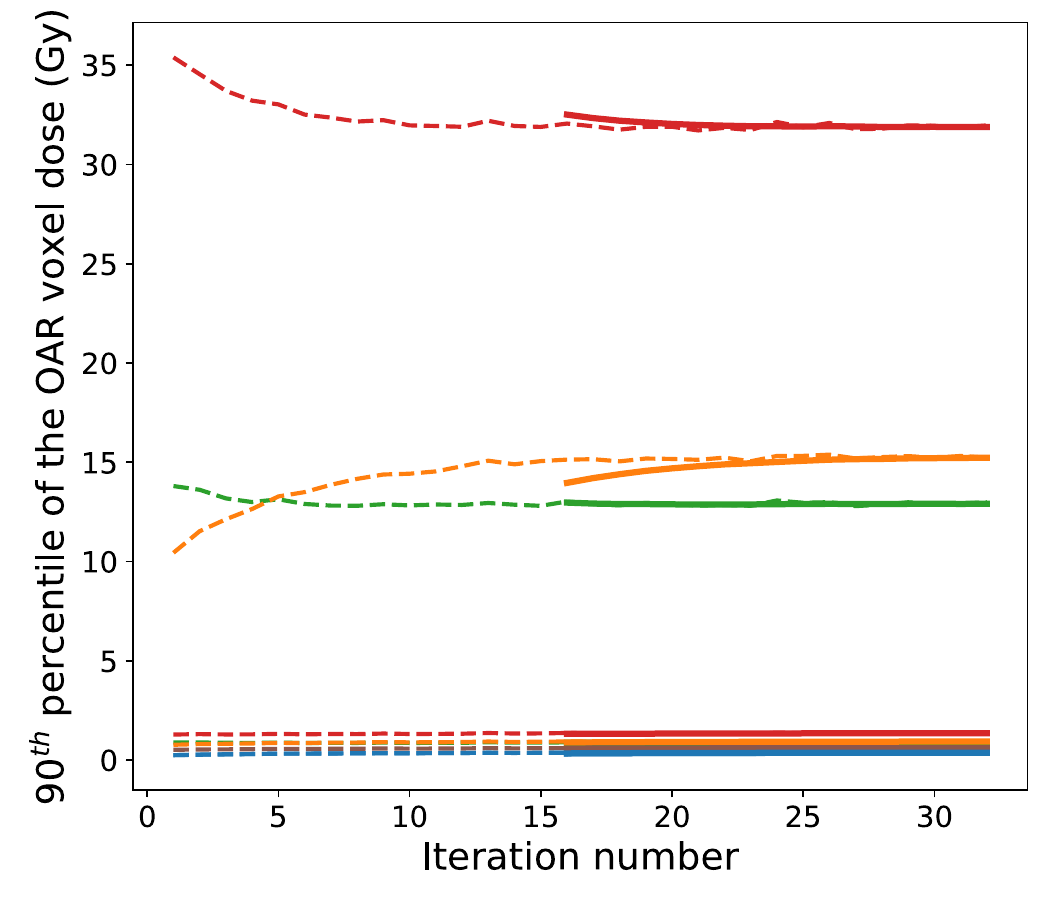}
        \caption{}
    \end{subfigure}

    %\vspace{1em}

    % Row 2
    \begin{subfigure}{0.32\textwidth}
        \centering
        \includegraphics[width=\linewidth]{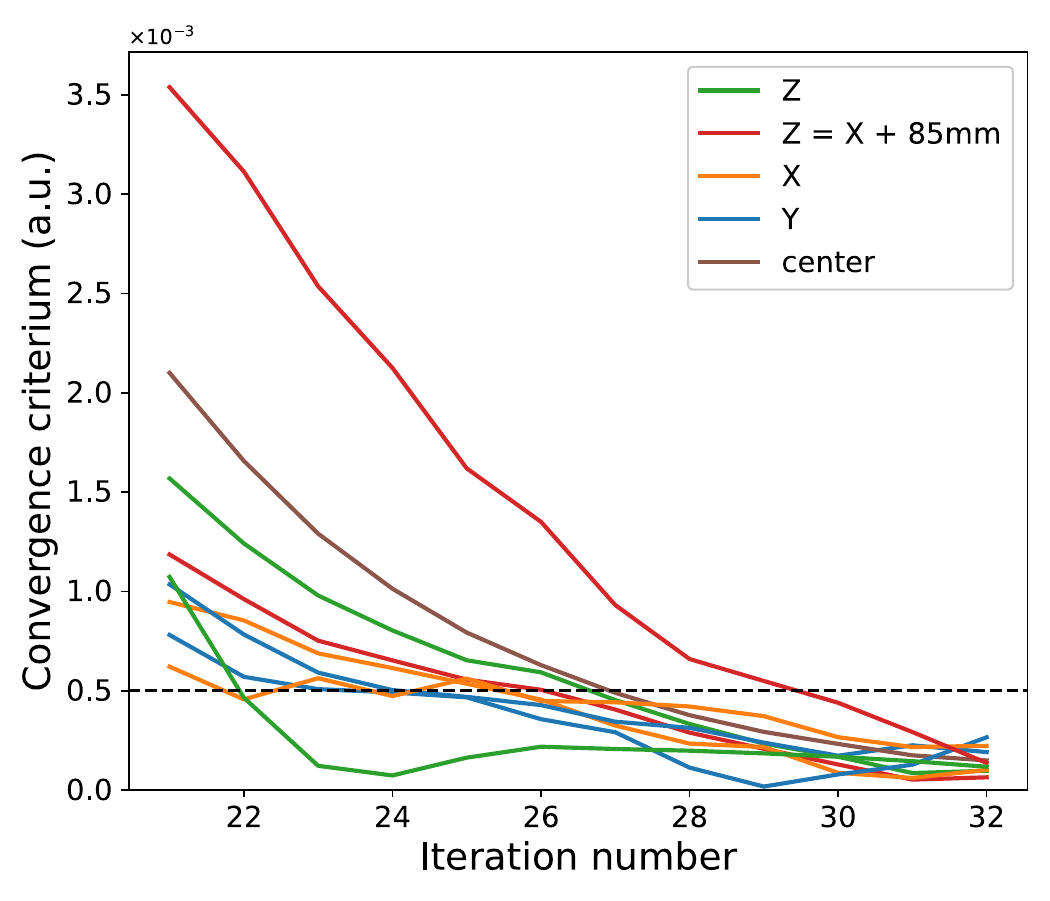}
        \caption{}
    \end{subfigure}
    %\hfill
    \begin{subfigure}{0.32\textwidth}
        \centering
        \includegraphics[width=\linewidth]{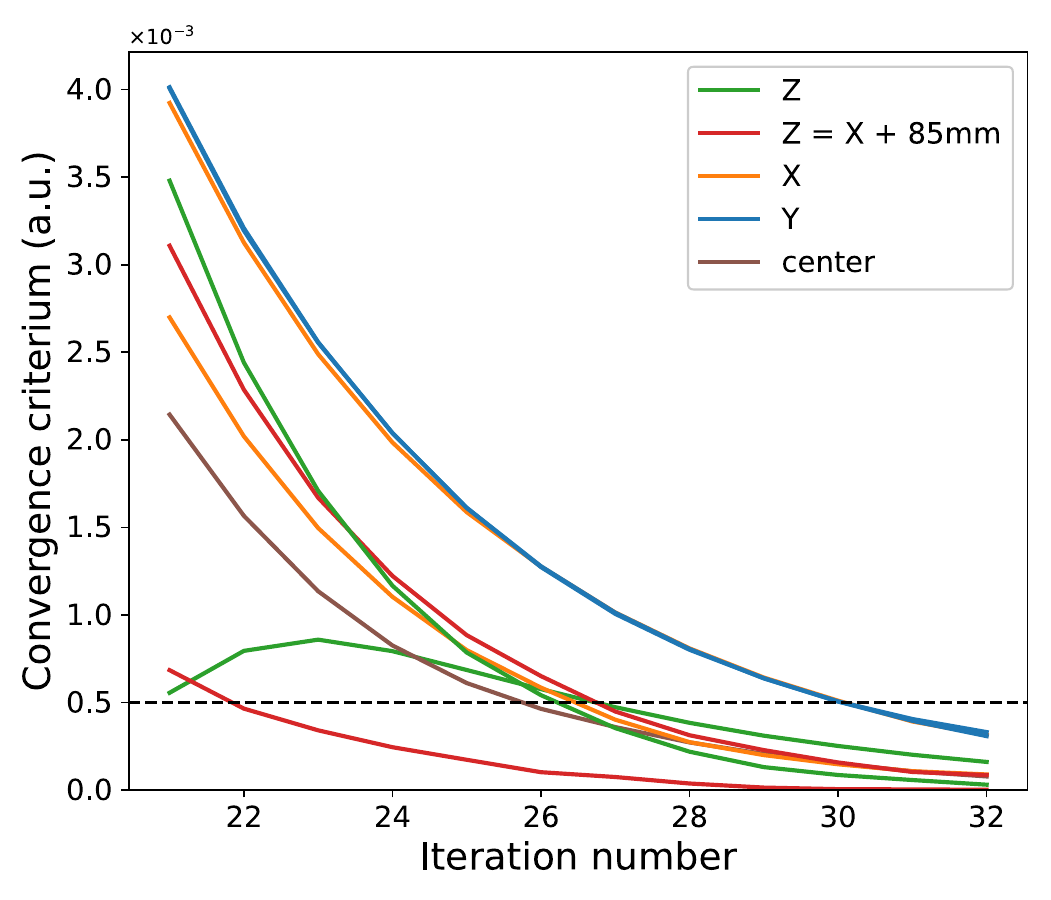}
        \caption{}
    \end{subfigure}
    %\hfill
    \begin{subfigure}{0.32\textwidth}
        \centering
        \includegraphics[width=\linewidth]{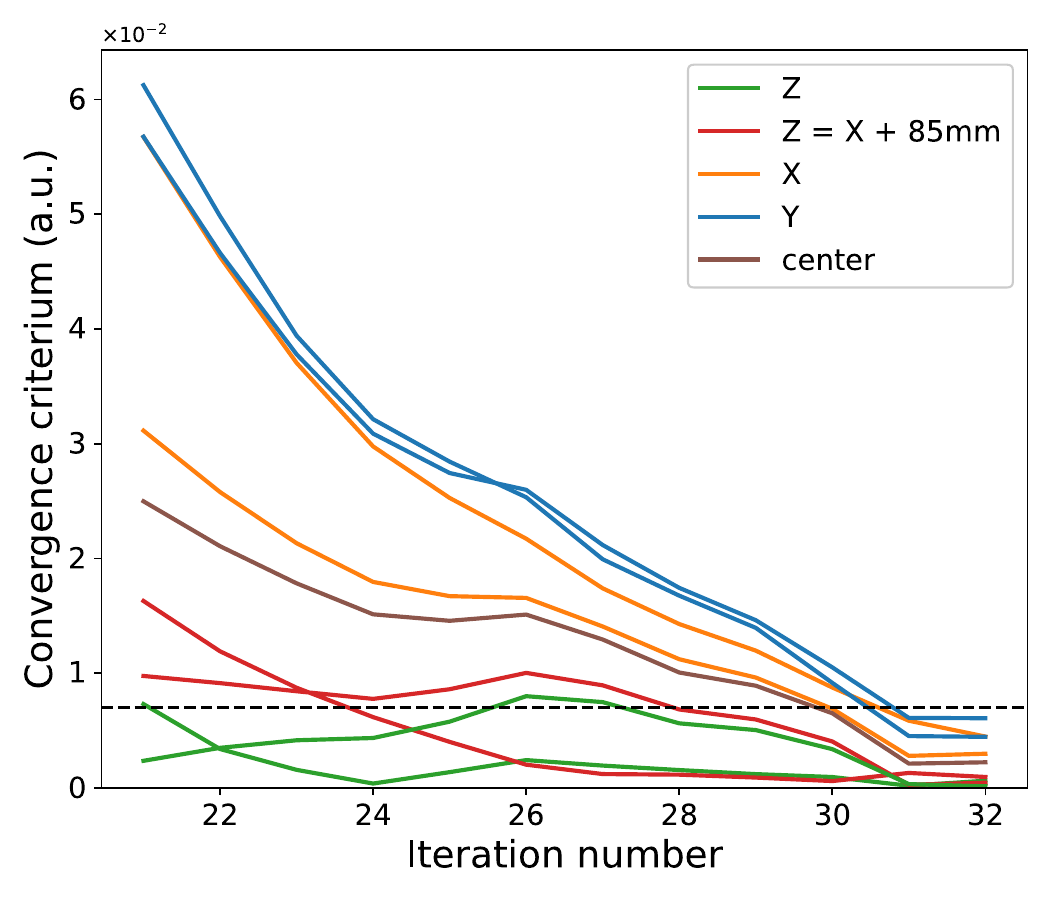}
        \caption{}
    \end{subfigure}

    \caption{The convergence behavior of ($10^{th}$ and $90^{th}$) CTV and ($90^{th}$) OAR voxel dose percentiles (dashed) for the XZ-displaced \textit{setupXYrange} probabilistic optimization, with their moving average (MA15, solid). Only representative outer voxels of both structures are shown, along the CTV and OAR centered $X$, $Y$, $Z$ and $Z = X + \qty{85}{\milli\meter}$ axes. The relative change in MA15 as in  Equation \ref{eq:Methods_convergenceCriteria_CTV} with $\Delta k = 5$ is shown (bottom) with the convergence tolerance (dashed black).}
    \label{fig:CTVandOARPercentileConvergence_XYr}
\end{figure}

Convergence criteria are defined in accordance with Section \ref{subsubsec:outerOptimization}. A representative example of the percentile convergence (corresponding to the damped beam weights) is shown on top in Figure \ref{fig:CTVandOARPercentileConvergence_XYr} for the spherical \textit{setupXYrange} plan, where the $10^{th}$- and $90^{th}$-percentiles of the CTV voxels and the $90^{th}$-percentiles of the OAR voxels are optimized for. The bottom of Figure \ref{fig:CTVandOARPercentileConvergence_XYr} shows the respective convergence criteria with the convergence thresholds (in dashed black).

\subsubsection{Probabilistic optimization times}
\label{app:probOptTimes}

\begin{figure}[]
    \centering
    \includegraphics[width=0.7\textwidth]{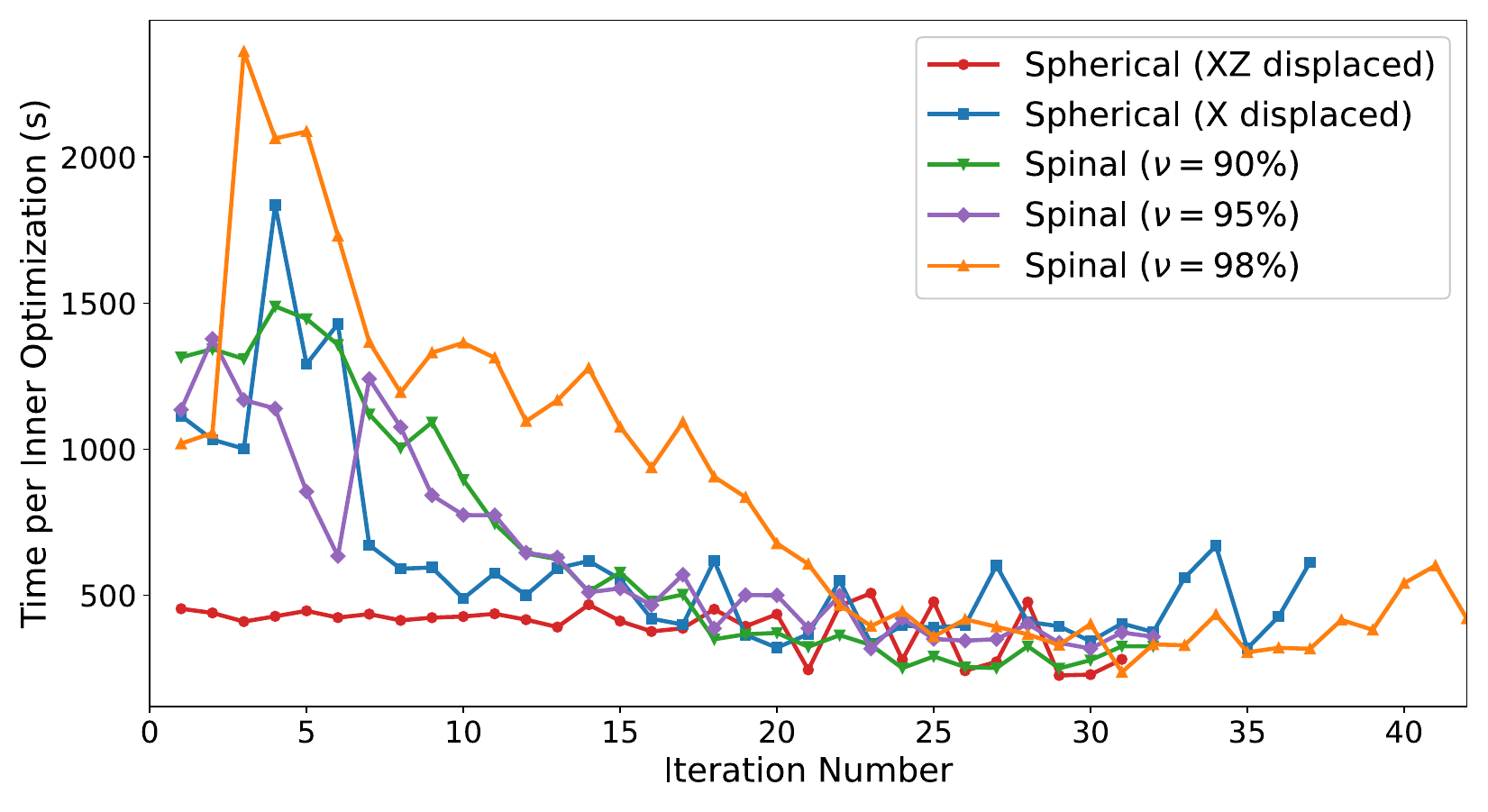}
    \caption{Inner optimization times of the probabilistic spherical and spinal plans.}
    \label{fig:optTimes_CTVandOAR} 
\end{figure}

Large computational times are primarily due to the inner optimization. Figure \ref{fig:optTimes_CTVandOAR} shows the inner optimization times for the spherical (CTV + OAR) and spinal plans, which decreases over the course of the optimization for all plans. This occurs because the probabilistic objectives are only evaluated for voxels that have not yet met the target probability level. Voxels that meet the target do not necessarily remain passing throughout the optimization; they can fall below the target again and be re-included in the probabilistic objective. However, as optimization continues, more voxels consistently satisfy the objectives. This reduces the number of voxels needing probabilistic optimization and thus lowers the overall computational load.

The $\nu = 98\%$ case took particularly longer than the other cases, mainly because of increased inner optimization times during the early iterations (before iteration 21). In addition to having slightly stricter convergence criteria compared to $\nu = 90\%$ and $\nu = 95\%$, stricter spinal overdose probabilities caused more voxels to not reach the target probability level, making them active in the optimization more often.

The fact that iterations take longer if more voxels have not reached the desired thresholds yet, means at the same time that optimization times are sensitive to the choice of initial beam weights. Inner optimization times may be improved by terminating the first iterations before fully converged beam weights are obtained. This can be done since the main purpose of these initial optimizations is not to obtain converged beam weights, but rather to direct the voxel dose percentiles (i.e., $\delta$-factors) to the desired levels. This may reduce the optimization time significantly, while the number of (outer loop) iterations is barely affected.

\subsection{Robust optimizations}
\label{app:robOpt}

All probabilistic plans in this work are compared to a composite-wise mini-max robust plan as in Equation \ref{eq:robOpt_general}. Table \ref{tab:robOptParameters} shows the resulting robust objective weights $\omega \in \Omega$, obtained from tuning the CTV coverage or OAR/spinal dose towards the probabilistic plan outcome. For the spherical CTV + OAR case, combinations of XZ-displaced or X-displaced and types of scaling (either by CTV coverage or OAR dose) are done. Three different spinal plans are compared, where the spinal overdosage probability is limited by different probability levels ($\nu_i = 90\%$, $\nu_i = 95\%$ and $\nu_i = 98\%, \forall i \in Spine$).

\begin{table}[t!]
\caption{\label{tab:robOptParameters} Robust optimization parameters for the XZ-displaced (XZ) and X-displaced (X) geometries, that are matched by either CTV coverage ($D_{98\%}^{10\text{th}}$) or OAR dose ($D_{2\%}^{90\text{th}}$). Prescribed dose $d_i^p = \qty{60}{\gray}$ and voxel weights for CTV, spine/OAR and tissue are $w_i^{CTV} = 100$, $w_i^{OAR} = 20$, $w_i^{Tissue} = 1$, respectively.} 
\begin{center}
\begin{tabular}{@{}*{7}{c}}
\toprule 
 & $\omega_{CTV}$ & $\omega_{OAR}$ & $\omega_{OAR}^{max}$ & $\omega_{CTV}^{nom}$ & $\omega_{Tissue}$ & $d_i^{maxOAR} (\unit{\gray})$ \\[0.5ex]
\midrule
CTV-only                            &  15   & ---   & ---   & ---   & 1      & 30 \cr
XZ ($D_{98\%}^{10\text{th}}$)            &  120  & 1     & 1     & ---   & 160    & 30 \cr
XZ ($D_{2\%}^{90\text{th}}$)             &  100  & 10    & 10    & ---   & 100    & 30 \cr
X ($D_{98\%}^{10\text{th}}$)             &  120  & 1     & 1     & ---   & 160    & 30 \cr
X ($D_{2\%}^{90\text{th}}$)              & 100   & 15    & 1     & ---   & 100    & 30 \cr
\midrule
spinal ($\nu = 90\%$)               &  2    &  1    & 1     & 4     &  1     & 54 \cr 
spinal ($\nu = 95\%$)               &  3    &  2    & 2     & 6     &  2     & 54 \cr
spinal ($\nu = 98\%$)               &  11   &  10   & 10    & 22    & 10     & 54 \cr
\bottomrule
\end{tabular}
\end{center}
\end{table}

\section{Comparison for the X-displaced spherical CTV + OAR case} 
\label{app:additionalComparisons_spherical}
This appendix presents additional results on the spherical CTV + OAR case, for the OAR that is X-displaced with respect to the CTV. Probabilistic optimization variables are identical to the XZ-displaced case (see Table \ref{tab:probOptParameters}). Robust optimization variables are shown in Table \ref{tab:robOptParameters}, which were obtained by manually tuning until the CTV coverage ($D_{98\%}^{10\text{th}}$) or OAR dose ($D_{2\%}^{90\text{th}}$) was similar as in the probabilistic plan. The tuned robust objective weights yielding similar CTV coverage and OAR dose are $\{ \omega_{CTV}, \omega_{OAR}, \omega_{OAR}^{max}, \omega_{Tissue} \} = \{120, 1, 1, 160\}$ and $\{100, 15, 1, 100\}$, respectively. Probabilistic scaling is done in accordance with Section \ref{subsec:Methods_scaling}. Table \ref{tab:DVHmetrics_CTVandOAR_Xshifts} presents the resulting DVH-metrics before and after scaling.

\begin{table}[b!]
\caption{\label{tab:DVHmetrics_CTVandOAR_Xshifts} Statistical DVH-metrics of CTV coverage ($D_{98\%}^{10\text{th}}$), OAR dose ($D_{2\%}^{90\text{th}}$) and $D_{50\%}^{50\text{th}}$ of the robust and probabilistic plans, before and after (with brackets) probabilistic scaling. The robust objective weights obtained after tuning are shown in brackets as $\{ \omega_{CTV}, \omega_{OAR}, \omega_{OAR}^{max}, \omega_{Tissue} \}$. } 

\begin{center}
\begin{tabular}{@{}*{4}{l}}
\toprule                             
 & $D_{98\%}^{10\text{th}}$ (\SI{}{Gy}) & $D_{50\%}^{50\text{th}}$ (\SI{}{Gy}) & $D_{2\%}^{90\text{th}}$ (\SI{}{Gy}) \\[0.5ex]
\midrule
Robust (\{120, 1, 1, 160\}) & 51.2 (51.3)  &  59.8 (60) & 48.3 (48.4) \cr
Robust (\{100, 15, 1, 100\}) & 42.8 (41.6)  &  59.7 (58.0) & 35.1 (34.1) \cr
Probabilistic &  51.4 (51.3)
   & 60.1 (60.0) & 34.1 (34.1) \cr 
\bottomrule
\end{tabular}
\end{center}
\end{table}

\subsection{Plan comparisons}
\begin{figure}[t!]
    \centering

    \begin{subfigure}[]{0.32\textwidth}
        \includegraphics[width=\linewidth]{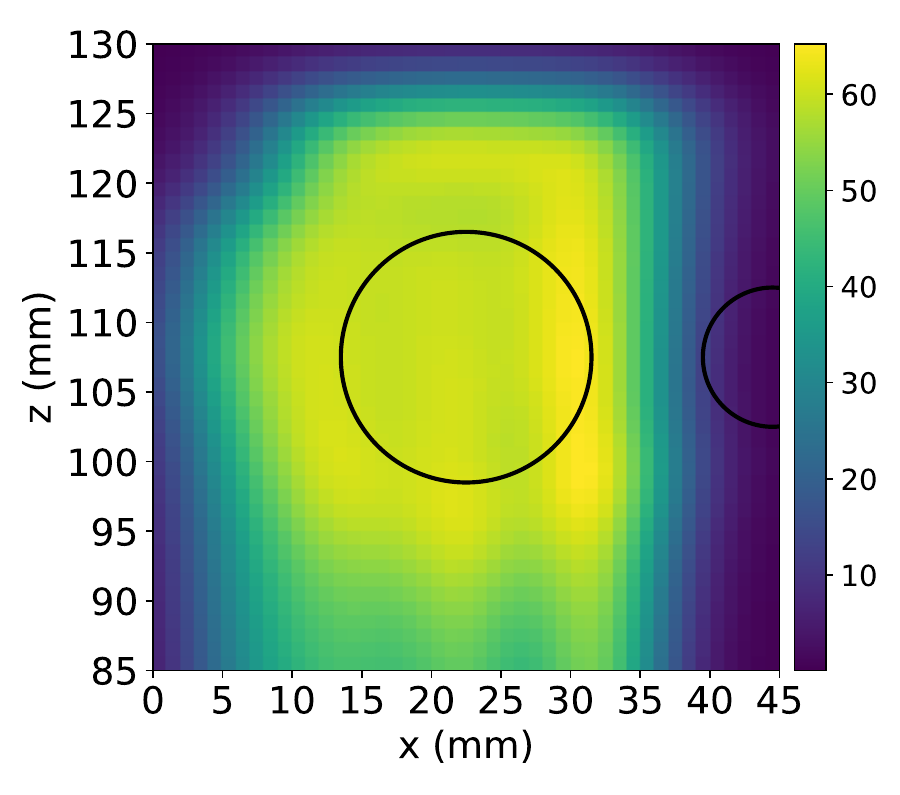}
        \caption{}
        \label{fig:CTVandOAR_XYr_Xshift_Prob_XZ}
    \end{subfigure}
    %\hfill
    \begin{subfigure}[]{0.32\textwidth}
        \includegraphics[width=\linewidth]{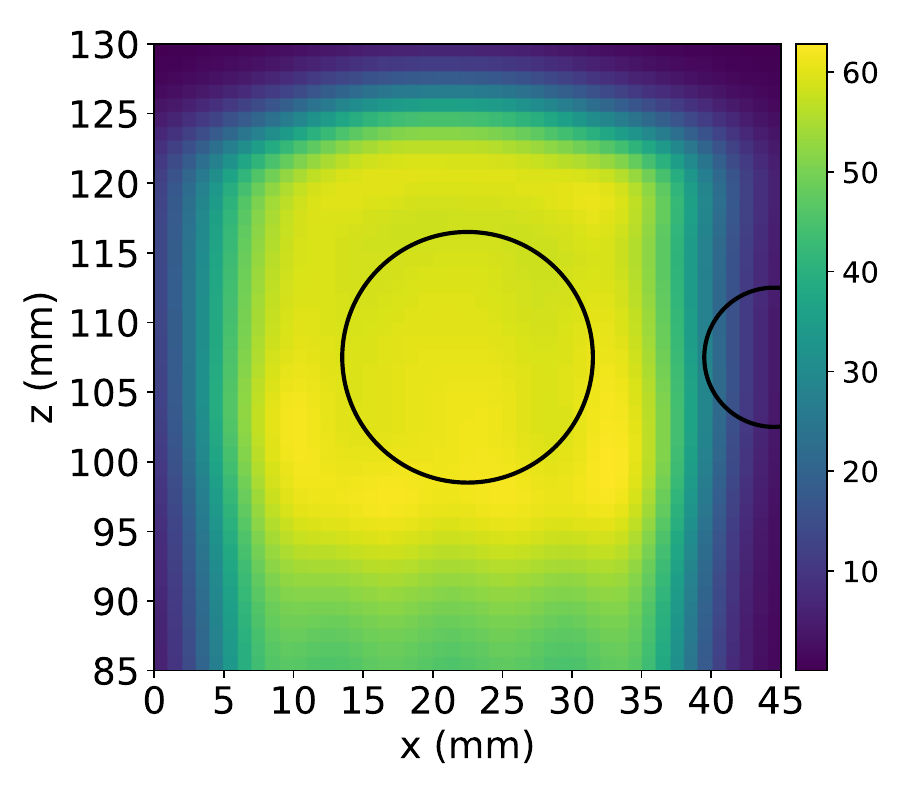}
        \caption{}
        \label{fig:CTVandOAR_XYr_Xshift_CTVcoverage_XZ}
    \end{subfigure}
    %\hfill
    \begin{subfigure}[]{0.32\textwidth}
        \includegraphics[width=\linewidth]{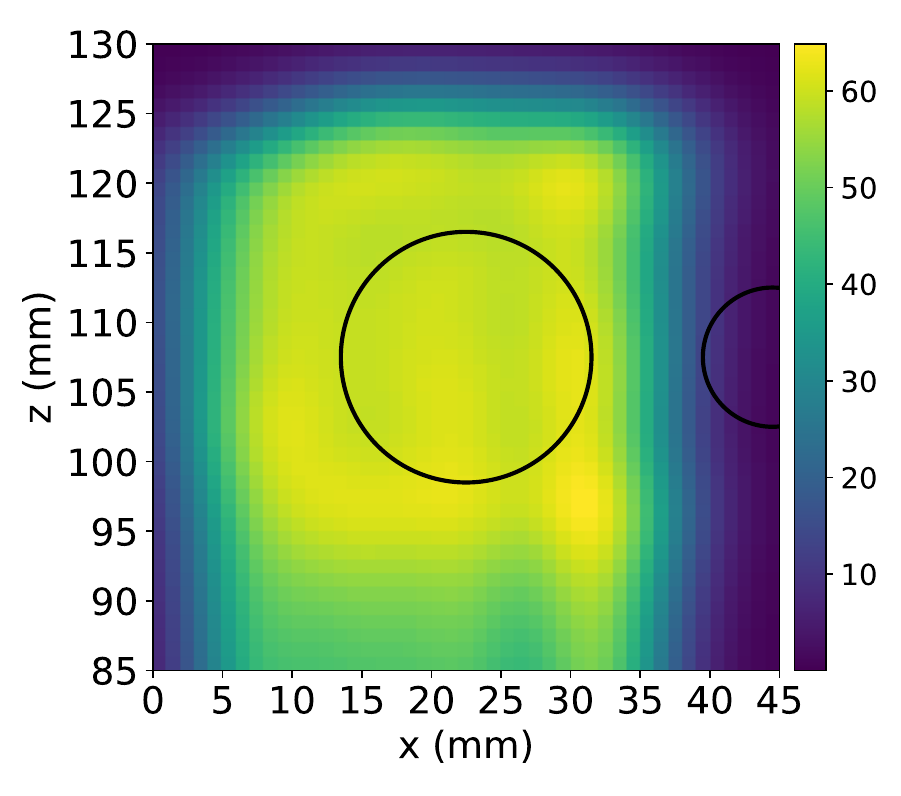}
        \caption{}
        \label{fig:CTVandOAR_XYr_Xshift_OARdose_XZ}
    \end{subfigure}

    %\vspace{0.5cm}

    \begin{subfigure}[]{0.32\textwidth}
        \includegraphics[width=\linewidth]{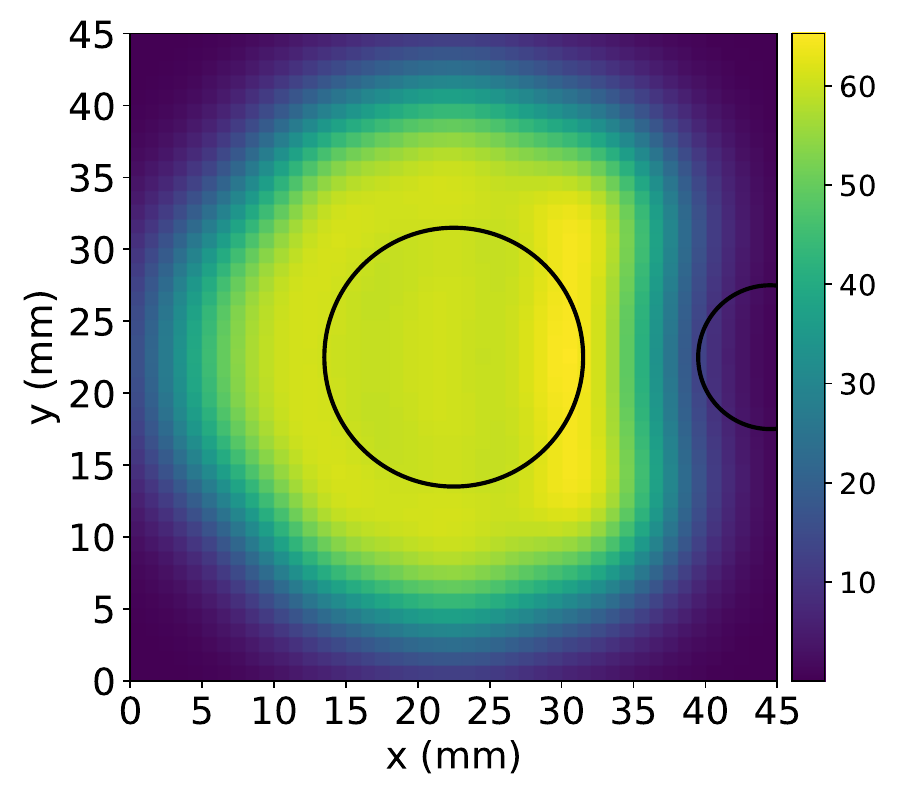}
        \caption{}
        \label{fig:CTVandOAR_XYr_Xshift_Prob_XY}
    \end{subfigure}
    %\hfill
    \begin{subfigure}[]{0.32\textwidth}
        \includegraphics[width=\linewidth]{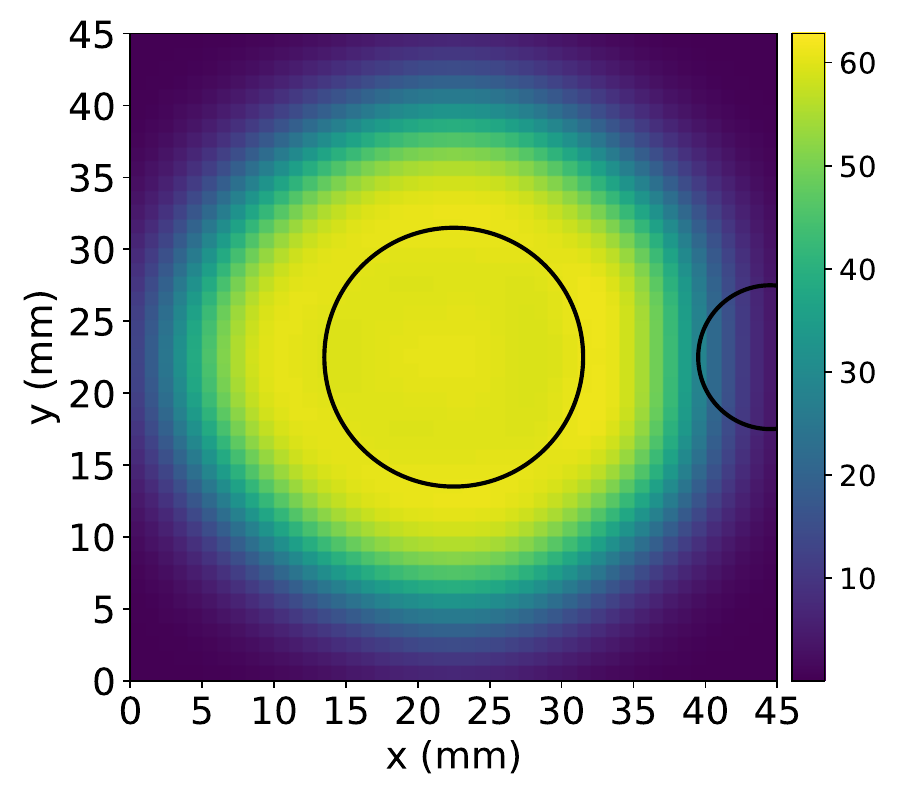}
        \caption{}
        \label{fig:CTVandOAR_XYr_Xshift_CTVcoverage_XY}
    \end{subfigure}
    %\hfill
    \begin{subfigure}[]{0.32\textwidth}
        \includegraphics[width=\linewidth]{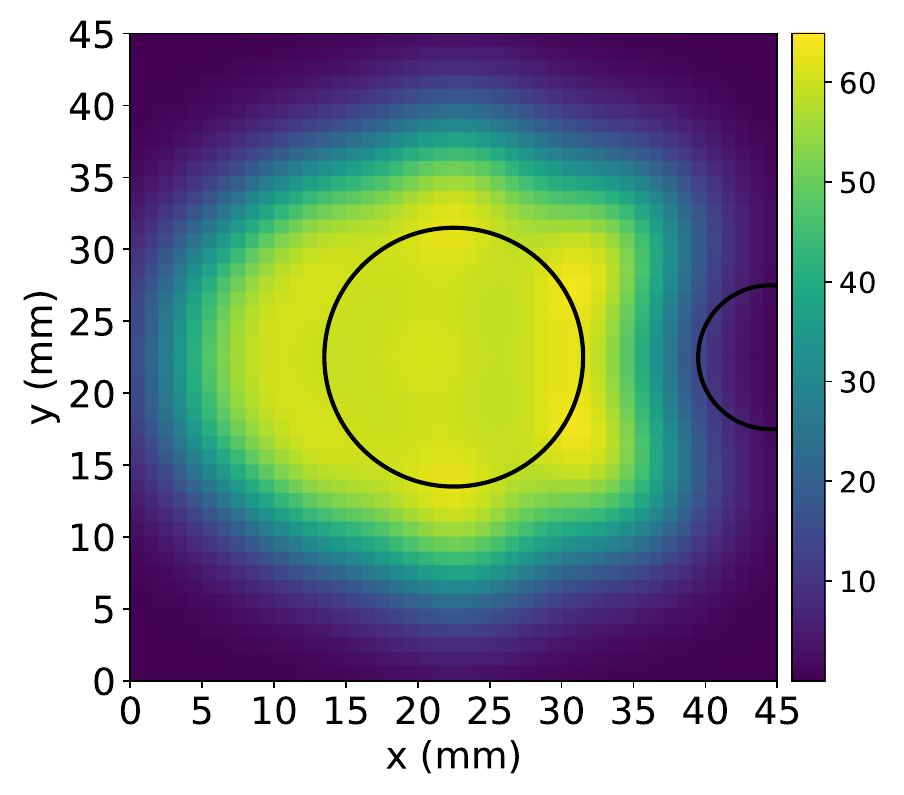}
        \caption{}
        \label{fig:CTVandOAR_XYr_Xshift_OARdose_XY}
    \end{subfigure}

    \caption{Comparison of nominal X-displaced dose distributions for the (left) probabilistic plan, with the robust plans that are matched based on the (middle) $D_{50\%}^{50\text{th}}$ and (right) $D_{2\%}^{90\text{th}}$ metrics. Both the (bottom) XY-plane and (top) XZ-plane through the CTV center are shown.}
    \label{fig:CTVandOAR_XYr_Xshift}
\end{figure}

The nominal dose distributions are shown in Figure \ref{fig:CTVandOAR_XYr_Xshift}. XZ-slices (top) and XY-slices (bottom) through the CTV center are shown for the probabilistic (left) and robust plans that are scaled by the (middle) $D_{50\%}^{50\text{th}}$ and (right) $D_{2\%}^{90\text{th}}$ metrics. OAR sparing in the probabilistic plan is achieved by a dose build-up at the OAR-side of the CTV. The robust plan with similar CTV coverage as the probabilistic plan shows a larger dose extension on the OAR-side of the CTV compared to the probabilistic plan, resulting in increased OAR dose. By increasing the relative OAR objective importance in the optimization (Figure \ref{fig:CTVandOAR_XYr_Xshift_OARdose_XZ} and Figure \ref{fig:CTVandOAR_XYr_Xshift_OARdose_XY}), the dose extension is indeed reduced, because scenarios that shift large dose values into the OAR are often worst-case. At the same time, shifts in the $y$-direction are uncommon to be the worst-case, leading to a significant reduction in dose extension in the $y$-direction, in turn reducing CTV coverage.

\begin{figure}[h!]
    \centering

    \begin{subfigure}[]{0.34\textwidth}
        \includegraphics[width=\linewidth]{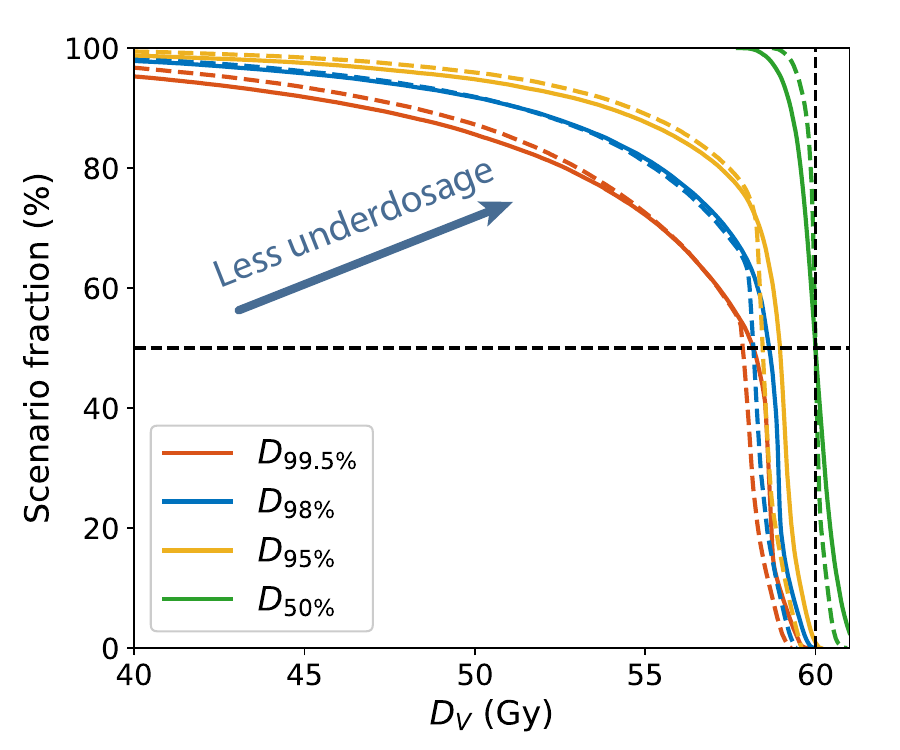}
        \caption{}
    \end{subfigure}
    %\hfill
    \begin{subfigure}[]{0.34\textwidth}
        \includegraphics[width=\linewidth]{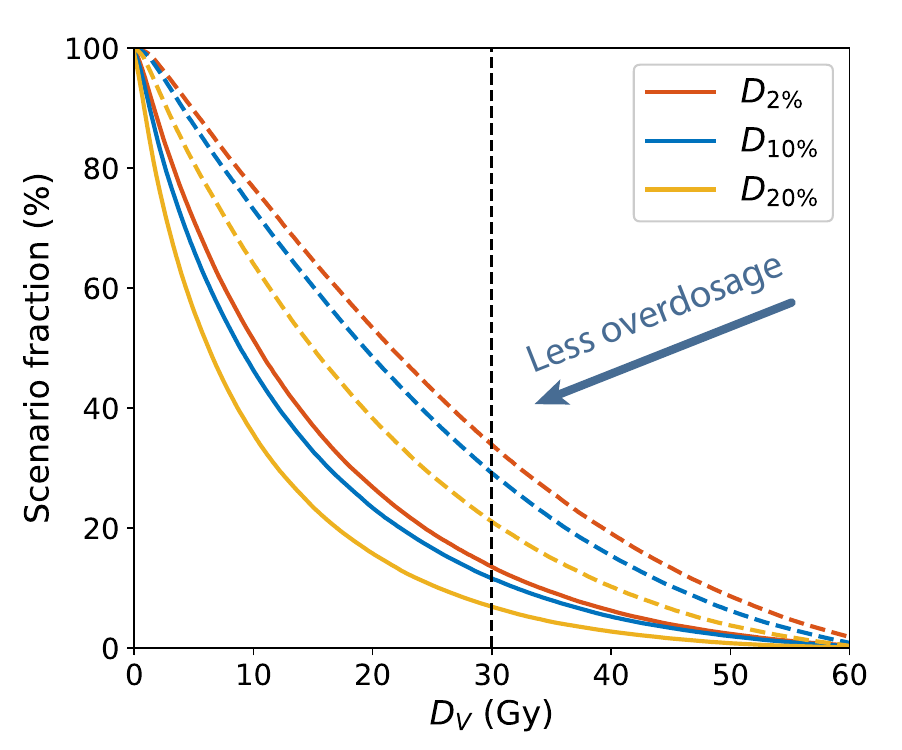}
        \caption{}
    \end{subfigure}

    %\vspace{0.5cm}

    \begin{subfigure}[]{0.34\textwidth}
        \includegraphics[width=\linewidth]{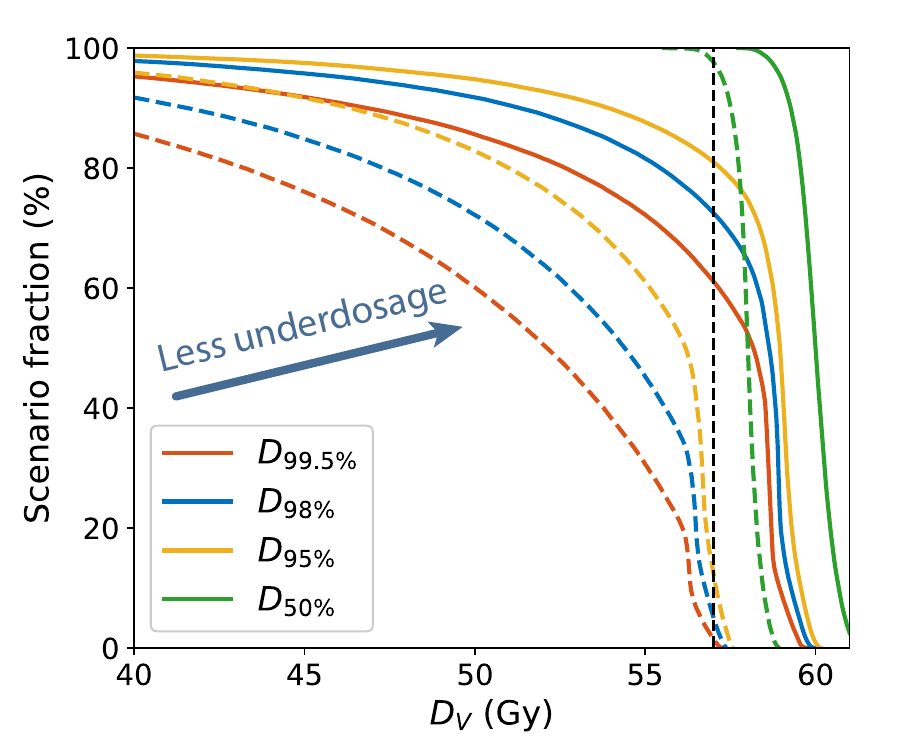}
        \caption{}
    \end{subfigure}
    %\hfill
    \begin{subfigure}[]{0.34\textwidth}
        \includegraphics[width=\linewidth]{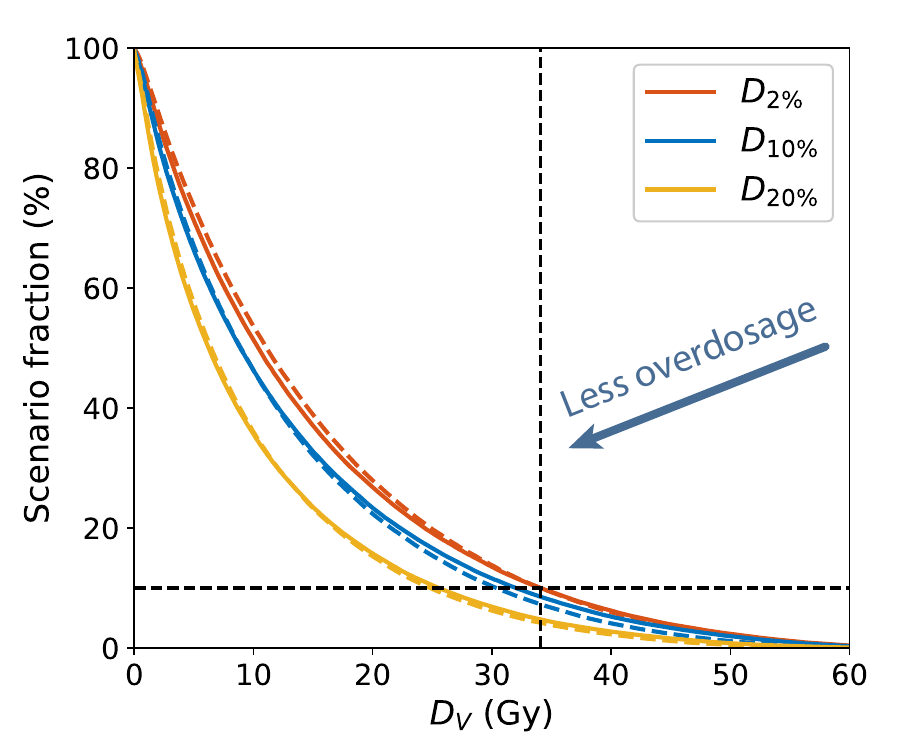}
        \caption{}
    \end{subfigure}

    \caption{Dose population histograms of various DVH metrics ($D_{V}$, e.g., $V=98\%$) for the CTV (left) and OAR (right), comparing the X-displaced probabilistic (solid) and robust (dashed) plans that are matched based on the (top) $D_{50\%}^{50\text{th}}$ and (bottom) $D_{2\%}^{90\text{th}}$ metrics.}
    \label{fig:CDFcomparisonSpheres_allDVHmetrics_Xshift}
\end{figure}

% DPH figures
Figure \ref{fig:CDFcomparisonSpheres_allDVHmetrics_Xshift} compares the dose population histograms (DPHs) of various DVH metrics between the probabilistic plan and the scaled robust plans based on the $D_{50\%}^{50\text{th}}$ percentile (top) and the $D_{2\%}^{90\text{th}}$ percentile (bottom). The results are similar to the XZ-displaced case: $P(D_{2\%} > \qty{30}{\gray})$ has decreased by about 20.5\% in the probabilistic $D_{50\%}^{50\text{th}}$ scaled plan, and $P(D_{98\%} > \qty{57}{\gray})$ has increased by about 67.5\% in the probabilistic $D_{2\%}^{90\text{th}}$ scaled plan.

% X shift (scaled 50thD50)
%               Probabilistic       robust
% D2 at 30Gy:   13.5%               33.9%   (reduction of about 20.5%)

% X shift (scaled 90thD2)
%               Probabilistic       robust
% D98 at 57Gy:  72.5%               5.2%    (increase of about 67.5%)

\begin{figure}[h!]
    \centering

    \begin{subfigure}[]{0.32\textwidth}
        \includegraphics[width=\linewidth]{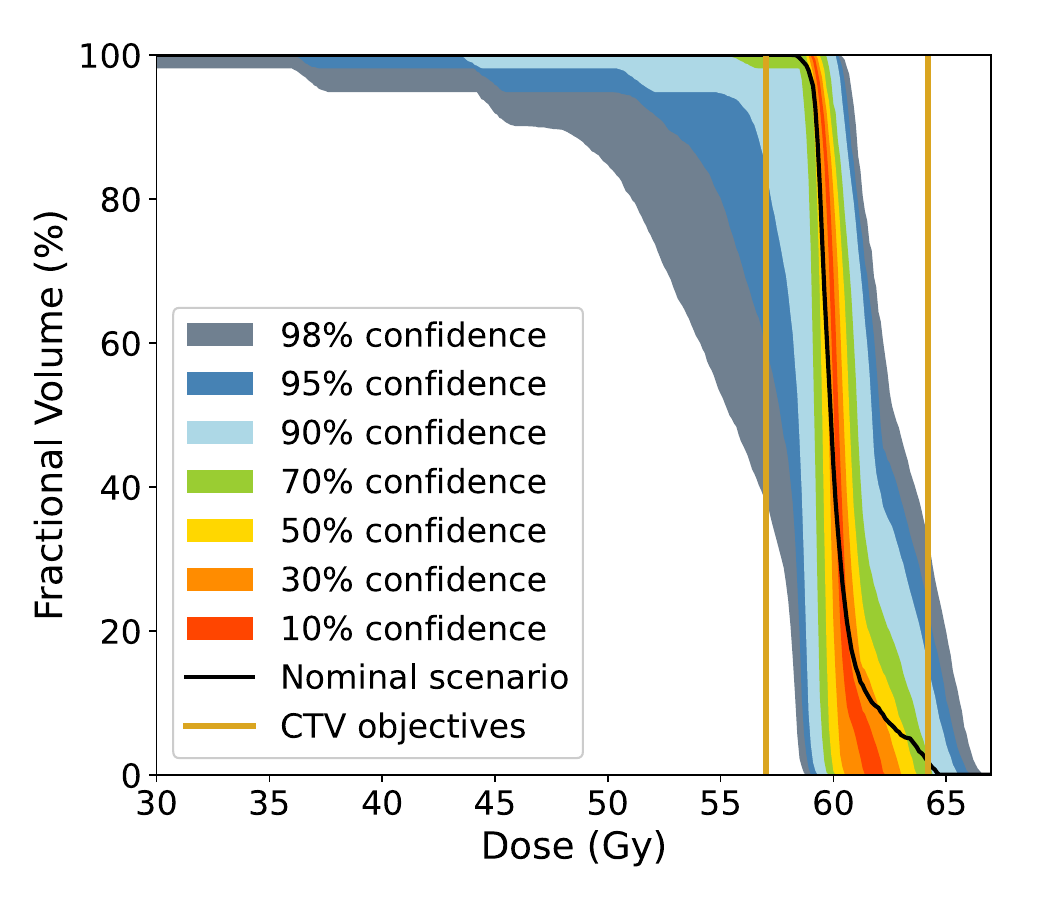}
        \caption{}
    \end{subfigure}
    %\hfill
    \begin{subfigure}[]{0.32\textwidth}
        \includegraphics[width=\linewidth]{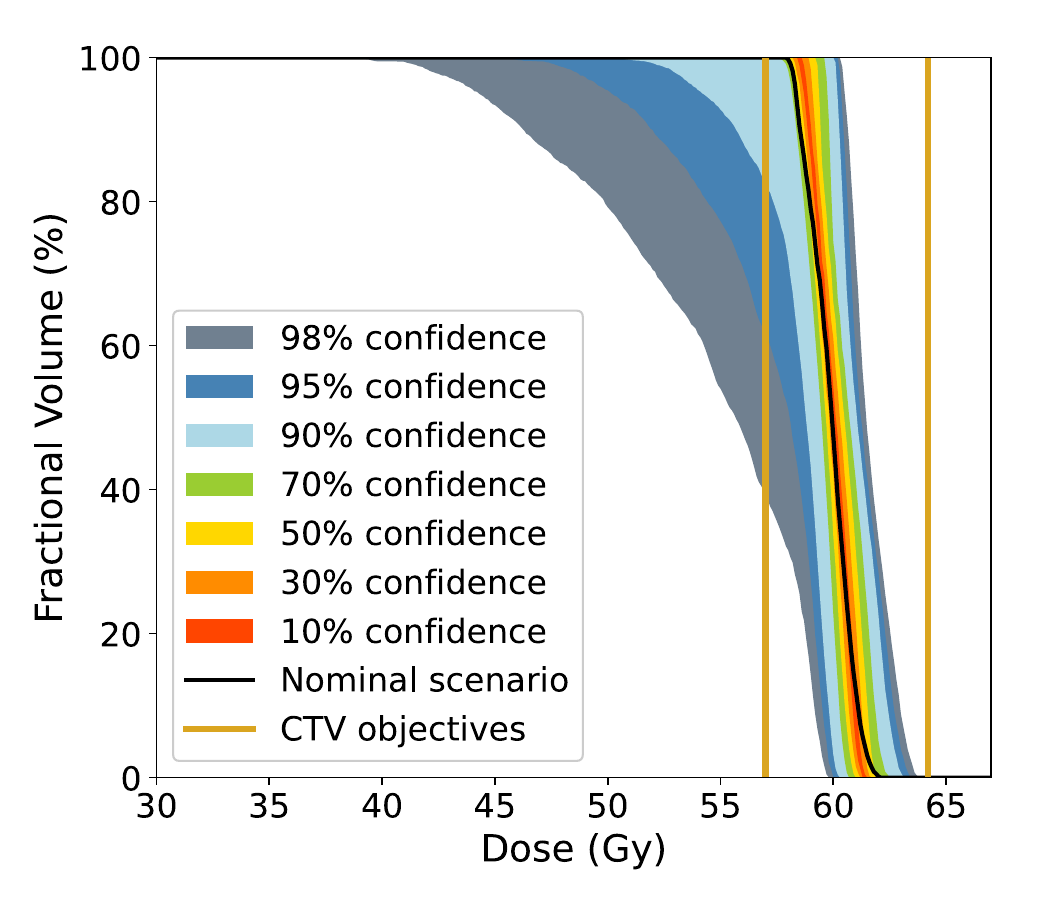}
        \caption{}
    \end{subfigure}
    \begin{subfigure}[]{0.32\textwidth}
        \includegraphics[width=\linewidth]{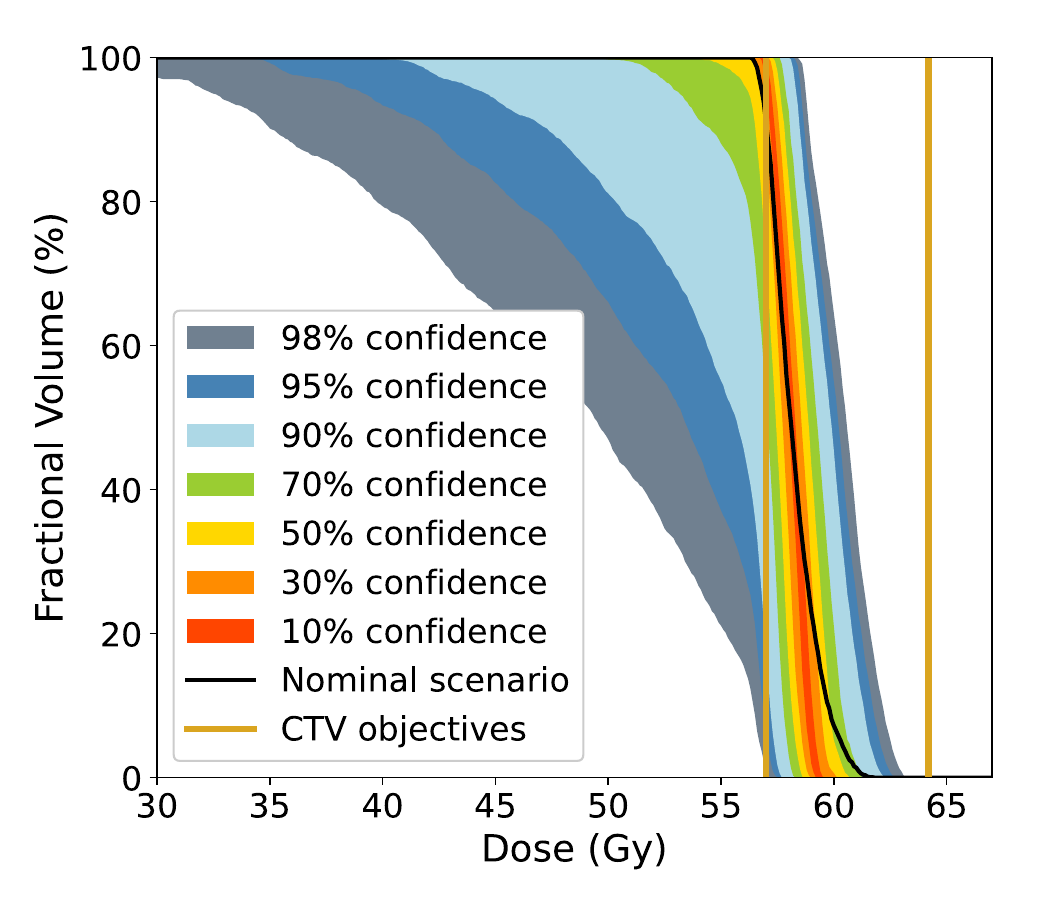}
        \caption{}
    \end{subfigure}

    %\vspace{0.5cm}

    \begin{subfigure}[]{0.32\textwidth}
        \includegraphics[width=\linewidth]{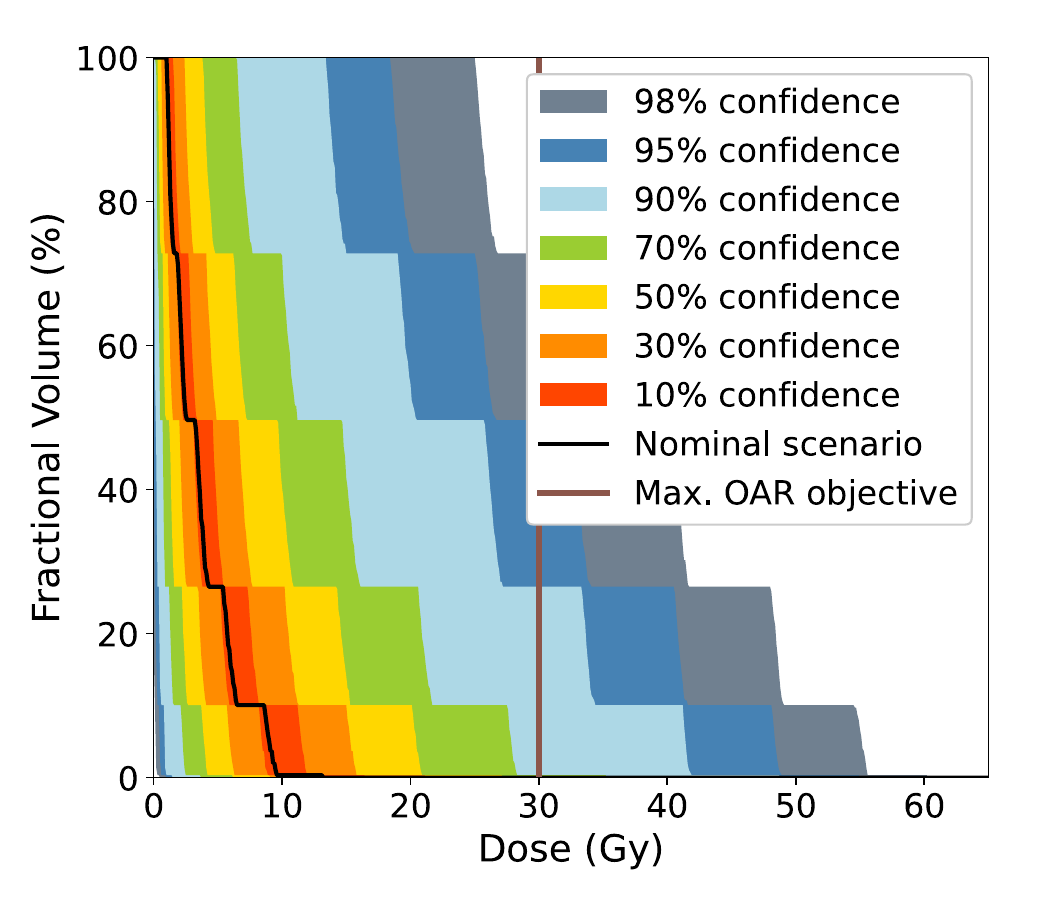}
        \caption{}
    \end{subfigure}
    %\hfill
    \begin{subfigure}[]{0.32\textwidth}
        \includegraphics[width=\linewidth]{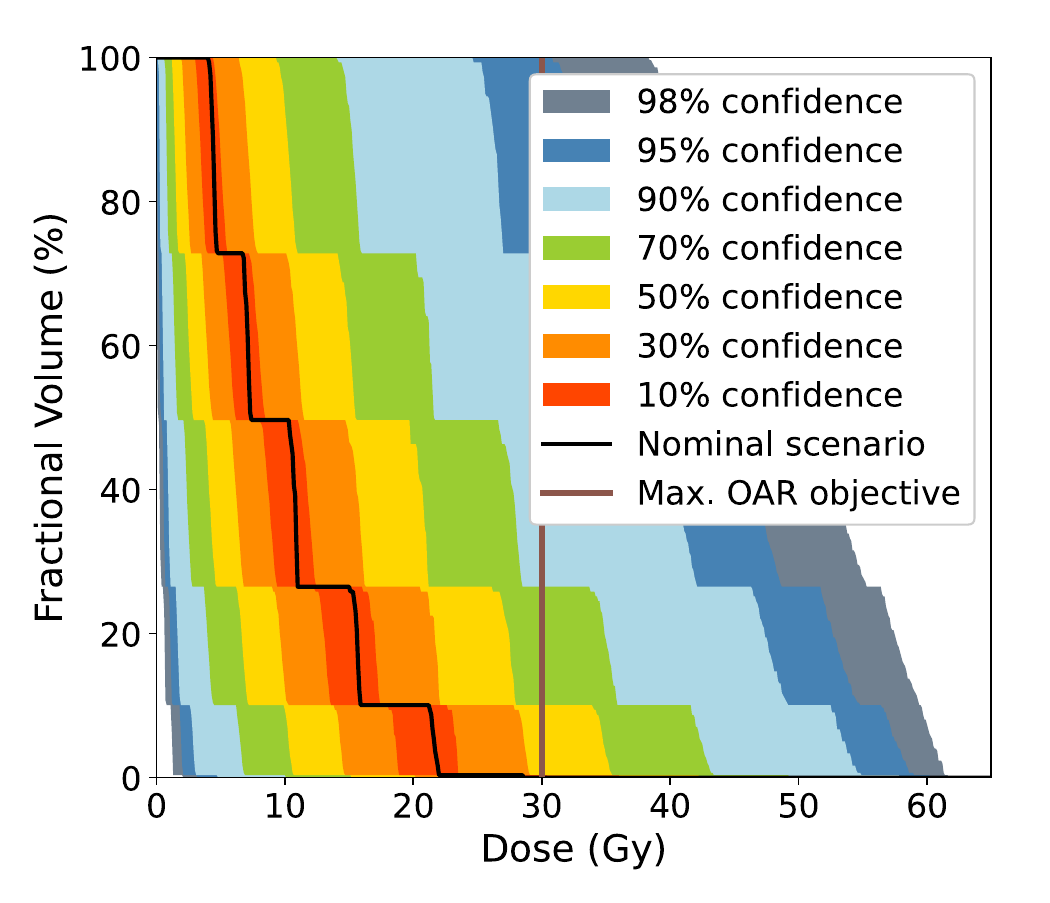}
        \caption{}
    \end{subfigure}
        \begin{subfigure}[]{0.32\textwidth}
        \includegraphics[width=\linewidth]{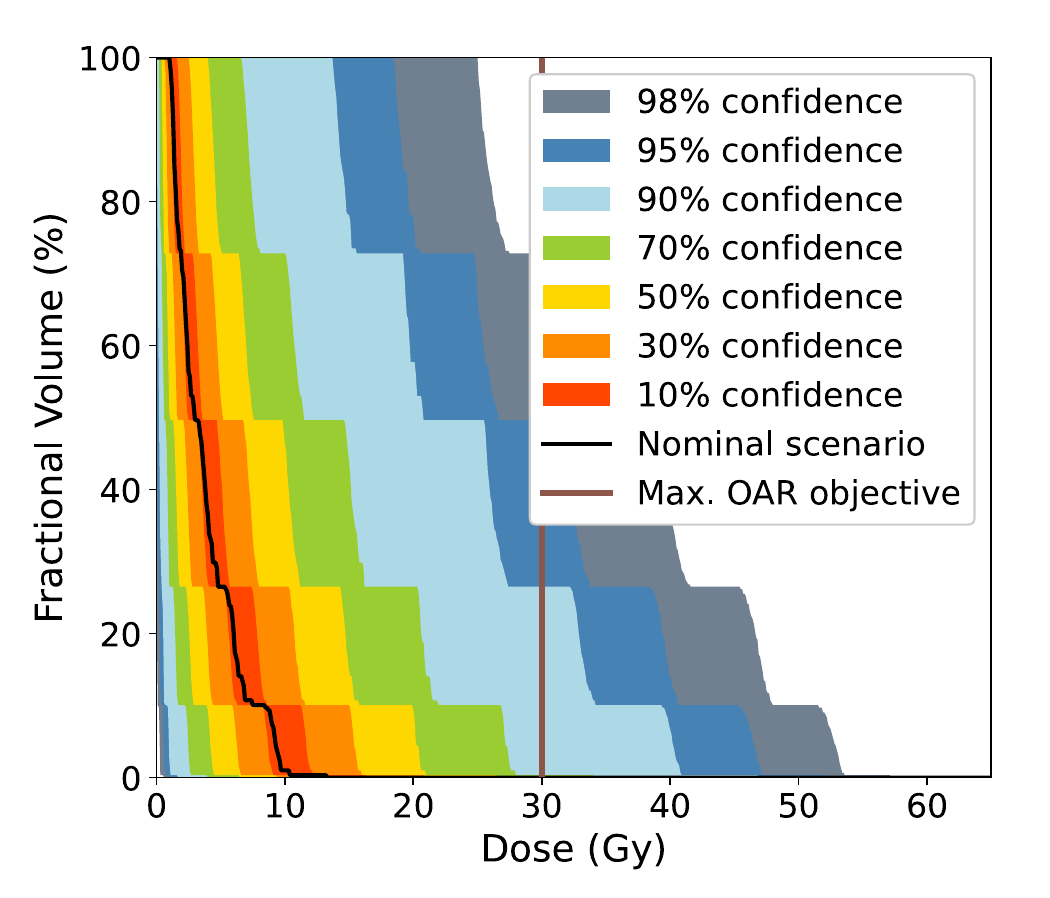}
        \caption{}
    \end{subfigure}
    
    \caption{Dose volume histogram distributions for the CTV (top) and OAR (bottom), comparing the X-displaced probabilistic (left) and robust plans that are matched based on the (middle) $D_{50\%}^{50\text{th}}$ and (right) $D_{2\%}^{90\text{th}}$ metrics. The nominal scenario (black) is shown together with various confidence bands. As the X-displaced OAR consists of only 298 voxels, the corresponding DVH bands are less continuous.}
    \label{fig:CTVandOAR_DVHbands_Xshift}
\end{figure}

In Figure \ref{fig:CTVandOAR_DVHbands_Xshift}, we compare the DVH bands of CTV (left) and OAR (right) between the X-displaced probabilistic and robust plans that are matched based on the (top) $D_{50\%}^{50\text{th}}$ and (bottom) $D_{2\%}^{90\text{th}}$ metrics. For the former comparison, the OAR has significantly improved dose values in the probabilistic case: the minimum OAR dose exceeds \qty{30}{\gray} in about 35\% of the cases for the robust plan (for 0\% volume fraction, the \qty{30}{\gray} is on the upper edge of the 30\% confidence band), whereas this is reduced to about 15\% in the probabilistic plan. As can be seen from the CTV DVH-distributions, this improvement is at the expense of having slightly more overdosage in a small part of the CTV and having longer tails in the near-maximum DVH-metrics. For the $D_{2\%}^{90\text{th}}$ scaled robust plan, the DVH distributions for the OAR are similar. The lower tails of the near-maximum dose distributions are less extreme, with the drawback of having more CTV overdosage probability in small volume fractions.

\begin{figure}[h!]
    \centering

    \begin{subfigure}[]{0.37\textwidth}
        \includegraphics[width=\linewidth]{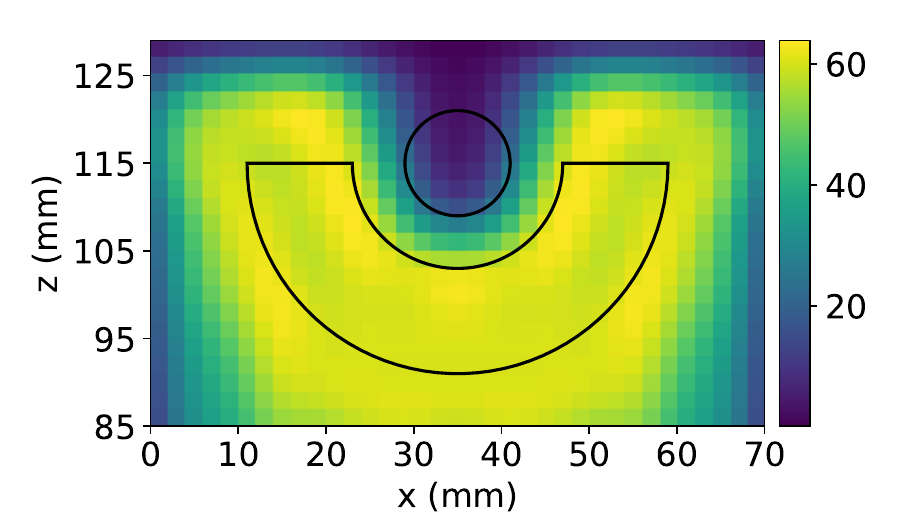}
        \caption{}
        \label{fig:spinal_nominal_prob95_98_95probCase}
    \end{subfigure}
    %\hfill
    \begin{subfigure}[]{0.37\textwidth}
        \includegraphics[width=\linewidth]{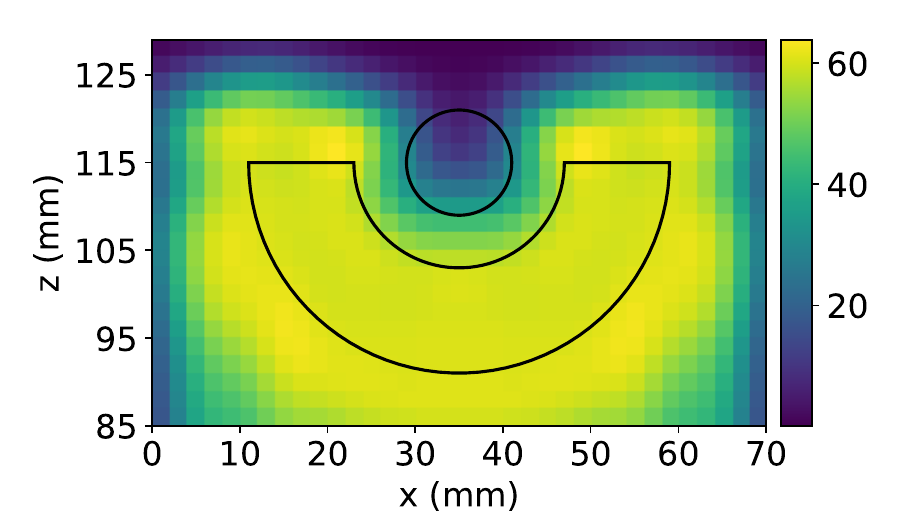}
        \caption{}
        \label{fig:spinal_nominal_prob95_98_95robCase}
    \end{subfigure}

    %\vspace{0.5cm}

    \begin{subfigure}[]{0.37\textwidth}
        \includegraphics[width=\linewidth]{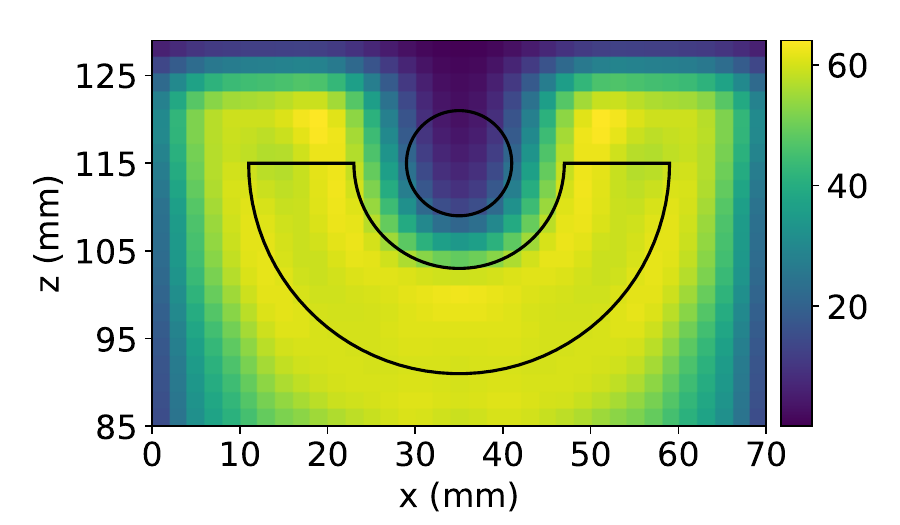}
        \caption{}
        \label{fig:spinal_nominal_prob95_98_98probCase}
    \end{subfigure}
    %\hfill
    \begin{subfigure}[]{0.37\textwidth}
        \includegraphics[width=\linewidth]{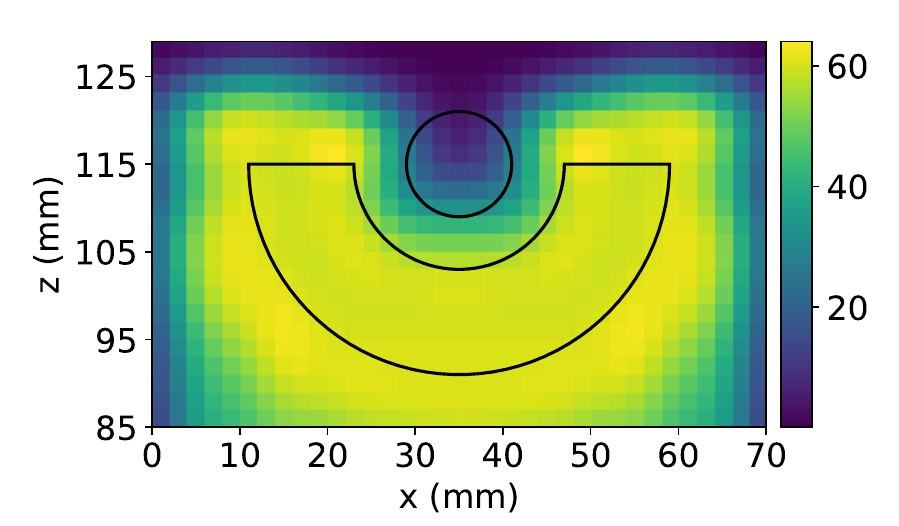}
        \caption{}
        \label{fig:spinal_nominal_prob95_98_98robCase}
    \end{subfigure}

    \caption{Comparison of the nominal dose distributions corresponding to the probabilistic (left) and robust (right) plans, for the $\nu = 95\%$ (top) and $\nu = 98\%$ (bottom) cases.}
    \label{fig:spinal_nominal_prob95_98}
\end{figure}

\section{Additional comparisons for the spinal case} 
\label{app:additionalComparisons_spinal}

In this appendix additional results are presented for the probabilistic spinal plans and the correspondingly matched robust plans (for $\nu = 95\%$ and $\nu = 98\%$; the $\nu = 90\%$ plans were compared in Section \ref{subsec:Results_spinal}). Plans are matched by their CTV coverage (and thus scaled by $D_{50\%}^{50\text{th}} = 100\% d^p$), resulting in the statistical DVH-metrics in Table \ref{tab:DVHmetrics_spinal}.

\begin{figure}[]
    \centering

    \begin{subfigure}[]{0.35\textwidth}
        \includegraphics[width=\linewidth]{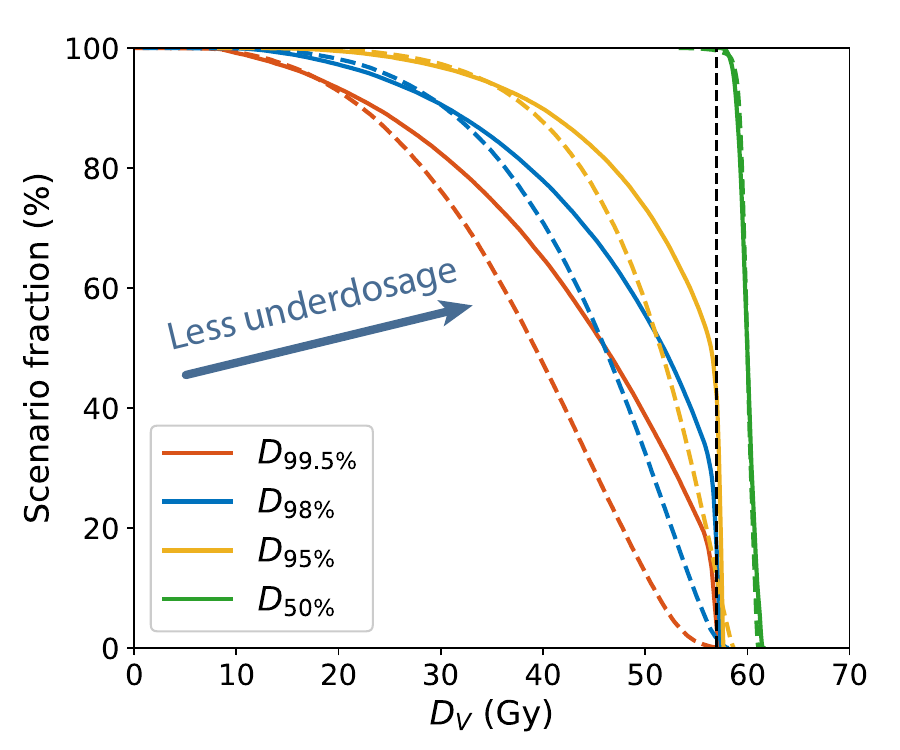}
        \caption{}
    \end{subfigure}
    %\hfill
    \begin{subfigure}[]{0.35\textwidth}
        \includegraphics[width=\linewidth]{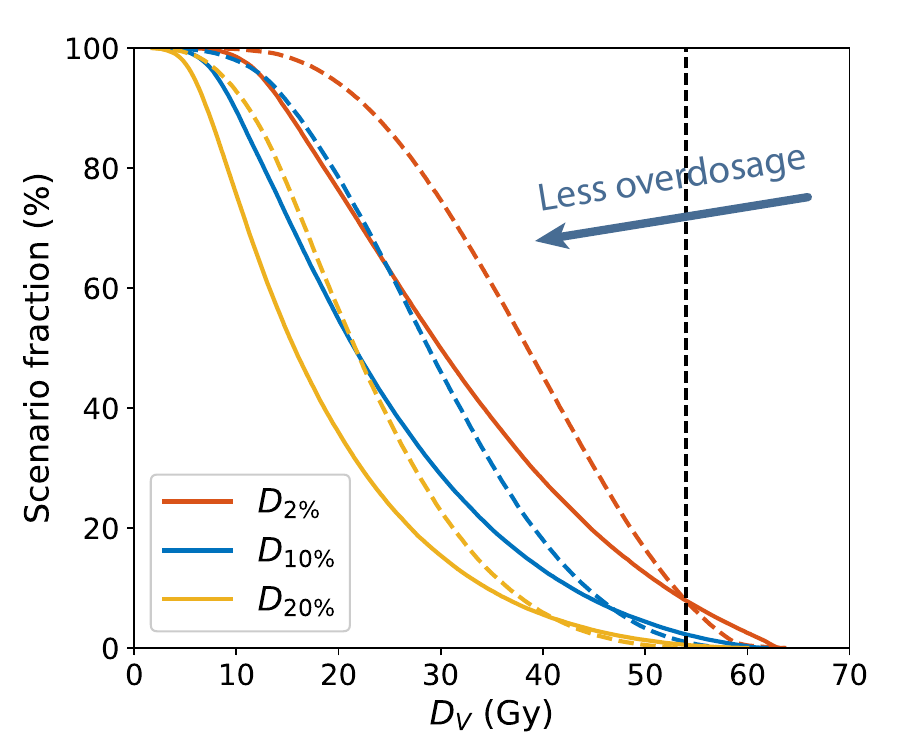}
        \caption{}
    \end{subfigure}

    %\vspace{0.5cm}

    \begin{subfigure}[]{0.35\textwidth}
        \includegraphics[width=\linewidth]{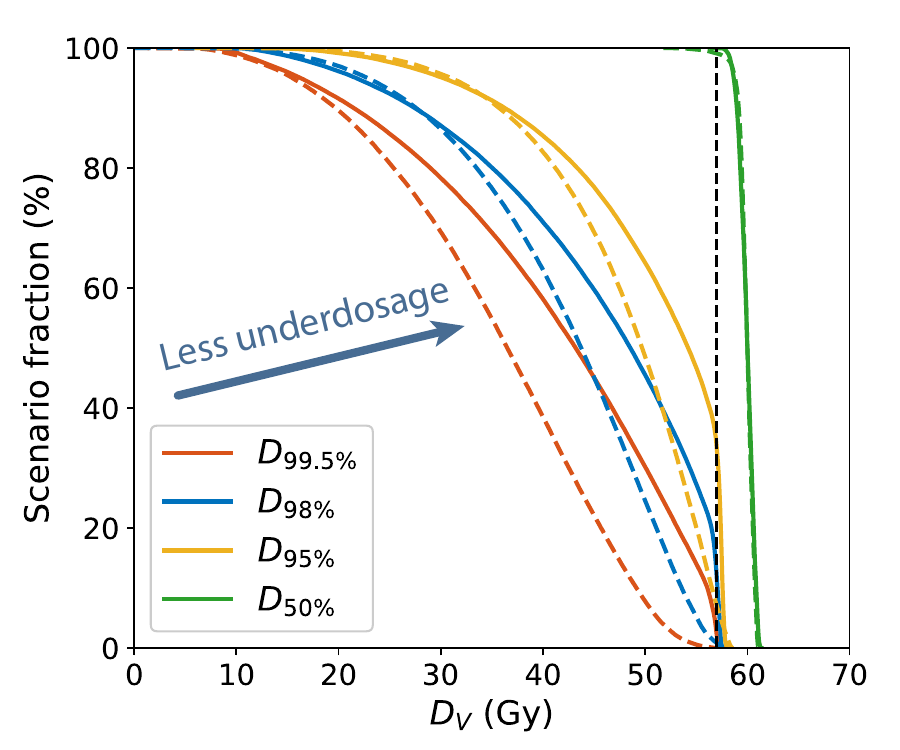}
        \caption{}
    \end{subfigure}
    %\hfill
    \begin{subfigure}[]{0.35\textwidth}
        \includegraphics[width=\linewidth]{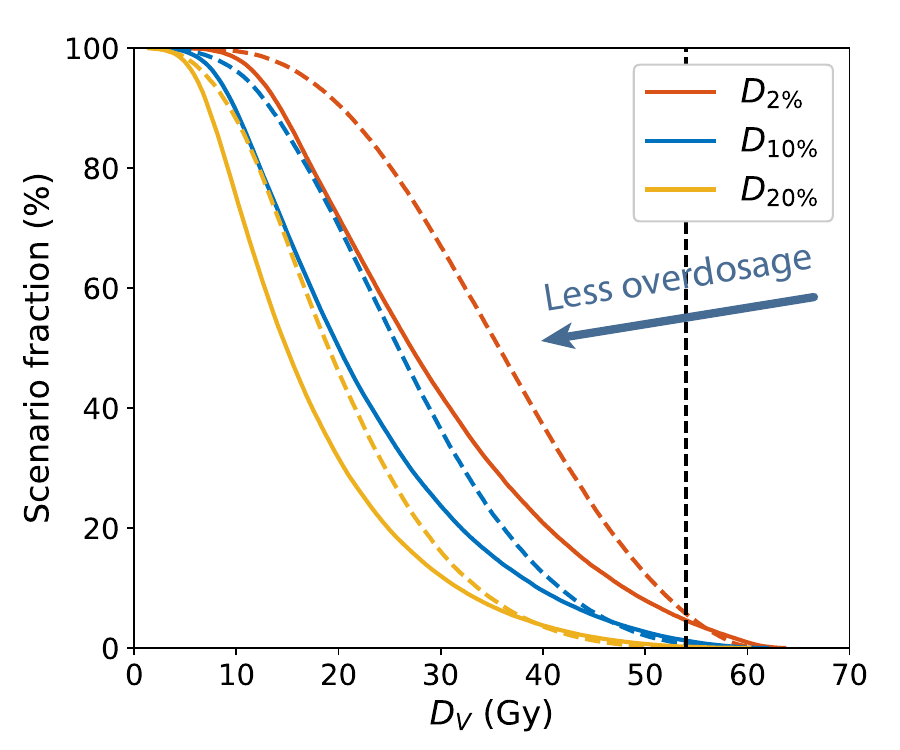}
        \caption{}
    \end{subfigure}

    \caption{Dose population histograms of various DVH metrics ($D_{V}$, e.g., $V=98\%$) comparing spinal probabilistic (solid) and robust (dashed) plans for (top) $\nu = 95\%$ and (bottom) $\nu = 98\%$, scaled by $D_{50\%}^{50\text{th}}$. The probabilistic plans show reduced spinal overdosage probability and reduced CTV underdosage probability in the same plan.}
    \label{fig:spinal_DPH_nu95_98}
\end{figure}

Nominal dose distributions of the $\nu = 95\%$ (top) and $\nu = 98\%$ (bottom) cases are shown in Figure \ref{fig:spinal_nominal_prob95_98}, for the probabilistic (left) and robust (right) plans. The corresponding DPHs are shown in Figure \ref{fig:spinal_DPH_nu95_98} for the $\nu = 95\%$ (top) and $\nu = 98\%$ (bottom) cases. The probabilistic plan reduces CTV underdosage probability and spinal overdosage probability in the same plan. Specifically, the probabilistic plan shows smaller probability of CTV underdosage probability (e.g., $P(D_{98\%} > \qty{57}{\gray})$ increased by about 14.5\% and by about 11\% in the $\nu = 95\%$ and $\nu = 98\%$, respectively), and spinal overdosage probability (e.g., $P(D_{98\%} > \qty{57}{\gray})$ increased by about 11\% in both the $\nu = 95\%$ and $\nu = 98\%$ plans) in the same plan.

% spinal v=95
%               Probabilistic       robust
% D98 at 57Gy:  16.1%               1.6%    (increase by about 14.5%)
% D2 at 30Gy:   49.9%               74.6%   (decrease by about 24.5%)

% spinal v=98
%               Probabilistic       robust
% D98 at 57Gy:  11.9%               0.82%   (increase by about 11%)
% D2 at 30Gy:   42.3%               67.0%   (decrease by about 24.5%)

\begin{figure}[h!]
    \centering
    
    % First row
    \begin{subfigure}{0.32\textwidth}
        \centering
        \includegraphics[width=\linewidth]{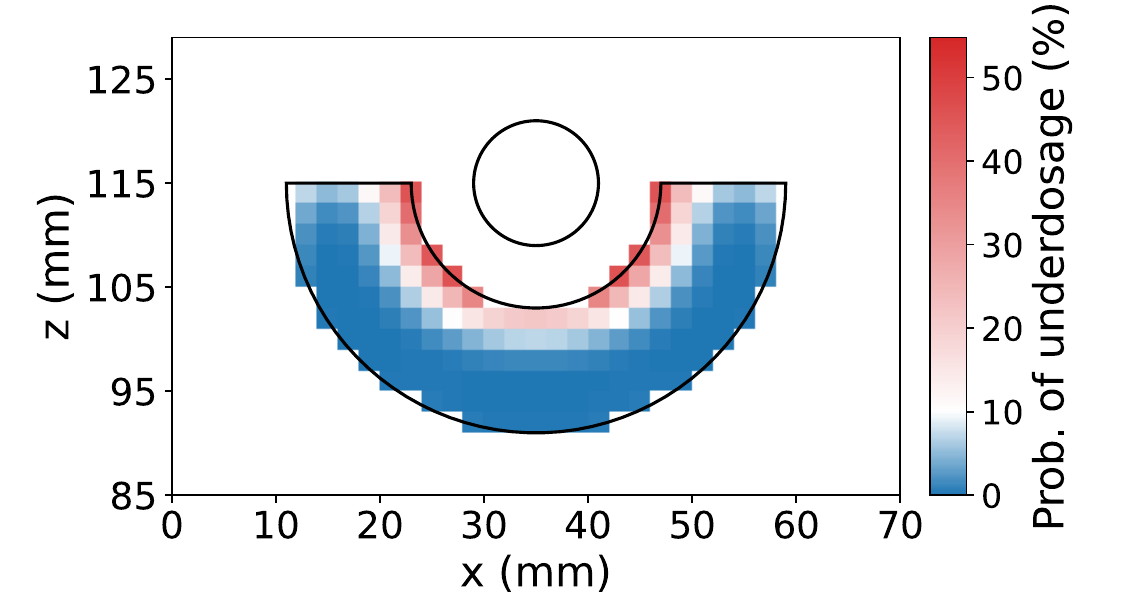}
        \caption{}
    \end{subfigure}
    %\hfill
    \begin{subfigure}{0.32\textwidth}
        \centering
        \includegraphics[width=\linewidth]{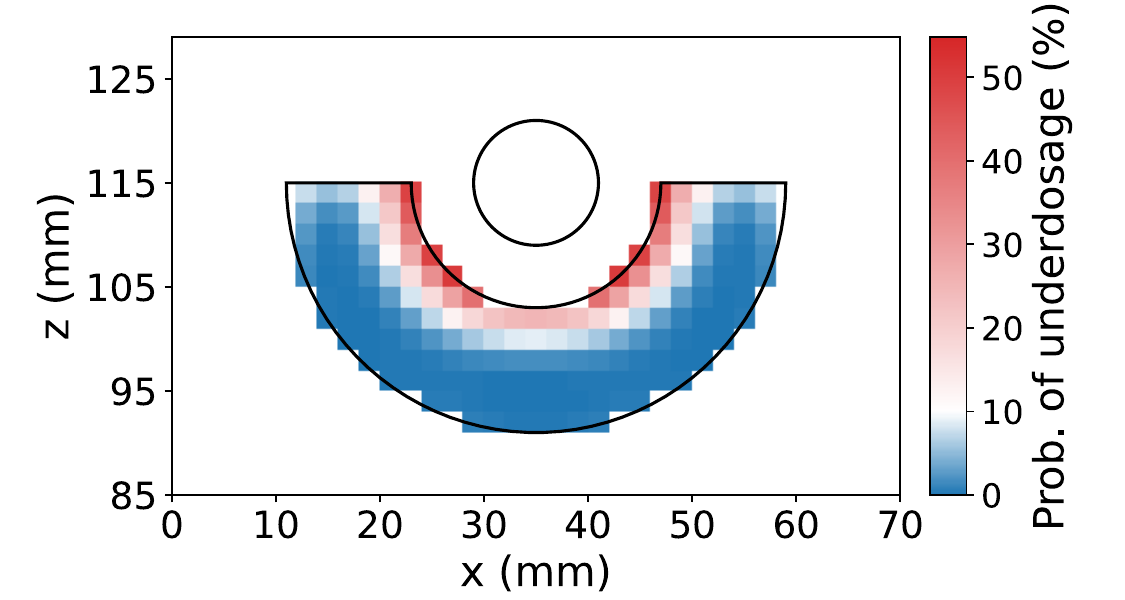}
        \caption{}
    \end{subfigure}
    %\hfill
    \begin{subfigure}{0.32\textwidth}
        \centering
        \includegraphics[width=\linewidth]{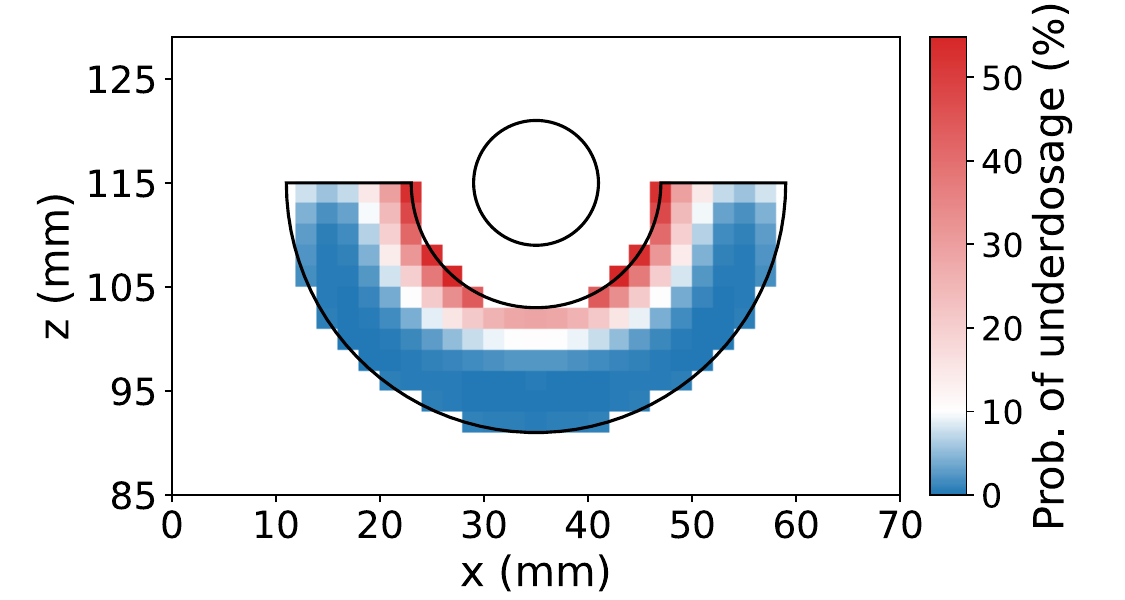}
        \caption{}
    \end{subfigure}

    %\vspace{1em}

    % Second row
    \begin{subfigure}{0.32\textwidth}
        \centering
        \includegraphics[width=\linewidth]{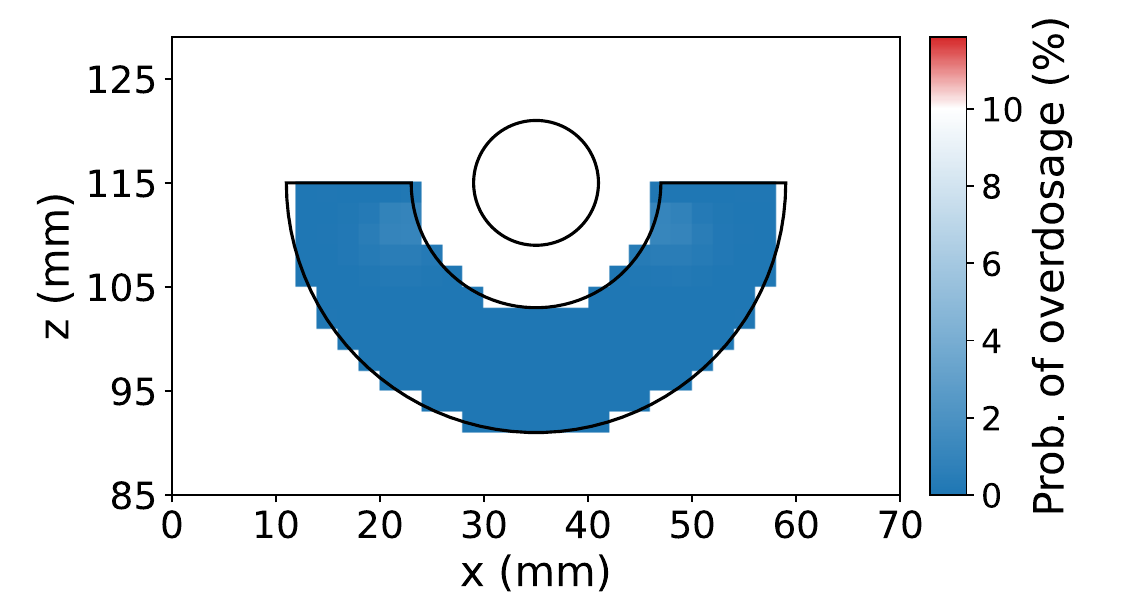}
        \caption{}
    \end{subfigure}
    %\hfill
    \begin{subfigure}{0.32\textwidth}
        \centering
        \includegraphics[width=\linewidth]{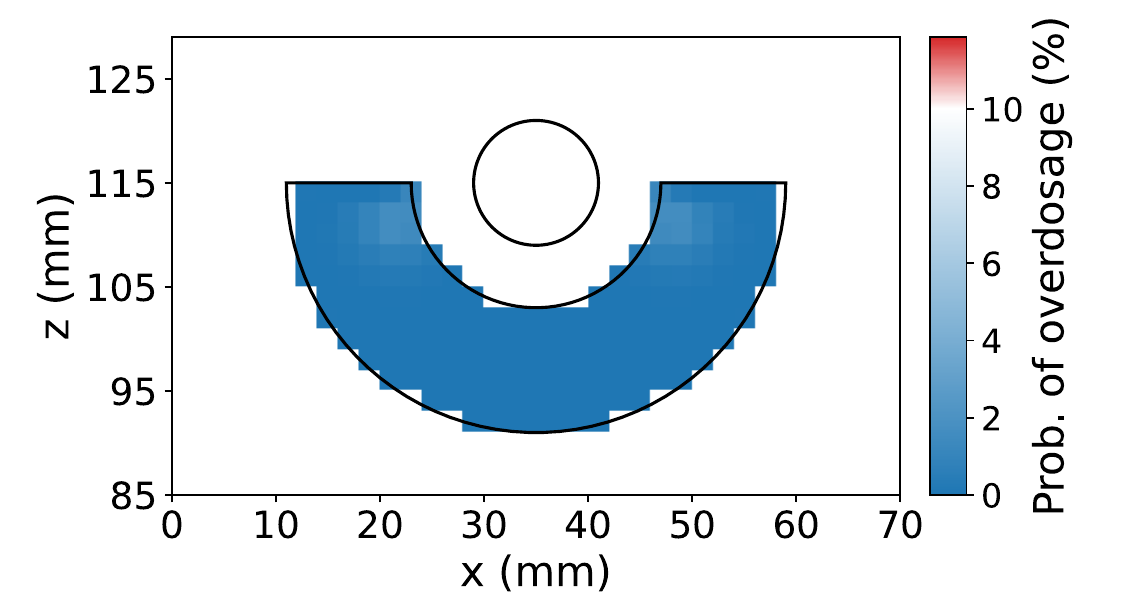}
        \caption{}
    \end{subfigure}
    %\hfill
    \begin{subfigure}{0.32\textwidth}
        \centering
        \includegraphics[width=\linewidth]{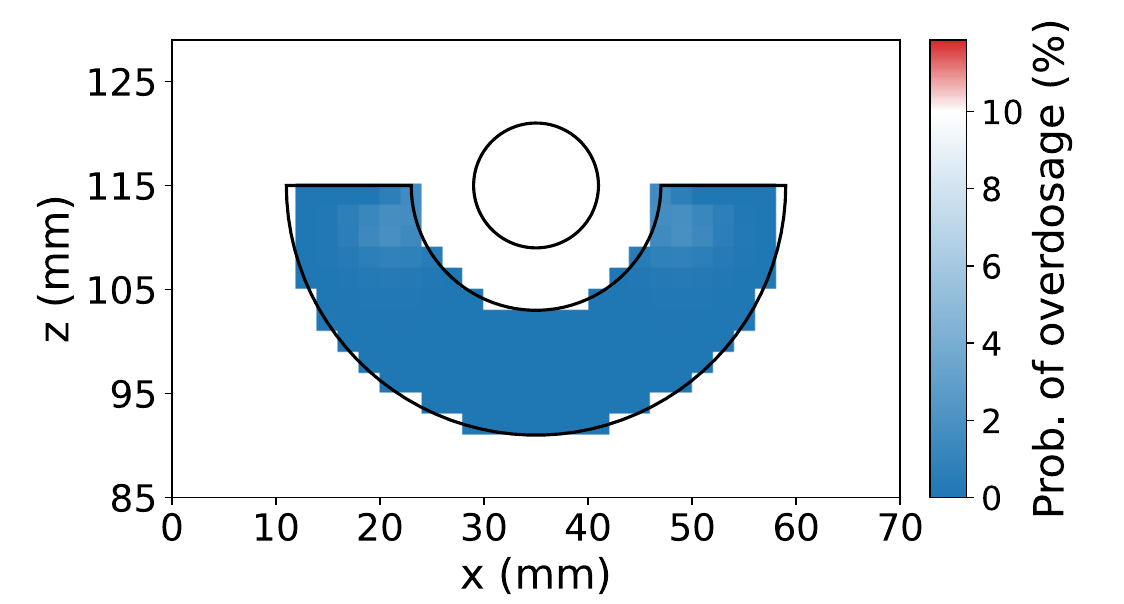}
        \caption{}
    \end{subfigure}

    %\vspace{1em}

    % Third row
    \begin{subfigure}{0.32\textwidth}
        \centering
        \includegraphics[width=\linewidth]{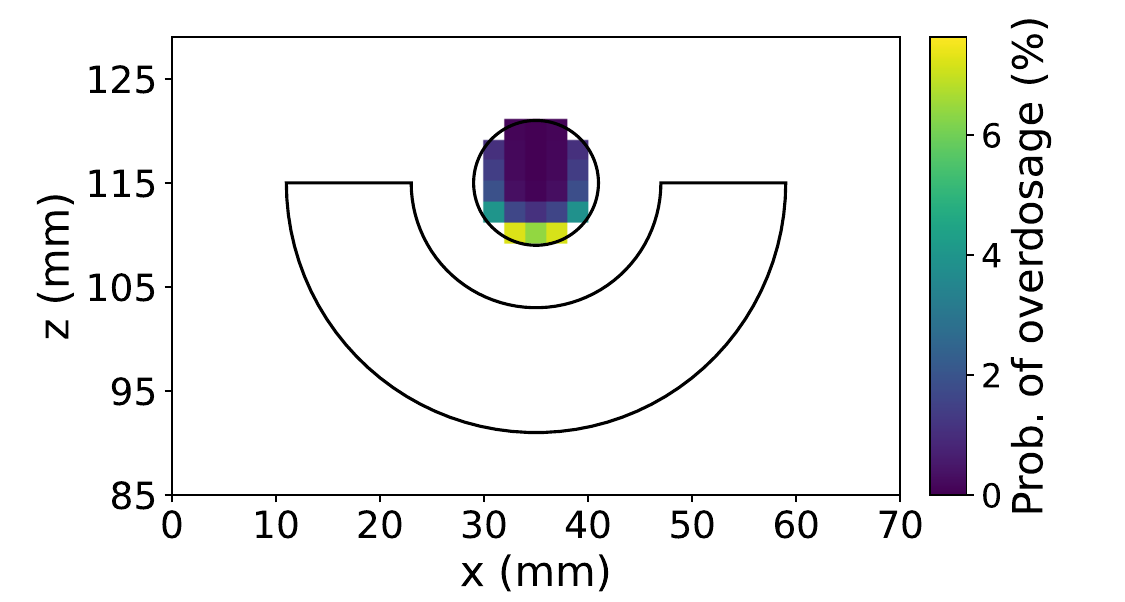}
        \caption{}
    \end{subfigure}
    %\hfill
    \begin{subfigure}{0.32\textwidth}
        \centering
        \includegraphics[width=\linewidth]{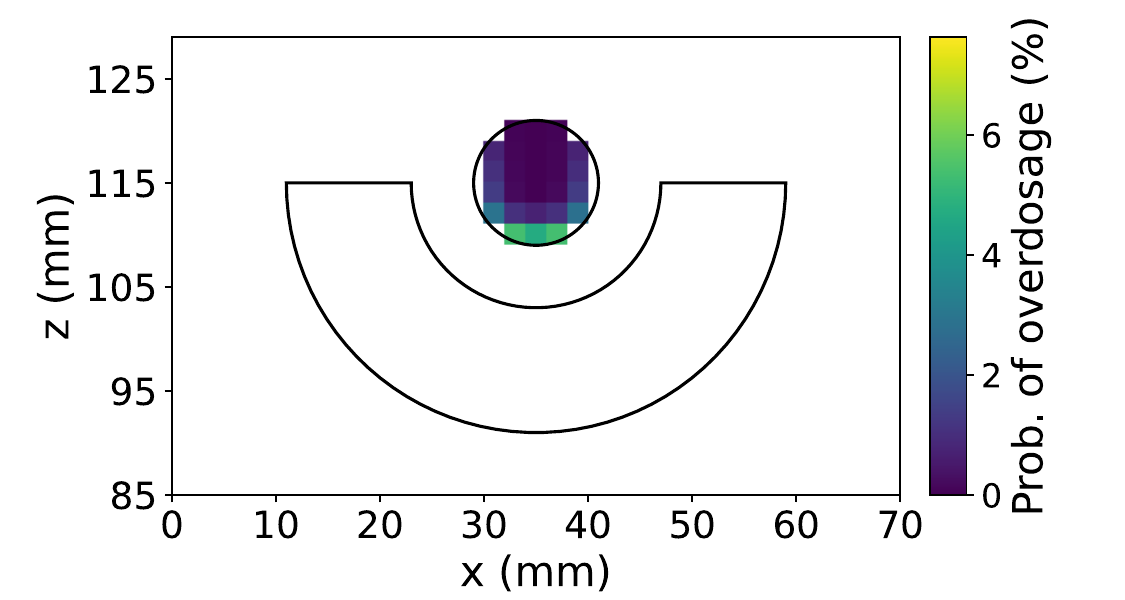}
        \caption{}
    \end{subfigure}
    %\hfill
    \begin{subfigure}{0.32\textwidth}
        \centering
        \includegraphics[width=\linewidth]{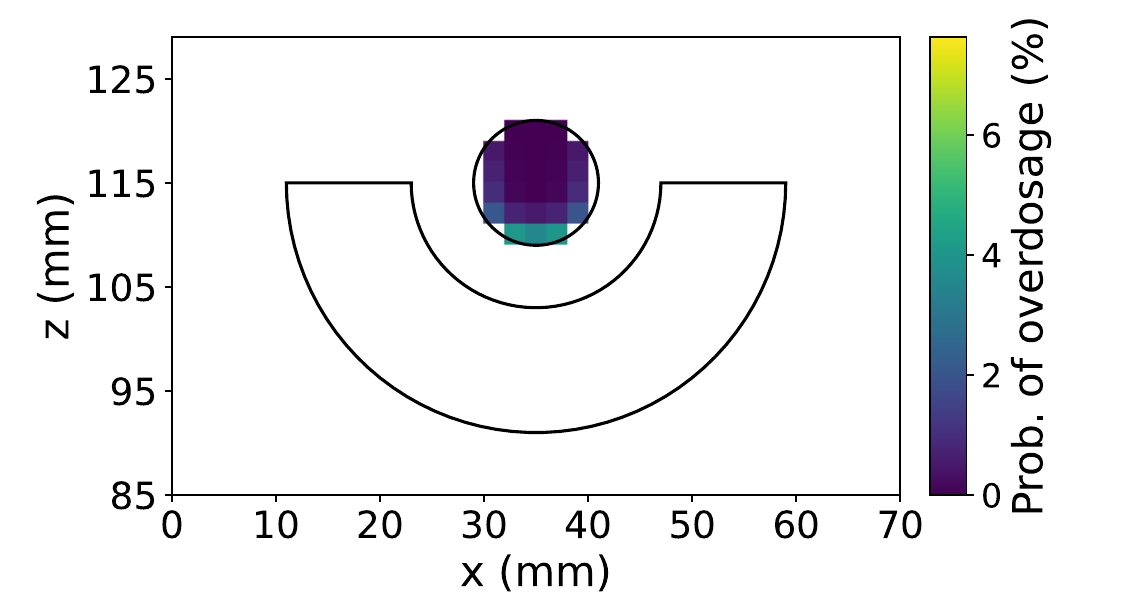}
        \caption{}
    \end{subfigure}

    \caption{The probability of CTV underdosage (top), CTV overdosage (middle) and spine overdosage (bottom) probability, for the robust plans, equivalent to the (left to right) $\nu = 90\%$, $\nu = 95\%$ and $\nu = 98\%$ cases.}
    \label{fig:spinal_probUnderOverDosage_robust}
\end{figure}

%%% Comparing under/overdosage between probabilistic and robust %%%
% prob under/overdosage prob versus robust comparison (prob10)
In Figure \ref{fig:spinal_probUnderOverDosage_robust}, we show the probabilities of CTV under- and overdosage and spinal overdosage probabilities of the robust treatment plans that are equivalent to the $\nu = 90\%$ (left), $\nu = 95\%$ (middle) and $\nu = 98\%$ (right) cases. These results can be compared to the outcomes of the probabilistic plans in Figure \ref{fig:spinal_probUnderOverDosage}), where the color-bars are identically scaled.

For the $\nu = 90\%$ case, the probability of CTV underdosage is reduced in the probabilistic plan, at the expense of having a slight overdosage and inhomogeneity in (part of) the CTV. Spinal overdosage is similar in the probabilistic and robust plan. Results for the $\nu = 95\%$ case are very similar to the $\nu = 90\%$ case. However, the CTV underdosage probabilities are larger compared to the $\nu = 90\%$ case, because the dose extension at the spinal side (see Figure \ref{fig:nominalDose_spinal_crossSections}) is reduced. As the dose build-up is less enhanced (i.e., the maximum nominal dose is smaller), the CTV overdosage probability is reduced as well. In the $\nu = 98\%$ plans, CTV under- and overdosage probability becomes even more similar between probabilistic and robust plans, while the probabilistic plan has about 1.5\% - 1.8\% reduction in spinal overdosage probability. \\

%%%%% EXTRA RESULTS ON v95 and v98 DPHs %%%%%

\begin{figure}[h!]
    \centering

    \begin{subfigure}{0.42\textwidth}
        \centering
        \includegraphics[width=\linewidth]{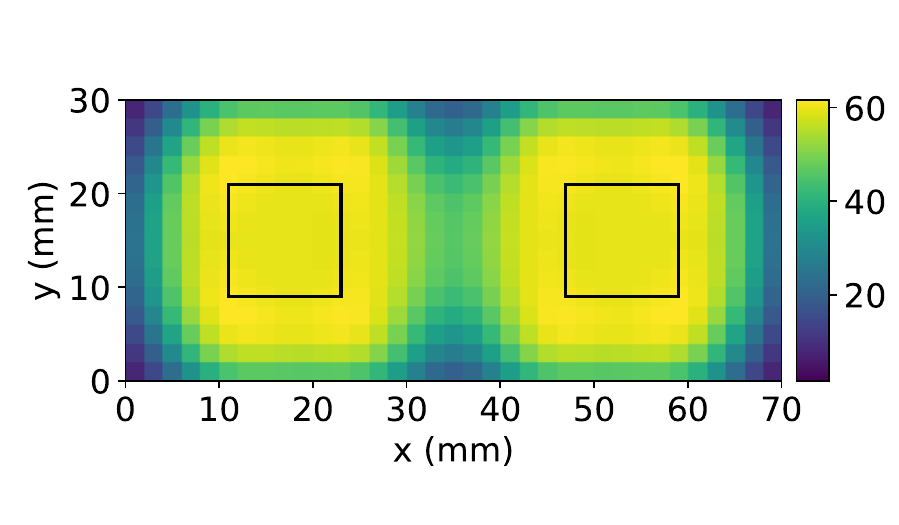}
        \caption{}
        \label{fig:left}
    \end{subfigure}
    %\hfill
    \begin{subfigure}{0.42\textwidth}
        \centering
        \includegraphics[width=\linewidth]{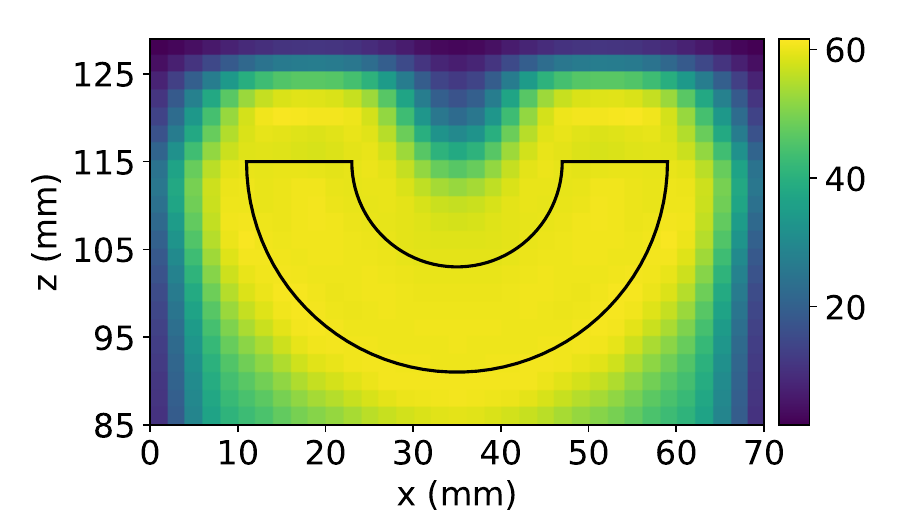}
        \caption{}
        \label{fig:right}
    \end{subfigure}

    \caption{Nominal dose distribution for probabilistically optimizing the horseshoe-shaped CTV for underdosage (of $P(d_i < \qty{57}{\gray}) < 2\%$) and overdosage (of $P(d_i > \qty{64.2}{\gray}) < 10\%$) probabilities, showing the (a) XY-plane and (b) XZ-plane through the CTV center, yielding $D_{98\%}^{10\text{th}} = 96\% d^p$.}
    \label{fig:horseshoe_CTVonly}
\end{figure}

In all spinal probabilistic optimizations, reducing spinal overdosage probability was prioritized. As a result, significant CTV underdosage probability was seen for all probabilistic and robust plans (see Figure \ref{fig:spinal_probUnderOverDosage} and Figure \ref{fig:spinal_probUnderOverDosage_robust}). For completeness, Figure \ref{fig:horseshoe_CTVonly} shows the probabilistic optimization of the horseshoe-shaped CTV without spine, optimizing for $P(d_i < \qty{57}{\gray}) \leq 2\%$ and $P(d_i > \qty{64.2}{\gray}) \leq 10\%$ with $\{ \pi_{CTV}^{\alpha}, \pi_{CTV}^{\beta}, \pi_{CTV}^{low}, \pi_{Tissue} \} = \{ 15, 15, 1, 1 \}$, resulting in (the clinically robust) $D_{98\%}^{10\text{th}} = 96\% d^p$. Convergence criteria were identical to the $\nu = 98\%$ case (see Table \ref{tab:probOptParameters}). This example is included to stress that the CTV underdosage probability that was seen in the spinal comparisons are due to the choice of probabilities and thresholds.

\begin{figure}[]
    \centering
    \includegraphics[width=\textwidth]{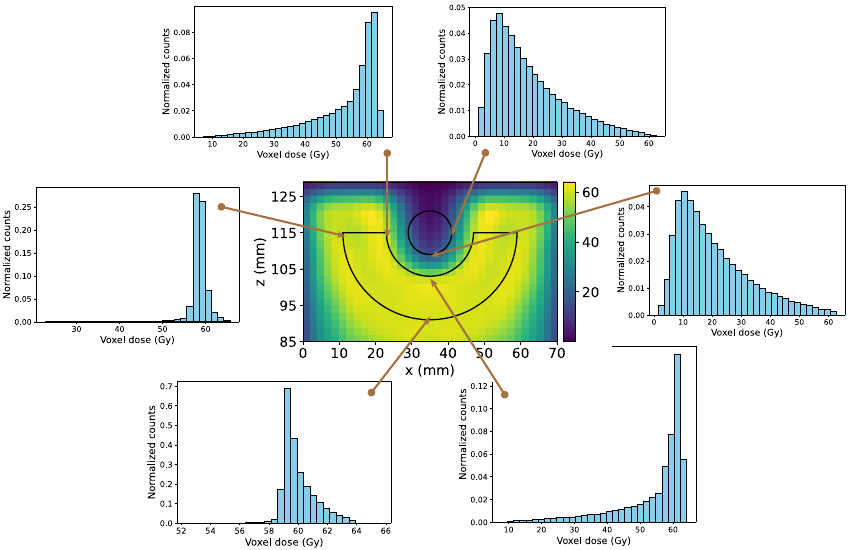}
    \caption{Probability density functions of some representative CTV and spinal voxels for the $\nu = 98\%$ probabilistic plan. The nominal plan is shown in the middle (identical to Figure \ref{fig:spinal_nominal_prob95_98_98probCase}).}
    \label{fig:prob2_PDFs}
\end{figure}

For the $\nu = 98\%$ probabilistic plan, we show PDFs of the voxel dose for some representative CTV and spinal voxels in Figure \ref{fig:prob2_PDFs}. Voxel dose PDFs at the outer parts of the CTV are narrower because of the larger dose extension, whereas PDFs of CTV voxels on the spinal side (as well as spinal voxels) have broader distributions. In general, PDFs are non-Gaussian and have long tails. \\

\clearpage

\section{Gradient and Hessian of the probabilistic objective} \label{app:GradientAndHessian}

In the following the gradient and Hessian of the objective function in Equation \ref{eq:objective} is derived for $f_{CTV}^{\alpha,\gamma}(\bi{x})$ only, as the other percentile objectives ($f_{CTV}^{\beta,\epsilon}(\bi{x})$ and $f_{OAR}^{\nu,\mu}(\bi{x})$) have a  similar form. For reference, $f_{CTV}^{\alpha,\gamma}(\bi{x})$ is repeated here as
\begin{equation} \label{eq:fCTV_repeat}
    f_{CTV}^{\alpha,\gamma}(\bi{x}) = \frac{1}{N_{CTV}} \sum_{i \in CTV} w_i^{CTV} \bigl[ \gamma_i - d_i^{\alpha \%}(\bi{x}) \bigr]_+^2.
\end{equation}
Its gradient and Hessian with respect to pencil-beam spot $p$ and $q$ are given by
\begin{align}
    \frac{\partial f_{CTV}^{\alpha,\gamma}(\bi{x})}{\partial x_q} & = \frac{-2}{N_{CTV}} \sum_{i \in CTV} w_i^{CTV} \bigl[ \gamma_i - d_i^{\alpha \%}(\bi{x}) \bigr]_+ \cdot \frac{\partial d_i^{\alpha \%}(\bi{x})}{\partial x_q}, \nonumber \\
    \frac{\partial^2 f_{CTV}^{\alpha,\gamma}(\bi{x})}{\partial x_p \partial x_q} & = \frac{2}{N_{CTV}} \sum_{i \in CTV} w_i^{CTV} \bigl[ \gamma_i - d_i^{\alpha \%}(\bi{x}) \bigr]_+ \cdot \Biggl[ \frac{\partial d_i^{\alpha \%}(\bi{x})}{\partial x_p} \frac{\partial d_i^{\alpha \%}(\bi{x})}{\partial x_q} - \frac{\partial^2 d_i^{\alpha \%}(\bi{x})}{\partial x_p \partial x_q} \Biggr], \nonumber
\end{align}
where
\begin{align}
    \frac{\partial d_i^{\alpha \%}(\bi{x})}{\partial x_q} & = \frac{\partial}{\partial x_q} \biggl[ \mathbb{E}[d_i(\bi{x},\bxi)] - \delta_i^{\alpha\%} SD[d_i(\bi{x},\bxi)] \biggr] \nonumber \\
    & = \frac{\partial \mathbb{E}[d_i(\bi{x},\bxi)]}{\partial x_q}  - \frac{\delta_i^{\alpha\%}}{2 SD[d_i(\bi{x},\bxi)]} \frac{\partial \text{Var}[d_i(\bi{x},\bxi))]}{\partial x_q}, \nonumber \\
    \frac{\partial^2 d_i^{\alpha \%}(\bi{x})}{\partial x_p \partial x_q} & = \frac{- \delta_i^{\alpha\%}}{2 SD[d_i(\bi{x},\bxi)]} \frac{\partial^2 \text{Var}[d_i(\bi{x},\bxi))]}{\partial x_p \partial x_q} + \nonumber \\
    & \quad \frac{\delta_i^{\alpha\%}}{4 SD[d_i(\bi{x},\bxi))]^3} \frac{\partial \text{Var}[d_i(\bi{x},\bxi))]}{\partial x_p} \frac{\partial \text{Var}[d_i(\bi{x},\bxi))]}{\partial x_q}, \nonumber
\end{align}
and
\begin{align}
    \frac{\partial \mathbb{E}[d_i(\bi{x},\bxi))]}{\partial x_q} & = \mathbb{E}[D_{iq}(\bxi)] = \int D_{iq}(\bxi) p(\bxi) d\bxi, \nonumber \\
    \frac{\partial \text{Var}[d_i(\bi{x},\bxi))]}{\partial x_q} & = \sum_{j \in \mathbb{B}} x_j \bigl( \mathbb{E}[D_{ij}(\bxi)D_{iq}(\bxi)] - \mathbb{E}[D_{ij}(\bxi)] \mathbb{E}[D_{iq}(\bxi)] \bigr), \nonumber \\
    \frac{\partial^2 \text{Var}[d_i(\bi{x},\bxi))]}{\partial x_p \partial x_q} & = \bigl( \mathbb{E}[D_{ip}(\bxi)D_{iq}(\bxi)] - \mathbb{E}[D_{ip}(\bxi)] \mathbb{E}[D_{iq}(\bxi)] \bigr), \nonumber \\
    \mathbb{E}[D_{ij}(\bxi) D_{ij'}(\bxi)] & = \int D_{ij}(\bxi) D_{ij'}(\bxi) p(\bxi) d\bxi. \nonumber
\end{align}
$\mathbb{E}[D_{ij}(\bxi) D_{ij'}(\bxi)]$ is memory heavy, but is symmetric so that it has at most $N_b + (N_b^2 - N_b)/2$ unique elements for voxel $i$. The objective in Equation \ref{eq:fCTV_repeat} only optimizes for CTV voxels that have $d_i^{\alpha \%}(\bi{x}) < \gamma_i$. As $d_i^{\alpha \%}(\bi{x})$ depends on the beam weights, the active voxel set is different for every iteration. Therefore, summation over the voxels must be done for every iteration, where we determine $\bigl[ \gamma_i - d_i^{\alpha \%}(\bi{x}) \bigr]_+$.

\end{document}